\documentclass[12pt,american,english]{article}
\usepackage{libertine}
\usepackage[T1]{fontenc}
\usepackage[latin9, utf8]{inputenc}
\usepackage{color}
\definecolor{lyxnotefontcolor}{rgb}{0.792969, 0.105469, 0.253906}
\usepackage{babel}
\usepackage{float}
\usepackage{units}
\usepackage{amsmath}
\usepackage{amsthm}
\usepackage{amssymb}
\usepackage{graphicx}
\usepackage{geometry}
\usepackage{setspace}
\usepackage{esint}
\usepackage[authoryear]{natbib}
\usepackage[pdfusetitle,
 bookmarks=true,bookmarksnumbered=false,bookmarksopen=false,
 breaklinks=false,pdfborder={0 0 1},backref=false,colorlinks=false]
 {hyperref}
\hypersetup{
 colorlinks,citecolor=darkblue,linkcolor=maroon,urlcolor=maroon,bookmarks=true,backref=page}

\makeatletter
\usepackage{tikz}
\usetikzlibrary{calc}
\usetikzlibrary{arrows, arrows.meta} 
\usetikzlibrary{patterns}

\usepackage{pgfplots}

\usepackage{color}
\definecolor{darkblue}{rgb}{0.0,0,.6}
\definecolor{lightblue}{rgb}{.0,0.1,1}
\definecolor{maroon}{rgb}{0.68,0,0}
\definecolor{darkgreen}{rgb}{0,0.369,0.086}
\definecolor{gray}{rgb}{.5,.5,.5}

\usepackage{titletoc}
\usepackage[page,header]{appendix}
\usepackage{bbm}
\usepackage{mathtools}

\newcommand{\E}{\mathbb{E}}

\ifdefined\showcaptionsetup
 \PassOptionsToPackage{caption=false}{subfig}
\fi
\usepackage{subfig}
\makeatother

\theoremstyle{plain}
\newtheorem{lem}{\protect\lemmaname}
\theoremstyle{remark}
\newtheorem{rem}{\protect\remarkname}
\theoremstyle{definition}
\newtheorem{defn}{\protect\definitionname}
\theoremstyle{plain}
\newtheorem{thm}{\protect\theoremname}
\newtheorem{cor}{\protect\corollaryname}
\theoremstyle{definition}
\newtheorem*{example*}{\protect\examplename}
\theoremstyle{plain}
\newtheorem{prop}{\protect\propositionname}
\newtheorem{assumption}{\protect\assumptionname}
\theoremstyle{definition}
\newtheorem{example}{\protect\examplename}
\addto\captionsamerican{\renewcommand{\assumptionname}{Assumption}}
\addto\captionsamerican{\renewcommand{\corollaryname}{Corollary}}
\addto\captionsamerican{\renewcommand{\definitionname}{Definition}}
\addto\captionsamerican{\renewcommand{\examplename}{Example}}
\addto\captionsamerican{\renewcommand{\lemmaname}{Lemma}}
\addto\captionsamerican{\renewcommand{\propositionname}{Proposition}}
\addto\captionsamerican{\renewcommand{\remarkname}{Remark}}
\addto\captionsamerican{\renewcommand{\theoremname}{Theorem}}
\addto\captionsenglish{\renewcommand{\assumptionname}{Assumption}}
\addto\captionsenglish{\renewcommand{\corollaryname}{Corollary}}
\addto\captionsenglish{\renewcommand{\definitionname}{Definition}}
\addto\captionsenglish{\renewcommand{\examplename}{Example}}
\addto\captionsenglish{\renewcommand{\lemmaname}{Lemma}}
\addto\captionsenglish{\renewcommand{\propositionname}{Proposition}}
\addto\captionsenglish{\renewcommand{\remarkname}{Remark}}
\addto\captionsenglish{\renewcommand{\theoremname}{Theorem}}
\providecommand{\assumptionname}{Assumption}
\providecommand{\corollaryname}{Corollary}
\providecommand{\definitionname}{Definition}
\providecommand{\examplename}{Example}
\providecommand{\lemmaname}{Lemma}
\providecommand{\propositionname}{Proposition}
\providecommand{\remarkname}{Remark}
\providecommand{\theoremname}{Theorem}

\begin{document}
\title{Divide and Confer: Aggregating Information without Verification}
\author{James Best, Daniel Quigley, Maryam Saeedi, Ali Shourideh\thanks{Best: Carnegie Mellon University Tepper School of Business (email:
\protect\href{mailto:jabest@andrew.cmu.edu}{jabest@andrew.cmu.edu}).
Quigley: Exeter College and Department of Economics, University of
Oxford (email: \protect\href{mailto:daniel.quigley@economics.ox.ac.uk}{daniel.quigley@economics.ox.ac.uk}).
Saeedi: Carnegie Mellon University Tepper School of Business (email:
\protect\href{mailto:msaeedi@andrew.cmu.edu}{msaeedi@andrew.cmu.edu}).
Shourideh: Carnegie Mellon University Tepper School of Business (email:
\protect\href{mailto:ashourid@andrew.cmu.edu}{ashourid@andrew.cmu.edu}).
We thank Nemanja Antic, Ian Ball, Marco Battaglini, Ben Brooks, Raphael
Boleslavsky, Gorkem Celik, Archishman Chakraborty, Yeon-Koo Che, Francois
Forges, Drew Fudenberg, Marina Halac, Johannes Hörner, Rick Harbaugh,
Ian Jewitt, Emir Kamenica, Elliot Lipnowski, Meg Meyer, Dimitri Migrow,
Paula Onuchic, Wolfgang Pesendorfer, Anna Sanktjohanser, Vasiliki
Skreta, Joel Sobel, Ina Taneva, Alex Wolitzky, Andriy Zapechelnyuk,
and audiences at various seminars and conferences, for their feedback.}}
\maketitle
\begin{abstract}
We study mechanisms for aggregating information divided across a large
population of biased senders. Each sender privately observes an unconditionally
independent signal about an unknown state, so no sender's report can
be verified against another\textquoteright s. A receiver makes a binary
accept/reject decision whose payoffs depend on the state. Even though
cross-verification is impossible, we show the receiver can benefit
from informational division. We introduce a novel \emph{incentive-compatibility-in-the-large}
approach that studies optimal design via the large-population limit.
For fixed population size, optimal mechanisms are in general complex.
However, we show that in the limit they converge to a simple mechanism
that depends only on the payoff from acceptance, and punishes excessive
consensus in the direction of the common bias. These surplus burning
punishments yield payoffs bounded away from the first best; the resulting
inefficiency demonstrates how our concept of informational division
is distinct from standard models of information in large populations.
\end{abstract}

\section{Introduction}

Decision makers often rely on information that is divided across individuals
with whom they have a conflict of interest. A familiar small-scale
example is a CEO evaluating a new product: he must elicit information
divided across the marketing manager (demand) and the production manager
(costs), who have different incentives from the CEO. This type of
elicitation problem is increasingly important in environments where
information is dispersed across large populations. For example, governments
use polling to assess the effectiveness of policies that depend on
local knowledge spread throughout large populations. Similarly, e-commerce
and social media platforms often need to aggregate information about
consumer products, political news, and opinions from a large number
of individuals whose interests differ from those of the platform or
society at large.

A standard solution is verification: comparing a source's report to
those of others and punishing them collectively in case of disagreement.
This strategy is especially effective when agents' information sets
overlap to such an extent that each individual's information is largely
dispensable.\footnote{For instance, \citet{krishna2001model} show how to extract information
perfectly when two senders have the same information. \citet{gerardi2009aggregation}
show a similar result with large populations. } However, in the aforementioned examples, it is not at all clear that
individual's information sets do overlap sufficiently to make cross-verification
effective. Indeed, when information is divided so each person has
some unique piece of indispensable information, it is impossible to
verify one person\textquoteright s report against that of another.
This raises our main question: how should a decision maker elicit
information when information is divided across a large population
and cross-verification is impossible?

To answer this question, we study the design of information aggregation
rules in a multi-sender cheap talk environment. A receiver faces a
binary decision: ``accept'' or ``reject''. While the payoff to
reject is known, all players' payoffs from accept increase in a stochastic
state variable. There is a population of $N$ senders, each observing
a private signal about the state, and sharing a common bias toward
acceptance. Signals are i.i.d. across senders and the state is the
sum of the signals divided by $\sqrt{N}$. This information structure
has two critical features for our analysis. First, information is
unverifiable: because the signals are unconditionally independent,
no sender's report can be verified against another's. Second, it captures
informational division: the collective private information (as measured
by the variance of the state variable) stays constant in $N$, while
the information held by each sender shrinks as $N$ grows. Our goal
is to characterize the transfer-free recommendation mechanism that
maximizes the receiver's payoff when the population is arbitrarily
large.

When there is no informational division, i.e., $N=1$, then the only
informative incentive compatible mechanism is the sender-preferred
mechanism which effectively delegates the decision to the sender.
Since he knows the state and is the only person reporting, his report
will maximize the probability of acceptance whenever he wants it,
and minimize it otherwise. This has a cost to the receiver: there
exists an intermediate region of states---the disagreement region---at
which the receiver would prefer reject but the mechanism recommends
accept.

Yet, even if information is divided among only two senders, we can
use one sender's report to discipline the other. An example in Section
\ref{sec:ternaryintuition} illustrates this point: a \emph{simple
mechanism}---one whose recommendations depend only on the reported
payoff-relevant state---can strictly improve the receiver's payoff.
This mechanism does not recommend accept in the disagreement region,
but has a reduced probability of acceptance for the higher values
of the state. The reduced acceptance at the top acts a ``punishment''
that deters a sender from over-reporting out of fear that this will
cause a rejection when the other sender's report is also high. While
this simple mechanism improves the receiver's payoff, the optimal
mechanism is more complex: it uses the full profile of reports, beyond
the payoff-relevant state, to screen each sender more effectively
according to his private information.

The focus of this paper is on aggregating information when it is divided
among a large population. As $N\rightarrow\infty$,\foreignlanguage{american}{
each sender becomes essentially uninformed and the impact of their
reports becomes negligible.} To characterize the asymptotics of incentive
compatibility, we introduce the concept of \textit{incentive compatibility
in the large} (ICL), which extends incentive compatibility continuously
to $N=\infty$. We say a mechanism is ICL if its induced distribution
over outcomes at $N=\infty$ can be approximated arbitrarily well
by a sequence of finite-$N$ incentive compatible mechanisms. Our
main technical result (Theorem \ref{thm:A-recommendation-mechanism})
characterizes the set of ICL mechanisms in terms of two finite sets
of linear constraints. The first treats each sender as if he were
uninformed but still able to have a small impact on outcomes via his
report; the second is a monotonicity condition that ensures the probability
of acceptance is weakly increasing in reports. Together, these constraints
characterize the set of mechanisms that can be approximated by incentive
compatible mechanisms for large enough $N$. Thus, they serve as the
limiting analogues of the familiar envelope formulation of incentive
compatibility and interim monotonicity constraints.

Theorem \ref{thm:The-optimum-is} applies this result to show that
the optimal ICL mechanism is a type of simple mechanism, which we
call an \emph{interval mechanism} (depicted in Figure \ref{fig:Optimal-mechanism-in}).
It recommends accept for an intermediate range of state values while
rejecting at both low and extremely high values. This interval mechanism
ensures incentive compatibility by trading off two scenarios in which
a sender's upward lie is pivotal: at the lower threshold, lying induces
acceptance when payoffs are low; at the upper threshold, lying induces
rejection when payoffs are high.

\begin{figure}[h]
\centering
\centering{}\begin{tikzpicture}

    \begin{axis}[
    axis lines=none,
    ticks=none,
	xmin= -5,
	xmax=5,
    samples=100,
    domain=-4:4,
    clip=false,
    width=12cm,
    height=5cm
  ]

    \addplot[blue!76!black, very thick, smooth, dashed] {1/sqrt(2*pi) * exp(-(x)^2/2)};
  \end{axis}

  \pgfmathsetmacro{\d}{5.3};
  \draw[stealth-stealth] (-5+\d,0) -- (5+\d,0) node[below] {\footnotesize receiver's payoff};

  \draw[ultra thick, color =red!85!black ] (-4.7+\d,0) -- (-2.75+\d,0)  node[midway, above, yshift=6] {$a=0$}; 
  \draw[ultra thick, color =red!85!black, dashed ] (-2.75+\d,0) -- (-2.75+\d,1.5);

  \draw[ultra thick, color =red!85!black ] (-2.75+\d,1.5) -- (3.5+\d,1.5) node[midway, above,xshift=-1.5] {$a=1$};
  \draw[ultra thick, color =red!85!black, dashed ] (3.5+\d,1.5) -- (3.5+\d,0);
  \draw[ultra thick, color =red!85!black ] (3.5+\d,0) -- (4.7+\d,0) node[midway, above, yshift=6] {$a=0$}; 

  \draw[thick] (-3.5+\d,0.05) -- (-3.5+\d,-.05);
  \draw[thick] (-1+\d,0.05) -- (-1+\d,-.05);

  \draw[<->] (-5+\d,-1.5) -- (-3.52+\d,-1.5) node[midway, above]{\footnotesize reject agreement};
  \draw[<->] (-3.48+\d,-1.5) -- (-1.02+\d,-1.5) node[midway, below]{\footnotesize disagreement};
  \draw[<->] (-.98+\d,-1.5) -- (5+\d,-1.5) node[midway, above]{\footnotesize accept agreement};
\end{tikzpicture}\caption{{\footnotesize Optimal mechanism in the large economy. In the disagreement
region, the preferred action of the senders is accept, and the receiver
prefers reject.}{\footnotesize\textit{\label{fig:Optimal-mechanism-in}}}{\footnotesize{}
The dashed blue line represents the distribution of the receiver's
payoff from accept at the infinite population limit.}}
\end{figure}

That optimal mechanisms are simple arises from each sender being essentially
uninformed in the limit. In particular, his signal gives him no further
information about the reports of the other senders. This is true even
conditional on the realized state. As a result, any optimal complex
ICL mechanism can be replaced by its simple counterpart---which induces
the same expected probability of acceptance conditional on the aggregate
state---without harming incentives. By contrast, such a replacement
does not preserve incentives when $N$ is small, as our example in
Section 3 illustrates.

To understand why the interval structure is optimal among simple mechanisms,
consider a perturbation of the sender-preferred mechanism which introduces
some rejection at the bottom of the disagreement region. This has
a large benefit for the receiver since it is where she wants to reject
most within the disagreement region. Moreover, since senders are nearly
indifferent at this boundary, the change minimally affects their incentives
to lie. However, this perturbation still violates incentive compatibility.
Perhaps surprisingly, the cheapest way to restore incentives is to
punish senders by rejecting when the payoff to acceptance is highest
for all players. While this punishment is very costly when it occurs,
it occurs with extremely low likelihood under truthful reporting.
At the same time, it has a very high relative likelihood when a sender
lies. Hence, punishing at the top restores incentives for the sender
at the smallest ex-ante cost to the receiver.

Nonetheless, these surplus-burning punishments imply that the receiver
is strictly bounded away from her first best in the large-economy
limit. Beyond the impossibility of crosschecking, this is because
our model features informational division. In the rest of the literature
on information aggregation---such as \citet{feddersen1997voting},
\citet{pesendorfer1997loser}, and \citet{gerardi2009aggregation}---information
is additive. There, enlarging the population adds information, so
even a vanishingly small fraction of the population can eventually
identify the payoff-relevant state, and the first best becomes attainable.
Indeed, even though crosschecking is impossible in our setting, if
information were additive, then conflicts of interest would disappear
in the limit and the first best would still be achievable. However,
because we divide information, the incentive problem does not disappear,
optimal aggregation remains nontrivial at the limit, and we cannot
achieve the first best. These considerations suggest that the concept
of ICL may be applied to gain new insights by reexamining the broader
information-aggregation literature through the lens of informational
division, with implications for voting, auctions, markets, and collective
action.

The rest of the paper is structured as follows. Section \ref{sec:Model}
introduces the model. Section \ref{sec:ternaryintuition} presents
an illustrative example showing how the receiver can benefit from
informational division. Section \ref{sec:The-Value-of_mediator} develops
the ICL approach, and Section \ref{sec:Optimality-in-Large} characterizes
the optimal large-economy mechanism. Section \ref{sec:Heterogeneous-Bias-and}
extends the analysis to heterogeneous bias. Section \ref{sec:Related-Literature}
provides an extensive discussion of the related literature, and Section
\ref{sec:Conclusion-and-Extensions} concludes. Proofs are in the
Appendix.

\section{Model\label{sec:Model}}

We consider a model of strategic information transmission involving
multiple senders (each ``he'') and a single receiver (``she'').
There are $N$ senders, denoted by $i\in\left\{ 1,2,\cdots,N\right\} $.
Each sender privately observes a random signal $s_{i}$ drawn \emph{independently}
from finite set $S=\left\{ t_{1},\cdots,t_{K}\right\} \subset\mathbb{R}$,
where $t_{k}$ is increasing in $k$, and the probability that signal
$s_{i}=t_{k}$ is$f_{k}$. We denote the ordered vector of possible
signal realizations by $\mathbf{t}=\left(t_{1},\cdots,t_{K}\right)\in\mathbb{R}^{K}$,
the probability distribution by $\mathbf{f}=\left(f_{1},\cdots,f_{K}\right)$,
and normalize the mean of $s_{i}$ to zero: $\E[s_{i}]=\mathbf{f}\cdot\mathbf{t}=0$.
We denote the profile of senders' signals by $\mathbf{s}=\left(s_{1},s_{2},...,s_{N}\right)$.
Note, as signals are drawn independently, $s_{i}$ contains no information
about any other element of $\mathbf{s}$. Hence, one sender's signal
cannot be used to evaluate the truthfulness of another sender's report.\footnote{This structure of private information also satisfies the statistical
properties of privacy developed in \citet{strack2024privacy} and
\citet{he2026private}. Hence, privacy of information and verifiability
are intimately related.}

The receiver chooses an action $a\in\left\{ 0,1\right\} $. We refer
to $a=0$ as \emph{reject}, and $a=1$ as \emph{accept}. The payoffs
of the receiver and the senders are, respectively
\begin{align*}
u_{R}\left(a,\omega\right) & =a\left(\omega+r\right)\\
u_{S}\left(a,\omega\right) & =a\left(\omega+b\right)
\end{align*}
where $\omega$ is the payoff-relevant \emph{state}, given by 
\begin{equation}
\omega=\sum^{N}_{i=1}\frac{s_{i}}{\sqrt{N}},\label{eq:=00003D000020omega}
\end{equation}
and $b>0$ and $r<b$ are parameters determining the senders' and
the receiver's preference toward accept, respectively. The flexibility
to vary preferences over $a$ via $b$ and $r$, allows us to treat
$\mathbb{E}[s_{i}]=0$ as a normalization without loss of generality.
The assumption $b>r$ captures the senders' ``bias'' toward accept.
In other words, a conflict of interest exists between the senders
and the receiver for realizations of the state $\omega$ between $-b$
and $-r$, where the senders prefer accept while the receiver prefers
reject. To emphasize the relationship between $\omega$ and $\mathbf{s}$,
we sometimes use $\omega\left(\mathbf{s}\right)$ instead of $\omega$;
we refer to the set $\{\mathbf{s}:\omega\left(\mathbf{s}\right)\in(-b,-r)\}$
as the \textit{disagreement region}.

Our definition of $\omega$ implies that the effect of each signal
$s_{i}$ on the payoff-relevant state shrinks at rate $\sqrt{N}$.
Hence, $\nicefrac{s_{i}}{\sqrt{N}}$ is the analogue of sender $i$'s
private type in the mechanism design literature: we say \emph{type}
$k$ to refer to $\nicefrac{t_{k}}{\sqrt{N}}$. We model payoffs in
this way to ensure that the variance of the aggregate private information
is constant in $N$, while the private information held by each individual
sender shrinks. For large $N$, this approach allows us to think of
increasing $N$ as dividing the same information among more agents.
Indeed, as $N\rightarrow\infty$, we may interpret the model as one
in which each sender observes a disjoint increment of a Brownian motion
whose terminal value determines the payoff from accept.

We study the design of transfer-free recommendation mechanisms that
maximize the receiver's payoff. By the revelation principle\textit{
}of \citet{myerson1982optimal}, it is without loss of generality
to focus on direct recommendation mechanisms (henceforth, mechanism):
these collect the senders' reported signals, $\tilde{\mathbf{s}}=(\tilde{s}_{1},\tilde{s}_{2}\cdots,\tilde{s}_{N})$,
and recommend an action to the receiver. Formally, a mechanism is
a mapping $\sigma:S^{N}\rightarrow\left[0,1\right]$, where $\sigma\left(\mathbf{s}\right)=\Pr\left(a=1|\mathbf{s}\right)$
is the conditional probability of recommending $a=1$.\footnote{This approach contrasts with a literature on dynamic multi-sender
cheap talk that assesses specific dynamic communication protocols
in which communication may not be truthful. For example, see \citet{Aumann2003},
\citet{krishna_morgan2004}, \citet{ambrus_etal_2013}, \citet{golosov_etal_2014},
\citet{migrow2021designing}, and \citet{antic2025subversive}.}

The revelation principle requires that the mechanism $\sigma$ satisfy
Bayesian incentive compatibility for the senders and obedience for
the receiver:
\begin{enumerate}
\item \emph{Incentive compatibility}: For each sender $i\in\left\{ 1,\cdots,N\right\} $,
and for all $t_{k},t_{l}\in S$, 
\begin{equation}
\mathbb{E}\left[\sigma\left(\mathbf{s}\right)\left(\omega\left(\mathbf{s}\right)+b\right)|s_{i}=t_{k},\tilde{s}_{i}=t_{k}\right]\geq\mathbb{E}\left[\sigma\left(\tilde{\mathbf{s}}\right)\left(\omega\left(\mathbf{s}\right)+b\right)|s_{i}=t_{k},\tilde{s}_{i}=t_{l}\right].\tag{IC}\label{eq:general_IC}
\end{equation}
\item \emph{Obedience}: 
\begin{equation}
\mathbb{E}\left[\sigma\left(\mathbf{s}\right)\left(\omega\left(\mathbf{s}\right)+r\right)\right]\geq0\geq\mathbb{E}\left[\left(1-\sigma\left(\mathbf{s}\right)\right)\left(\omega\left(\mathbf{s}\right)+r\right)\right].\tag{Ob}\label{eq:general_OB}
\end{equation}
\end{enumerate}
Inequality (\ref{eq:general_IC}) requires that each sender prefer
to truthfully report his own signal to the mediator, given that the
other senders do the same and the receiver obeys the mediator's recommendation.
The obedience constraint (\ref{eq:general_OB}) states that the receiver
should be willing to follow the recommendation of the mechanism, given
that the senders are truthful.

Since we are interested in mechanisms that maximize the receiver's
payoff and the receiver can take only two actions, it is sufficient
to focus on incentive compatibility and ignore obedience. To see this,
note that if a mechanism $\sigma$ violated either of the inequalities
in (\ref{eq:general_OB}), then it would be dominated by the (trivially
obedient) uninformative mechanism which always recommends the receiver's
ex ante optimal action $a=\mathbf{1}(r\geq0)$. We thus have the following
lemma:\footnote{The formal proof of Lemma \ref{lem:If--achieves} is omitted for brevity
and is available on request. When there are more than two actions,
dropping obedience may not be without loss. See \citet{whitmeyer2024bayesian}
for a counterexample. A similar result holds in \citet{ballscoring}.
We thank Ian Ball for raising this point.}
\begin{lem}
\label{lem:If--achieves}If $\sigma$ achieves the highest payoff
for the receiver among all mechanisms that satisfy incentive compatibility
(\ref{eq:general_IC}), then $\sigma$ satisfies obedience (\ref{eq:general_OB}).
\end{lem}
Lemma \ref{lem:If--achieves} implies that the problem of finding
the receiver's best mechanism boils down to the following: 
\begin{equation}
\max_{\sigma:S^{N}\to[0,1]}\mathbb{E}\left[\left(\omega\left(\mathbf{s}\right)+r\right)\sigma\left(\mathbf{s}\right)\right]\tag{P}\label{eq:P}
\end{equation}
subject to incentive compatibility (\ref{eq:general_IC}).
\begin{rem}
\label{rem:voting}In light of Lemma \ref{lem:If--achieves}, the
problem of finding the best mediated mechanism has an alternative
interpretation. It is equivalent to a design problem in which the
receiver can commit to her actions as a function of the senders' reports.
Hence, the analysis extends naturally to other environments such as
voting.
\end{rem}

\subsection{Simple Mechanisms}

We call a mechanism \emph{simple} if it only depends on the reported
state $\tilde{\omega}$ rather than on the entire signal profile $\tilde{\mathbf{s}}$.
With a small abuse of notation, we denote these mechanisms by $\sigma(\tilde{\omega})$.
The following three benchmark mechanisms are all simple. First is
the \emph{receiver-preferred} mechanism, $\sigma^{R}\left(\tilde{\omega}\right)=\mathbf{1}\left[\tilde{\omega}+r\geq0\right]$.
This attempts to give the receiver her first best. However, it is
incentive compatible only in the uninteresting case where the disagreement
region is empty. Otherwise, facing $\sigma^{R}$, a sender with signal
$s_{i}=t_{k}$ prefers to report $\tilde{s}_{i}=t_{k+1}$ as this
lie will only induce $a=1$ for some marginal state $\omega>-b$,
i.e., when he strictly prefers acceptance. Second is the \textit{sender-preferred}
mechanism, $\sigma^{S}\left(\tilde{\omega}\right)=\mathbf{1}\left[\tilde{\omega}+b>0\right]$,
which is trivially incentive compatible as it gives the senders their
preferred outcome, but is not always obedient. Third is the\textit{
uninformative} mechanism, $\sigma^{U}\left(\tilde{\omega}\right)=\mathbf{1}\left[r>0\right]$,
which trivially satisfies incentive compatibility and obedience.

Whenever there is no informational division, i.e., when $N=1$, all
mechanisms must be simple since $s_{1}=\omega$. Moreover, one of
these three simple benchmarks solves the problem (\ref{eq:P}). In
the uninteresting case where the disagreement region is empty, $\sigma^{R}$
obviously solves (\ref{eq:P}). Outside this case, incentive compatibility
severely limits what can be done. Condition (\ref{eq:general_IC})
reduces to 
\[
\sigma\left(\omega\right)\left(\omega+b\right)\geq\sigma\left(\tilde{\omega}\right)\left(\omega+b\right),\forall\omega,\tilde{\omega}\in S.
\]
As the sender strictly prefers $a=1$ if and only if $\omega>-b$,
he will report whatever maximizes (minimizes) $\sigma$ whenever $\omega>-b$
($\omega\leq-b$). Thus, $\sigma\left(\omega\right)$ must be constant
on the sets $\left(-\infty,-b\right]$ and $\left(-b,\infty\right)$,
and non-decreasing in $\omega$. If $\E[\omega+r\mid\omega>-b]\geq0>\E[\omega+r\mid\omega\leq-b]$,
then the receiver clearly prefers $\sigma^{S}$ among all such mechanisms.
Otherwise, she must prefer $\sigma^{U}$. That is, the optimal mechanism
is $\sigma^{S}$ if and only if the information that $\omega\geq-b$
is valuable for the receiver.\footnote{While this is the case in our environment, there is a literature on
single sender cheap talk \citep{Crawford1982} showing conditions
under which a mediator can improve payoffs. See for instance, \citet{Goltsman_etal_2009},
\citet{salamanca2021value}, \citet{corrao2023bounds}, and \citet{best2024persuasion}.}

The reason we cannot do better with one sender is because he knows
the state and is the only person reporting. This implies that to reduce
the probability of accept in the disagreement region, we also have
to pay the cost of equally reducing the probability of accept when
the receiver agrees, i.e., wherever $\omega\geq-r$. However, when
$N>1$, each sender has fewer deviations available and less information;
thus, there may be scope to use informational division to improve
incentives.\footnote{In the online appendix we also provide results on the optimality of
the sender preferred allocation with a rich (continuous) signal space
for $N>1$. If the bias is not too large and the senders are not too
many, then it is optimal.} In the next section, we demonstrate that such improvements can be
achieved by simple mechanisms, though the restriction to such simple
mechanisms can itself be costly.

\section{An Illustrative Example \label{sec:ternaryintuition}}

This section presents an example to develop intuition for a) how we
can improve the receiver's payoff with a simple mechanism when information
is divided across multiple strategic senders; and b) why the optimal
mechanism may be complex, i.e., depend on the whole vector $\mathbf{s}$
beyond $\omega(\mathbf{s})$. We provide supporting calculations in
Online Appendix \ref{subsec:Simple-Example-Appendix}.

Let $S=\left\{ -\sqrt{2},0,\sqrt{2}\right\} $ where $\mathbf{f}$
is uniform on $S$ and $N=2$. Hence, each sender has type $\nicefrac{s_{i}}{\sqrt{2}}$,
which we denote by $l=-1$ (low), $m=0$ (medium), and $h=1$ (high),
and the state is
\[
\omega(\mathbf{s})=\frac{s_{1}+s_{2}}{\sqrt{2}}\in\left\{ -2,-1,0,1,2\right\} .
\]
Finally, we let $r=0$ and $b=3$ so that $u_{R}(a,\omega)=a\omega$
and $u_{S}(a,\omega)=a(\omega+3)$.

In this example, the sender preferred mechanism $\sigma^{S}$ is uninformative,
since $\omega+b>0$ for all $\omega$. Hence, if a \emph{single} sender
observed both $s_{1}$ and $s_{2}$ (equivalently, observed $\omega$
perfectly), then the only truthful and obedient mechanism would be
$\sigma^{S}=\sigma^{U}$ with expected payoff of 0 for the receiver.
However, when the information is divided between two senders, the
following truthful and obedient simple mechanism (also illustrated
in Figure \ref{fig:The-optimal-state-dependent}) increases receiver
payoffs:
\begin{equation}
\sigma^{\dagger}(\omega)=\begin{cases}
0 & \text{ if }\omega\in\{-2,-1\}\\
1 & \text{ if }\omega\in\{0\}\\
\frac{1}{3} & \text{ if }\omega\in\{1,2\}
\end{cases}.\label{eq:ternary_sdpmech}
\end{equation}
 Obviously, this is a strict improvement on no information, giving
a receiver payoff of $\frac{4}{27}>0$. This mechanism takes advantage
of the fact that when action $a=1$ is most costly to the receiver,
i.e., when $\omega\in\{-2,-1\}$, it is also least desirable to the
sender. As a result, a surplus-burning reduction in the probability
of $a=1$ at the top of the distribution---where the sender values
acceptance the most---acts as a sufficiently large punishment to
deter upwards lies.

To see this, notice that an upward lie by a sender ($l$ reporting
$m$, or $m$ reporting $h$) shifts the reported state from $\omega$
to $\omega+1$. Hence, sender $i$'s lie is \emph{pivotal} only when
$\sigma^{\dagger}(\omega)\neq\sigma^{\dagger}(\omega+1)$. In this
example, this occurs in two states, $\omega\in\{-1,0\}$. When the
true state is $\omega=-1$ the lie benefits the sender by increasing
the probability of acceptance; but when $\omega=0$, the lie decreases
its probability. The expected benefit of the lie is $\Pr(\omega=-1\mid s_{i})[\sigma^{\dagger}(0)-\sigma^{\dagger}(-1)](-1+b)=2/3$,
which is offset by the expected cost $\Pr(\omega=0\mid s_{i})[\sigma^{\dagger}(1)-\sigma^{\dagger}(0)](0+b)=-2/3$,
so the mechanism is incentive compatible.\footnote{Notice, this binding local incentive constraint is identical for both
types $l$ and $m$ because $\mathbf{f}$ is uniform in this example.
Moreover, the interim probability of acceptance is increasing in the
sender's report $\tilde{s}_{i}$. Hence, by standard arguments these
local incentive constraints imply global incentive compatibility.}

\begin{figure}[H]
\subfloat[\label{fig:The-optimal-state-dependent}The optimal simple mechanism]{\begin{centering}
\begin{tikzpicture}[scale=1.5]
\definecolor{crimson}{RGB}{165,15,15}
\pgfmathsetmacro{\sq}{1.414}

\draw[black, thick, dashed, rounded corners=14pt,
rotate around={-45:(0,0)}]
(-2.2,-0.37) rectangle (2.2,0.37);

\draw[crimson, thick, dashed, rounded corners=15pt]
(-1.68, 0.70) -- (-1.68, -1.68) -- (.7, -1.68) -- cycle;
\draw[blue, thick, dashed, rounded corners=15pt]
(1.68, -0.70) -- (1.68, 1.68) -- (-.7, 1.68) -- cycle;

\draw[black, very thick, ->]
(-2.45,0) -- (2.65,0) node[right]{$s_1$};
\draw[black, very thick, ->]
(0,-2.30) -- (0,2.65) node[above]{$s_2$};

  \fill[black] (0,0) circle (2.3pt);

\fill[black] (-1.38, 0) circle (2.3pt);
\fill[black] (1.38, 0) circle (2.3pt) node[below]{\small $\sqrt{2}$};
\fill[black] (0,-1.38) circle (2.3pt);
\fill[black] (0,1.38) circle (2.3pt) node[right]{\small $\sqrt{2}$};
\fill[black] (-1.38, -1.38) circle (2.3pt);
\fill[black] (1.38, -1.38) circle (2.3pt);
\fill[black] (-1.38,1.38) circle (2.3pt);
\fill[black] (1.38,1.38) circle (2.3pt);

\node[black] at (-1.88, 1.88) {$\sigma=1$};
\node[blue] at ( .88, 1.88) {$\sigma=\tfrac{1}{3}$};
\node[crimson] at (-0.88,-1.88) {$\sigma=0$};
\end{tikzpicture}
\par\end{centering}
}\subfloat[\label{fig:A-fully-optimal}A fully optimal mechanism]{\begin{centering}
\begin{tikzpicture}[scale=1.4]
					\definecolor{crimson}{RGB}{165,15,15}
					\pgfmathsetmacro{\sq}{1.414}
					
					\draw[blue!75!black, thick, dashed, rounded corners=14pt,shift={(0.7,0.7)}, rotate=-45]
				(-1.35, -0.30) rectangle (1.35, .3);
					
					\draw[crimson, thick, dashed, rounded corners=15pt]
					(-1.68, 0.70) -- (-1.68, -1.68) -- (.7, -1.68) -- cycle;
					
					\draw[crimson, thick, dashed] (1.38, 1.38) circle[radius=8pt];

                    \draw[black, thick, dashed] (-1.38,1.38) circle[radius=8pt];
                     \draw[black, thick, dashed] (1.38,-1.38) circle[radius=8pt];
                     \draw[cyan, thick, dashed] (0,0) circle[radius=8pt];
                     
					\draw[black, very thick, ->]
					(-2.45,0) -- (2.65,0) node[right]{$s_1$};
					\draw[black, very thick, ->]
					(0,-2.30) -- (0,2.65) node[above]{$s_2$};

					\fill[black] (0,0) circle (2.3pt);
					
					\fill[black] (-1.38, 0) circle (2.3pt);
					\fill[black] (1.38, 0) circle (2.3pt);
					\fill[black] (0,-1.38) circle (2.3pt);
					\fill[black] (0,1.38) circle (2.3pt);
					\fill[black] (-1.38, -1.38) circle (2.3pt);
					\fill[black] (1.38, -1.38) circle (2.3pt);
					\fill[black] (-1.38,1.38) circle (2.3pt);
					\fill[black] (1.38,1.38) circle (2.3pt);

					\node[black] at (-1.73, 1.73) {$\sigma=1$};
					\node[black] at (1.73, -1.73) {$\sigma=1$};
					\node[cyan] at (-.4,.4) {$\sigma=\tfrac{3}{7}$};
					\node[blue!75!black] at (-.4, 1.88) {$\sigma=\tfrac{5}{7}$};
					\node[crimson] at (-0.88,-1.88) {$\sigma=0$};
					\node[crimson] at (1.88,1.88) {$\sigma=0$};
				\end{tikzpicture}

\par\end{centering}
}

\caption{Illustration of mechanisms: Figure \ref{fig:The-optimal-state-dependent}
shows the best incentive compatible simple mechanism for the receiver
in this environment; Figure \ref{fig:A-fully-optimal} presents a
mechanism which is optimal in the class of all truthful direct mechanisms.
No simple mechanism achieves the fully optimal payoff associated with
the mechanism in Figure \ref{fig:A-fully-optimal}.}
\end{figure}

This mechanism is optimal among all simple mechanisms and illustrates
how informational division allows us to improve payoffs by reducing
the set of deviations available to each sender.\footnote{Notice from Figure \ref{fig:The-optimal-state-dependent}, when $N=2$,
each sender only chooses a row (or column) of acceptance probabilities
indexed by the other sender's type. By contrast, if a single sender
could observe and report $(s_{1},s_{2})$, he can select both the
row and the column to secure acceptance, e.g., by reporting $(m,m)$.} However, the restriction to simple mechanisms is costly. In particular,
Figure \ref{fig:A-fully-optimal} illustrates the optimal mechanism
$\sigma^{\star}(\mathbf{s})$ which achieves a strictly greater payoff
of $\frac{10}{63}>\frac{4}{27}$ for the receiver. Clearly, $\sigma^{\star}(\mathbf{s})$
is not simple: it provides a lower acceptance probability for the
report $(m,m)$ than for $(l,h)$ and $(h,l)$ despite all corresponding
to $\omega=0$. Relative to $\sigma^{\dagger}(\omega)$, reducing
$\sigma(m,m)$ costs the receiver nothing---she is indifferent over
$a$ when $\omega=0$. However, it serves as a punishment targeted
at an $m$-report that strictly deters $l$ from an upward lie. This
allows us to increase the probability of $a=1$ when $\omega=1$ while
still respecting $l$'s incentives. Finally, while these changes do
make an upward lie more appealing for the $m$-types, the mechanism
recovers incentive compatibility with a punishment targeted at the
top, $\tilde{\mathbf{s}}=(h,h)$, where the sender values $a=1$ most.

This example makes clear why restricting to simple mechanisms can
be costly: they cannot independently assign acceptance probabilities
across reports. In particular, a simple mechanism could only reduce
$\sigma$ at $(m,m)$ if it also made identical reductions at $(l,h)$
and $(h,l)$. While this would reduce the payoff from reporting $m$,
it would for $l$ and $h$ too. In this example, such a reduction
would actually encourage the $l$ type to lie because it would reduce
his truth-telling payoff by more than his payoff from masquerading
as $m$. By contrast, $\sigma^{\star}(\mathbf{s})$ used the freedom
to independently reduce the payoffs for higher reports (i.e., without
also reducing the truth-telling payoffs for the lower types) to more
effectively screen senders according to their type.

The optimality of complex mechanisms is not unique to this example.
Indeed, in richer type spaces, effective screening of types can require
mechanisms to be even more complex.\footnote{See Online Appendix \ref{sec:Numerical-Solution-of} for a numerical
example.} However, in the next section, we show that the importance of each
sender's type becomes negligible as $N\rightarrow\infty$, which simplifies
the characterization of incentive constraints. Then in Section \ref{sec:Optimality-in-Large},
we use this result to show that in the large-$N$ limit, the optimal
mechanism is in fact simple. Moreover, this simple mechanism shares
two important properties with $\sigma^{\dagger}(\omega)$ above: First,
it uses surplus-burning punishments for high aggregate reports---where
the sender wants $a=1$ most---to buy much greater reductions in
the probability of acceptance in the disagreement region. Second,
an upward lie is pivotal only in two states: increasing the probability
of accept in the lower state and decreasing it in the higher state.

\section{Incentive Compatibility in Large Economies\label{sec:The-Value-of_mediator}}

The focus of this paper is on aggregating information when it is divided
among a large population. To examine this, we study the solution to
problem (\ref{eq:P}) as $N\rightarrow\infty$. To make this problem
amenable to limiting arguments, we first reformulate it as a choice
over mechanisms that map frequencies of signal realizations into the
probability of recommending accept. This allows us to establish our
main technical result (Theorem \ref{thm:A-recommendation-mechanism}),
which characterizes the limit set of incentive compatible mechanisms.
Theorem \ref{thm:A-recommendation-mechanism} shows that the right
way to extend incentives to the limit, is to treat each sender as
if he is uninformed but can nevertheless have a small impact on the
distribution of reports.

In section \ref{sec:Optimality-in-Large}, we characterize the optimal
mechanism at infinity subject to this characterization of incentive
compatibility in the limit (Theorem \ref{thm:The-optimum-is}). As
this mechanism is the limit point to which all optimal mechanisms
converge, it characterizes the asymptotic properties of optimal aggregation
when information is divided across large finite populations.

\subsection{Mechanisms in the Frequency Domain}

Recall a direct mechanism is a mapping $\sigma:S^{N}\rightarrow[0,1]$.
Hence, the domain of direct mechanisms changes in $N$ and its dimensionality
explodes as $N\rightarrow\infty$. To avoid these issues, we reframe
the problem in terms of mechanisms that operate on the frequency domain---which
is invariant in $N$. Formally, we define the normalized deviation
of the sample frequency of signal $t_{k}$ from its population frequency
by

\begin{equation}
h_{k}\left(\mathbf{s}\right)=\sqrt{N}\left(\frac{\left|\left\{ i|s_{i}=t_{k}\right\} \right|}{N}-f_{k}\right),\label{eq: empiricaldist}
\end{equation}
and the vector of these deviations by $\mathbf{h}^{N}\left(\mathbf{s}\right)=\left(h_{1}\left(\mathbf{s}\right),\cdots,h_{K}\left(\mathbf{s}\right)\right)$.
We refer to $\mathbf{h}^{N}\left(\mathbf{s}\right)$ as \textit{normalized
empirical frequencies} \textit{(NEF),} and to $h_{k}$ as $k$-NEF.
From here on, we consider mechanisms that are functions of the empirical
frequencies $\mathbf{h}^{N}(\mathbf{s})$, i.e., $\sigma:\mathbb{R}^{K}_{0}\rightarrow[0,1]$,
where $\mathbb{R}^{K}_{0}$ is the subset of $\mathbb{R}^{K}$ whose
elements sum to zero (because $\sum^{K}_{k=1}h^{N}_{k}\left(\mathbf{s}\right)=0$).
With a slight abuse of notation, we denote this function by $\sigma(\mathbf{h}^{N})$
and suppress the dependence of $\mathbf{h}^{N}$ on $\mathbf{s}$
hereafter.\footnote{Notice, the object $\mathbf{h}^{N}(\mathbf{s})$ has a familiar interpretation.
Each $h_{k}$ is very similar to the classical test statistic (a ``$Z$-test'')
for the population proportion of type $k$. Thus, one could also view
the design of $\sigma$ through the lens of hypothesis testing. Our
approach is also closely related to the `method of types' used in
information theory; see, for instance, \citet{cover2006elements}.
We use different language to avoid the obvious confusion.}

It is without loss of generality to focus on mechanisms expressed
on the $K-1$-dimensional frequency domain. To see this, consider
that it is without loss to focus on symmetric recommendation mechanisms,
for which $\sigma\left(\mathbf{s}\right)=\sigma\left(\pi\left(\mathbf{s}\right)\right)$
for all $\pi\in\Pi$, where $\Pi$ is the set of all permutations
of $\left\{ 1,\cdots,N\right\} $. This is because, for any incentive
compatible mechanism $\sigma$, the mechanism given by $\hat{\sigma}\left(\mathbf{s}\right)=\sum_{\pi\in\Pi}\sigma\left(\pi\left(\mathbf{s}\right)\right)/N!$
is incentive compatible, symmetric, and delivers the same payoff to
the receiver. Finally, notice that any symmetric mechanism is equivalent
to one that depends on a simple count of each report, and the NEF
maps one-to-one onto these counts.

As it is without loss to analyze problem (\ref{eq:P}) using mechanisms
on the frequency domain, we use the fact that

\[
\omega=\sum\dfrac{s_{i}}{\sqrt{N}}=\mathbf{h}^{N}\cdot\mathbf{t}.
\]
to express incentive compatibility directly in terms of $\mathbf{h}^{N}$,
as
\begin{equation}
\mathbb{E}\left[\sigma\left(\mathbf{h}^{N}\right)\left(\mathbf{h}^{N}\cdot\mathbf{t}+b\right)|s_{i}=t_{k},\tilde{s}_{i}=t_{k}\right]\geq\mathbb{E}\left[\sigma\left(\tilde{\mathbf{h}}^{N}\right)\left(\mathbf{h}^{N}\cdot\mathbf{t}+b\right)|s_{i}=t_{k},\tilde{s}_{i}=t_{l}\right],\tag{F-IC}\label{eq:freqIC}
\end{equation}
for all $t_{k},t_{l}\in S$, where $\tilde{\mathbf{h}}^{N}$ is the
reported NEF. Notice how the sender's expected payoffs depend on his
signal and report. First, since his signal enters directly into the
true NEF, it gives him private information about $\mathbf{h}^{N}$
and therefore the value of acceptance. Second, since his report enters
directly into the reported NEF, $\tilde{\mathbf{h}}^{N}$, he can
influence its distribution: if a $k$ type lies and reports $t_{l}$,
then he decreases $\tilde{h}^{N}_{k}$ and increases $\tilde{h}^{N}_{l}$.
Of course, this is not profitable in an incentive compatible mechanism.

A benefit of reframing mechanisms in this way is that it ensures the
domain of the mechanism is invariant in $N$. Moreover, the multidimensional
central limit theorem implies the argument of the mechanism, the NEF,
has appealing large sample properties:
\begin{lem}
\textbf{(Multidimensional Central Limit Theorem) }\label{lem:(Donsker.-CLT)-Let}Let
$\mathbf{h}^{N}\left(\mathbf{s}\right)$ be the normalized empirical
frequencies of $\mathbf{s}$ as defined in (\ref{eq: empiricaldist}).
Then, as $N\rightarrow\infty$,
\begin{equation}
\mathbf{h}^{N}\overset{d}{\longrightarrow}N\left(\mathbf{0},\mathbf{\Sigma}\right),\label{eq: adjusteddist}
\end{equation}
with $\mathbf{\Sigma}_{kl}=f_{k}\left(\mathbf{1}\left(k=l\right)-f_{l}\right)$.
Furthermore, $\omega\overset{d}{\longrightarrow}N\left(0,\text{Var}(s_{i})\right)$.
\end{lem}
Since the signals are independent, it is not surprising that a central
limit theorem applies. However, given this independence, it may be
surprising that there is (negative) correlation between the $k$-NEF
and $l$-NEF ($\mathbf{\Sigma}_{kl}<0$ for $k\neq l$). Yet, as in
the discussion following condition (\ref{eq:freqIC}), more of one
signal must result in less of another. Indeed, this correlation reflects
both the sender's private information about $\mathbf{h}^{N}$, and
the influence of his reports on $\tilde{\mathbf{h}}^{N}$. In the
next section, we use Lemma \ref{lem:(Donsker.-CLT)-Let} to analyze
how incentive compatibility changes as the population becomes large
and solve for the optimal mechanism in the limit.

Finally, in the $N$-sender problem, notice that a mechanism $\sigma^{N}$
and the random variable $\mathbf{h}^{N}$ jointly induce a joint distribution
over $(\mathbf{h}^{N},a^{N})$, where $a^{N}$ is the recommended
action and $\Pr\left[a^{N}=1\mid\mathbf{h}^{N}\right]=\sigma^{N}(\mathbf{h}^{N})$;
we refer to the joint distribution induced by the $N$-sender mechanism
$\sigma^{N}$ as its \emph{outcome distribution}. Similarly, we define
the outcome distribution of a mechanism $\sigma$ in the large-$N$
limit as the joint distribution over $(\mathbf{h},a)$ induced by
$\sigma(\mathbf{h})=\Pr[a=1\mid\mathbf{h}]$ and $\mathbf{h}\sim N(\mathbf{0},\mathbf{\Sigma})$.
For any incentive compatible direct mechanism, all payoff- and incentive-relevant
information is captured by its outcome distribution; in particular,
the outcome distribution determines all the expected values present
in both the objective and the constraints of problem (\ref{eq:P}).

\subsection{Incentive Compatibility in the Large \label{subsec:Optimal-Mechanisms-in}}

We are interested in the limiting properties of optimal mechanisms
as $N$ grows large. To characterize this limit, we study a design
problem in which feasible mechanisms must satisfy a notion of \emph{incentive
compatibility in the large}.
\begin{defn}
\label{def:A-recommendation-mechanism}A recommendation mechanism
$\sigma:\mathbb{R}^{K}_{0}\rightarrow[0,1]$ is \emph{incentive compatible
in the large (ICL)} if there exists a sequence of $N$-sender recommendation
mechanisms $\{\sigma^{N}\}_{N\geq1}$ satisfying the finite-$N$ incentive
compatibility condition (\ref{eq:freqIC}) whose induced outcome distributions
converge
\[
\left(\mathbf{h}^{N},a^{N}\right)\overset{d}{\longrightarrow}\left(\mathbf{h},a\right).
\]
\end{defn}
Thus, a mechanism $\sigma$ is ICL if its outcome distributions can
be approximated arbitrarily well by incentive compatible mechanisms
$\sigma^{N}$ in large, finite economies. Though the large-$N$ limit
in which $\mathbf{h}\sim N(\mathbf{0},\mathbf{\Sigma})$ is a limiting
construct (it does not correspond to any finite $N$-sender economy),
understanding the set of ICL mechanisms is useful because it approximates
what can be achieved when information is divided among sufficiently
many senders. The following theorem provides a simple characterization
of this set:
\begin{thm}
\label{thm:A-recommendation-mechanism}A recommendation mechanism
$\sigma:\mathbb{R}^{K}_{0}\rightarrow\left[0,1\right]$ is ICL if
and only if it satisfies

\begin{align}
\mathbb{E}\left[\sigma\left(\mathbf{h}\right)\left(\mathbf{h}\cdot\mathbf{t}+b\right)h_{k}/f_{k}\right] & =t_{k}\mathbb{E}[\sigma\left(\mathbf{h}\right)]\,\text{ for all }k,\tag{ENV}\label{eq:env}\\
\mathbb{E}\left[\sigma\left(\mathbf{h}\right)(\frac{h_{k}}{f_{k}}-\frac{h_{k-1}}{f_{k-1}})\right] & \geq0\text{ for all }k\geq2,\tag{MON}\label{eq:mon}
\end{align}
with $\mathbf{h}\sim\mathcal{N}\left(\mathbf{0},\mathbf{\Sigma}\right)$.
\end{thm}
Theorem \ref{thm:A-recommendation-mechanism} is our main technical
result and it will greatly simplify the analysis of optimal mechanisms
in large economies. In essence, it proves that conditions (\ref{eq:env})
and (\ref{eq:mon}) are the limiting analogues of the envelope formulation
of incentive compatibility and the interim monotonicity constraints
common in the mechanism design literature. As such, they fully characterize
the set of ICL mechanisms, i.e., those whose outcome distributions
can be approximated by an incentive compatible mechanism for large
enough $N$.

For some intuition, consider how a sender's incentives vary with $N$.
Incentive compatibility requires that type $k$ does not want to report
$t_{l}$. Defining the truthful utility of type $k$ as $U^{N}_{k}=\mathbb{E}\left[\left(\mathbf{h}\cdot\mathbf{t}+b\right)\sigma^{N}(\mathbf{h})\mid s_{i}=\tilde{s}_{i}=t_{k}\right]$,
this condition can be written as

\begin{equation}
U^{N}_{k}-U^{N}_{l}\geq\frac{t_{k}-t_{l}}{\sqrt{N}}\mathbb{E}\left[\sigma^{N}\left(\mathbf{h}^{N}\right)\mid\tilde{s}_{i}=t_{l}\right].\label{eq:no_masquerade}
\end{equation}
Inequality (\ref{eq:no_masquerade}) is the standard incentive constraint
that arises in mechanism design problems where senders' types are
independent. To fix ideas, let $t_{l}<t_{k}$. If type $k$ reports
$t_{l}$, he can induce the allocation offered to type $l$. However,
since type $k$ enjoys a higher private value for $a=1$, inducing
such an allocation would earn him an incremental information rent
over type $l$'s payoff, given by the right side of (\ref{eq:no_masquerade}).
In an incentive compatible mechanism, this payoff must be no more
than type $k$'s truthful utility $U^{N}_{k}$.

Since type $l$ must not want to report $t_{k}$ either, incentive
compatibility implies:

\begin{equation}
\mathbb{E}\left[\sigma^{N}\left(\mathbf{h}^{N}\right)|\tilde{s}_{i}=t_{k}\right]\geq\sqrt{N}\frac{U^{N}_{k}-U^{N}_{l}}{t_{k}-t_{l}}\geq\mathbb{E}\left[\sigma^{N}\left(\mathbf{h}^{N}\right)\mid\tilde{s}_{i}=t_{l}\right].\tag{N-ENV}\label{eq:finite_N_envelope_bounds}
\end{equation}
Notice that for adjacent types $k\geq2$ and $l=k-1$, this is the
discrete version of the standard envelope condition: it bounds the
\textit{rate} at which the sender's interim utility increases in his
type $(\nicefrac{t_{k}}{\sqrt{N}})$. Moreover, (\ref{eq:finite_N_envelope_bounds})
also implies that the usual monotonicity condition on the interim
allocation, $\mathbb{E}\left[\sigma^{N}\left(\mathbf{h}^{N}\right)\mid\tilde{s_{i}}=t_{k}\right]$,
holds. By the standard Spence-Mirrlees arguments, these envelope and
monotonicity conditions are also sufficient for global incentive compatibility.\footnote{Notice how this compares to mechanism design problems with transfers.
There, every nondecreasing allocation would be incentive compatible,
since transfers can always be constructed to appropriately bound the
marginal utilities. In our setting, we do not have the additional
tool of transfers to ensure that condition (\ref{eq:finite_N_envelope_bounds})
is satisfied by every monotone mechanism; accordingly, the bounds
(\ref{eq:finite_N_envelope_bounds}) on marginal utilities impose
additional restrictions on $\sigma$ beyond those imposed by monotonicity.}

To see how condition (\ref{eq:finite_N_envelope_bounds}) behaves
in the limit, we must establish how the sender's private information
about $\mathbf{h}^{N}$ and the influence of his report on $\tilde{\mathbf{h}}^{N}$
depend on $N$. To do this, we update beliefs about $\mathbf{h}^{N}$
and $\tilde{\mathbf{h}}^{N}$ via Bayes' rule. For $\mathbf{h}^{N}$,
this is\footnote{One can verify (\ref{eq:conditional_h_prob}) after noting that (i)
the population frequency of type $t_{l}$ is $f_{l}$, and (ii) the
realized sample frequency $\Pr\left(s_{i}=t_{l}\mid\mathbf{h}^{N}\right)$
is $\nicefrac{\left|\left\{ i:s_{i}=t_{l}\right\} \right|}{N}$. The
prior $\Pr\left(\mathbf{h}^{N}(\tilde{\mathbf{s}})\right)$ is updated
by their ratio.}

\begin{equation}
\Pr\left(\mathbf{h}^{N}|s_{i}=t_{l}\right)=\Pr\left(\mathbf{h}^{N}\right)\frac{\left|\left\{ i:s_{i}=t_{l}\right\} \right|}{f_{l}N}=\Pr\left(\mathbf{h}^{N}\right)\left(\frac{h^{N}_{l}}{\sqrt{N}f_{l}}+1\right).\label{eq:conditional_h_prob}
\end{equation}
An identical expression holds for $\Pr\left(\tilde{\mathbf{h}}^{N}|\tilde{s}_{i}=t_{l}\right)$.
Equation (\ref{eq:conditional_h_prob}) is a reflection of the sender's
private information, which diminishes in the number of senders $N$.
Indeed, as $N\rightarrow\infty$ the sender's posterior collapses
to the prior and he becomes essentially uninformed.

Using (\ref{eq:conditional_h_prob}) and $U^{N}_{k}=\mathbb{E}\left[\left(\mathbf{h}\cdot\mathbf{t}+b\right)\sigma(\mathbf{h})\mid s_{i}=\tilde{s}_{i}=t_{k}\right]$,
the components of condition (\ref{eq:finite_N_envelope_bounds}) can
be expressed:
\begin{eqnarray}
\mathbb{E}\left[\sigma^{N}(\mathbf{h}^{N})\mid\tilde{s}_{i}=t_{k}\right] & = & \mathbb{E}\left[\sigma^{N}(\mathbf{h}^{N})\left(1+\frac{h^{N}_{k}}{f_{k}\sqrt{N}}\right)\right],\label{eq:allocation}\\
\sqrt{N}\frac{U^{N}_{k}-U^{N}_{l}}{t_{k}-t_{l}} & = & \frac{1}{t_{k}-t_{l}}\mathbb{E}\left[\left(\mathbf{h}^{N}\cdot\mathbf{t}+b\right)\sigma^{N}(\mathbf{h^{N}})\left(\frac{h^{N}_{k}}{f_{k}}-\frac{h^{N}_{l}}{f_{l}}\right)\right].\label{eq:derivative}
\end{eqnarray}
We show in the proof that these expectations also converge appropriately,
so that as $N\rightarrow\infty$ (\ref{eq:finite_N_envelope_bounds})
becomes 
\begin{equation}
\mathbb{E}\left[\left(\mathbf{h}\cdot\mathbf{t}+b\right)\sigma(\mathbf{h})\left(\frac{h_{k}}{f_{k}}-\frac{h_{l}}{f_{l}}\right)\right]=\left(t_{k}-t_{l}\right)\mathbb{E}\left[\sigma(\mathbf{h})\right].\label{eq:diff_env}
\end{equation}
Intuitively, as $N$ gets large and the sender's information shrinks,
the impact of his report on the probability of acceptance becomes
negligible (observe $1+\frac{h^{N}_{k}}{f_{k}\sqrt{N}}\rightarrow1$),
so that (\ref{eq:allocation}) converges to $\mathbb{E}\left[\sigma(\mathbf{h})\right]$.
Nonetheless, we see in (\ref{eq:derivative}) that the sender's utility
does vary with his type $\frac{s_{i}}{\sqrt{N}}$\textbf{ }for all
$N$: his report always has a \emph{first-order} impact $\frac{h^{N}_{k}}{f_{k}}-\frac{h^{N}_{l}}{f_{l}}$
on the NEF, and so (\ref{eq:derivative}) becomes $\frac{1}{t_{k}-t_{l}}\mathbb{E}\left[\left(\mathbf{h}\cdot\mathbf{t}+b\right)\sigma(\mathbf{h})\left(\frac{h_{k}}{f_{k}}-\frac{h_{l}}{f_{l}}\right)\right]$
in the limit.\footnote{One can see this by examining how the rate at which $\Pr\left(\mathbf{h}^{N}\left(\mathbf{s}\right)|s_{i}\right)$
changes in type $\frac{s_{i}}{\sqrt{N}}$, using (\ref{eq:conditional_h_prob}).} Hence, even as $N$ becomes large, the sender's type continues to
have a first-order impact on payoffs.

Condition (\ref{eq:diff_env}) is equivalent to (\ref{eq:env}). Notice,
as incentive compatibility holds for all $l$, (\ref{eq:diff_env})
must also hold when taking expectations with respect to $l$ (and
fixing $k$). Moreover, since $\mathbb{E}[s_{i}]=0$ and $\sum h_{l}=0$,
the terms in $l$ drop out: equation (\ref{eq:env}) therefore expresses
the constraint in terms of deviations of the sender's payoff from
average. On the other hand, taking the difference of equation (\ref{eq:env})
for $k$ and $l$ yields (\ref{eq:diff_env}).

Equation (\ref{eq:mon}) is motivated by similar (albeit simpler)
reasoning. For finite $N$, it is easy to see from (\ref{eq:allocation})
that the monotonicity requirement is simply $\mathbb{E}\left[\sigma^{N}(\mathbf{h})\left(\frac{h^{N}_{k}}{f_{k}\sqrt{N}}-\frac{h^{N}_{k-1}}{f_{k-1}\sqrt{N}}\right)\right]\geq0$.
After multiplying by $\sqrt{N}$, we show the resulting expression
converges to (\ref{eq:mon}) as $N\rightarrow\infty$. Hence, (\ref{eq:env})
and (\ref{eq:mon}) can be viewed as the limiting analogues of the
envelope formulation of incentive compatibility and the interim monotonicity
constraints.

The necessity of (\ref{eq:env}) and (\ref{eq:mon}) for a mechanism
to be ICL follows almost immediately from the discussion above. Establishing
sufficiency is more challenging. Though it is not difficult to find
a sequence of finite-$N$ mechanisms $\sigma^{N}$ whose outcomes
$(\mathbf{h}^{N},a^{N})$ converge in distribution to $(\mathbf{h},a)$,
we must also ensure it satisfies incentive compatibility condition
(\ref{eq:freqIC}) for all $N$. The proof shows how to adjust any
such sequence to ensure both that (\ref{eq:freqIC}) is respected
and that the new sequence of induced outcomes still converges appropriately.
The key observation is that, since (\ref{eq:env}) and (\ref{eq:mon})
are limiting analogues of (\ref{eq:freqIC}), the required adjustments
to ensure incentive compatibility become negligible as $N\rightarrow\infty$,
and hence so do their impact on outcomes. This ensures that the outcomes
of the adjusted sequence of mechanisms also converge in distribution
to $(\mathbf{h},a)$.
\begin{rem}
\label{rem:ICL} Because types converge, $\nicefrac{s_{i}}{\sqrt{N}}\rightarrow0$,
each sender's preference over allocations is independent of their
type in the limit. Thus, one can view (\ref{eq:env}) as reflecting
the incentive compatibility of an essentially uninformed sender who
can nonetheless have a small, first-order impact on outcomes via his
report: we describe a deviation by a sender whose misreport shifts
the $k$-NEF $h_{k}$ by an incremental amount $dh_{k}$ as a `\emph{nudge}'
in `direction' $k$. Indeed, (\ref{eq:env}) is equivalent to the
first-order condition of an uninformed sender with the ability to
nudge the distribution of the NEF $\mathbf{h}$ in `direction' $k$;
when this condition holds, the sender has no incentive to make such
a nudge.\footnote{We provide a formal version of this uninformed-sender interpretation
of the problem (and its associated first-order conditions) in Online
Appendix \ref{subsec:remarkproofICL}.} Under this interpretation, (\ref{eq:mon}) is a condition under which
mechanisms obeying (\ref{eq:env}) are robust to the introduction
of infinitesimal amounts of private information.
\end{rem}
The convergence in types also makes clear why it was important for
our analysis to frame incentive constraints in terms of rates of change.
Had we expressed condition (\ref{eq:finite_N_envelope_bounds}) in
terms of differences $U^{N}_{k}-U^{N}_{l}$, the limit would yield
$\lim U^{N}_{k}=\lim U^{N}_{l}$. But this trivially applies to \textit{all}
mechanisms at $N=\infty$ since the sender's types collapse to $0$,
and therefore $\lim U^{N}_{k}=\lim U^{N}_{l}$ would impose no discipline
on incentives at all. For instance, this limit would permit the receiver-preferred
allocation, yet obviously this mechanism is not incentive compatible
for any finite $N$. By contrast, framing the constraint on the margins
imposes meaningful restrictions. For instance, it is not hard to see
that the receiver-preferred mechanism fails (\ref{eq:env}). While
the first-best solution is therefore not feasible, Theorem \ref{thm:A-recommendation-mechanism}
gives us the tools to solve the second-best problem (\ref{eq:P})
in the large.

\section{Optimality in Large Economies\label{sec:Optimality-in-Large}}

Theorem \ref{thm:A-recommendation-mechanism} implies that we can
characterize the optimal mechanism in the limiting economy as $N\rightarrow\infty$
by solving the much simplified problem
\begin{equation}
\max_{\sigma:\mathbb{R}^{K}_{0}\rightarrow\left[0,1\right]}\mathbb{E}\left[\sigma\left(\mathbf{h}\right)\left(\mathbf{h}\cdot\mathbf{t}+r\right)\right]\tag{P1}\label{eq: P1}
\end{equation}
subject to
\begin{equation}
\mathbb{E}\left[\sigma\left(\mathbf{h}\right)\left(\left(\mathbf{h}\cdot\mathbf{t}+b\right)h_{k}/f_{k}-t_{k}\right)\right]=0,\mathbb{E}\left[\sigma\left(\mathbf{h}\right)\frac{h_{k}}{f_{k}}\right]\geq\mathbb{E}\left[\sigma\left(\mathbf{h}\right)\frac{h_{k-1}}{f_{k-1}}\right],\label{eq: LIC}
\end{equation}
where $\mathbf{h}\sim\mathcal{N}\left(\mathbf{0},\mathbf{\Sigma}\right)$.

The following theorem contains our main result on optimal mechanisms
for large economies:
\begin{thm}
\label{thm:The-optimum-is}The optimum in (\ref{eq: P1}) is achieved
by a recommendation mechanism that is a function of only the sample
mean, $\omega=\mathbf{h}\cdot\mathbf{t}$, and satisfies
\[
\sigma^{*}\left(\mathbf{h}\right)=\begin{cases}
1 & \omega=\mathbf{h}\cdot\mathbf{t}\in\left[\underline{\omega},\overline{\omega}\right],\\
0 & \omega\notin\left[\underline{\omega},\overline{\omega}\right].
\end{cases}
\]
Furthermore, when $r\in\left(\frac{b-\sqrt{b^{2}+4}}{2},b\right)$,
then the cutoffs $\underline{\omega},\overline{\omega}$ satisfy $\underline{\omega}\in\left(-b,-r\right)$,
$\overline{\omega}\in\left(-r,\infty\right)$. If $r<\frac{b-\sqrt{b^{2}+4}}{2}$,
then $\underline{\omega}=\overline{\omega}$ and thus $\sigma^{*}\left(\mathbf{h}\right)=0$
almost surely.

Moreover, $\sigma^{*}$ is the unique recommendation mechanism (up
to measure zero changes) that achieves the optimum in (\ref{eq: P1}).
\end{thm}
Theorem \ref{thm:The-optimum-is} shows that a simple mechanism is
optimal. It recommends accept when the aggregate reported state lies
within a bounded interval. We call such mechanisms \emph{interval
mechanisms}. Figure \ref{fig:Optimal-mechanism-in} in the introduction
depicts its key features: relative to the sender-preferred allocation
$\sigma^{S}(\omega)=\mathbf{1}[\omega>-b]$, the mechanism (i) increases
the threshold at which acceptance begins; and (ii) introduces a surplus-burning
region of rejection beyond the second, higher threshold, $\overline{\omega}$.
Despite all parties wanting accept when $\omega>\overline{\omega}$,
this surplus burning at the top acts as a punishment for over-reporting
that allows us to take the receiver's preferred action, reject, when
$\omega\in[-b,\underline{\omega})$. Still, the receiver's payoff
from $\sigma^{*}$ is clearly bound away from the first-best payoff
induced by $\sigma^{R}$. In addition, Theorem \ref{thm:The-optimum-is}
establishes that when $r$ is negative, the bias \textbf{$b-r$} may
be so large that the uninformative mechanism is optimal.

Next we provide intuition for the optimality of the interval mechanism
in three parts: why it is incentive compatible; why it is optimal
among simple mechanisms; and why it is without loss to restrict attention
to simple mechanisms in the large. We conclude the section by showing
that the essential uniqueness of the solution to problem (\ref{eq: P1})
implies that the optimal mechanisms for large finite populations converge
to $\sigma^{*}$ in terms of their outcome distributions.\footnote{Notice, while convergence in outcome distributions does not require
that the $\{\sigma^{N}\}$ converge pointwise to $\sigma^{*}$, it
does imply that the conditional expectation of $\sigma^{N}$ converges
to that of $\sigma$ on every hypercube in $\mathbb{R}^{K}_{0}$---including
all arbitrarily small ones.}

\subsection{Incentive Compatibility and Pivotality\label{subsec:ICandPivot}}

The structure of incentive compatibility for the interval mechanism
$\sigma^{*}$ mirrors the logic of the optimal simple mechanism in
Section \ref{sec:ternaryintuition}. Recall that in the limit, we
can view a misreport by the essentially uninformed sender as a nudge
of the reported $k$-NEF by an incremental amount $dh_{k}$ in direction
$k$ (see Remark \ref{rem:ICL}). However, in a simple mechanism,
this nudge can only affect outcomes via its first-order impact on
the reported state. That is, an upward lie nudges the reported state
by some small increment $d\tilde{\omega}>0$. Under $\sigma^{*}$,
this nudge is pivotal in only two states: the lower and upper thresholds
$\underline{\omega}$ and $\bar{\omega}$. At $\omega=\underline{\omega}$,
such a lie benefits the sender because it nudges the reported state
into the acceptance region. By contrast, at $\omega=\overline{\omega}$,
the same lie induces the punishment of rejection. This reduces the
incentive compatibility condition in (\ref{eq:env}) to a simple condition
\begin{equation}
\phi(\underline{\omega})(\underline{\omega}+b)=\phi(\bar{\omega})(\bar{\omega}+b),\label{eq:pivotlarge}
\end{equation}
where $\phi$ is just the standard normal density, and we have normalized
$\text{var}(\omega)=1$ for notational simplicity. As in the example,
the sender values the action more at $\bar{\omega}$ than at $\underline{\omega}$
but the relative likelihood of $\bar{\omega}$ is lower, ensuring
that when the sender is pivotal, the expected cost of the lie equals
the benefit.

While this pivotality argument is intuitive, formally it follows from
equation (\ref{eq:env}) applied to simple mechanisms:\footnote{To see how (\ref{eq:easy_IC}) follows from (\ref{eq:env}), note
that $\mathbb{E}\left[(\omega+b)\sigma(\omega)\frac{h_{k}}{f_{k}}\right]=\mathbb{E}\left[(\omega+b)\sigma(\omega)\mathbb{E}\left[\frac{h_{k}}{f_{k}}\mid\omega\right]\right]$
by the law of iterated expectations. Moreover, by Lemma \ref{lem:(Donsker.-CLT)-Let}
$h_{k}$ and $\omega$ are jointly Normal with $\text{cov}(h_{k},\omega)=\sum t_{l}\text{cov}(h_{k},h_{l})=t_{k}f_{k}$.
Hence, Normal updating gives $\mathbb{E}\left[\frac{h_{k}}{f_{k}}\mid\omega\right]=t_{k}\frac{\omega}{\text{var}(\omega)}$.
Plugging in to (\ref{eq:env}), noting that the standard Normal density
satisfies $\phi^{\prime}(\frac{\omega}{\text{\text{var}(\ensuremath{\omega})}})=-\frac{\omega}{\text{var}(\omega)}\phi(\frac{\omega}{\text{var}(\omega)})$,
and recalling the normalization $\text{var}(\omega)=1$, establishes
the claim.}

\begin{equation}
-\int^{\infty}_{-\infty}\left(\omega+b\right)\sigma\left(\omega\right)\phi^{\prime}\left(\omega\right)d\omega-\mathbb{E}\left[\sigma\left(\omega\right)\right]=0.\tag{\ensuremath{\omega}-ENV}\label{eq:easy_IC}
\end{equation}
In light of Remark (\ref{rem:ICL}), note that equation (\ref{eq:easy_IC})
is the first-order condition ensuring that the uninformed sender does
not wish to nudge the distribution of the reported state $\tilde{\omega}$
upward. The first term represents the sender's utility from a true
upward shift in the distribution of $\omega$, and the second term
the marginal information rent.\footnote{Formally, the first term is the total derivative of utility with respect
to $\omega$. The second is the partial derivative of utility with
respect $\omega$, i.e., holding the reported distribution constant.
Their difference isolates the partial derivative of utility with respect
to the reported state $\tilde{\omega}$.} Obviously, when the difference is zero, the sender has no incentive
to misreport. We can derive the pivotal incentive compatibility condition
(\ref{eq:pivotlarge}) from (\ref{eq:easy_IC}) using integration
by parts and the fact that $\sigma(\omega)$ is constant on $[\underline{\omega},\bar{\omega}]$.

Condition (\ref{eq:pivotlarge}) also provides a simple intuition
for why $\sigma^{*}$ satisfies the monotonicity condition (\ref{eq:mon}):
since the first-order effect of an increase in the sender's type is
to shift the distribution of $\omega$ rightward by $d\omega$, $\phi(\underline{\omega})\geq\phi(\bar{\omega})$
implies that the probability of acceptance weakly increases in type
(to the first-order). When $\sigma^{*}$ is informative, $\phi(\underline{\omega})>\phi(\bar{\omega})$
and (\ref{eq:mon}) is in fact slack; otherwise, if $\sigma^{*}$
is uninformative ($\underline{\omega}=\overline{\omega}$), then $\phi(\underline{\omega})=\phi(\overline{\omega})$
and (\ref{eq:mon}) is binding.

\subsection{The Value of Punishing at the Top\label{subsec:punishtop}}

Unlike the example in Section \ref{sec:ternaryintuition}, it is not
immediately obvious that this interval mechanism is an improvement
on the sender-preferred allocation $\sigma^{S}$. The total cost of
the surplus-burning region could outweigh the benefit of reducing
acceptance in the disagreement region. Indeed, there are many interval
mechanisms whose thresholds satisfy incentive compatibility and yet
reduce payoffs.\footnote{For instance, if $r=0$, then we can maintain incentive compatibility
while taking both thresholds to $0$, and thus an uninformative payoff.} Moreover, it is even less obvious that interval mechanisms should
be optimal among all simple mechanisms in the large. Fortunately,
clear intuitions for both can be found in the simple economics of
problem (\ref{eq: P1}).

To explain why the mechanism described in Theorem \ref{thm:The-optimum-is}
improves on the sender-preferred mechanism $\sigma^{S}$, consider
an incentive compatible perturbation from $\sigma^{S}$ to a `nearby'
interval mechanism $\sigma^{\prime}=\mathbf{1}[\underline{\omega}^{\prime},\overline{\omega}^{\prime}]$
whose thresholds satisfy $\underline{\omega}^{\prime}=-b+\varepsilon$
and $\overline{\omega}^{\prime}<\infty$, where $\varepsilon>0$ is
chosen arbitrarily small and $\overline{\omega}^{\prime}$ is correspondingly
large. Consider first the effects of increasing the lower threshold
from $-b$ to $\underline{\omega}^{\prime}$. Doing so clearly benefits
the receiver: she is spared a cost of approximately $b-r$ when $\omega\in[-b,\underline{\omega}^{\prime}]$,
an event which occurs with probability close to $\phi(-b)\varepsilon$.
However, this change in isolation also violates incentive compatibility,
because a lie would now be pivotal only when the sender strictly prefers
$a=1$.

Indeed, increasing the lower threshold increases the left side of
the incentive constraint (\ref{eq:easy_IC}) by approximately

\begin{equation}
\left((\omega+b)\phi^{\prime}(\omega)+\phi(\omega)\right)\varepsilon,\label{eq:IC_cross_partial}
\end{equation}
evaluated at $\omega=\underline{\omega}^{\prime}$. This expression
reflects the effect of the change on the sender's payoff from an upward
lie of magnitude $d\tilde{\omega}>0$ relative to truth-telling. To
see this, notice that it reduces the sender's payoffs from both truth-telling
and lying. Under truth-telling, it reduces the sender's payoff by
approximately $(\underline{\omega}^{\prime}+b)\phi(\underline{\omega}^{\prime})\varepsilon$.
On the other hand, the change reduces the payoffs from an upwards
lie by $(\underline{\omega}^{\prime}-d\tilde{\omega}+b)\phi(\underline{\omega}^{\prime}-d\tilde{\omega})\varepsilon$
because the true state at which $\tilde{\omega}=\underline{\omega}^{\prime}$
is now $\underline{\omega}^{\prime}-d\tilde{\omega}$. Notice, the
reduction in $\sigma$ affects the net value of an upward lie in two
ways: it reduces the value of the state at which $a=1$ is foregone
by $d\tilde{\omega}$, and it shifts the likelihood of forgoing $a=1$.
By the product rule, the overall effect on incentives is (\ref{eq:IC_cross_partial}).
Of course, because $\underline{\omega}^{\prime}\approx-b$, the effect
of the change on the sender's incentives reduces to approximately
$\phi(-b)\varepsilon>0$, showing that the net value of a lie has
increased. Thus, the increase in the lower threshold induces a benefit
of approximately $b-r=\frac{(b-r)\phi(-b)\varepsilon}{\phi(-b)\varepsilon}$
per unit of distortion in the sender's incentives.

Of course, the upper threshold $\overline{\omega}^{\prime}$---determined
by condition (\ref{eq:pivotlarge})---is introduced as a deterrent
to lying that restores the sender's incentives. The key observation
is that the unit cost of restoring incentives in this way is negligible,
so that overall the receiver benefits from the perturbation. Indeed,
for a large threshold $\overline{\omega}^{\prime}$, the unit cost
is approximately\footnote{This ratio arises from an exercise very similar to that described
for calculating the unit benefit of raising the lower threshold, albeit
using l'Hopital's rule to approximate the ratio of cost to incentives
when $\overline{\omega}^{\prime}$ is large.}

\begin{equation}
\frac{-(\omega+r)\phi(\omega)}{(\omega+b)\phi^{\prime}(\omega)+\phi(\omega)}=\frac{(\omega+r)}{(\omega+b)\omega-1},\label{eq:unit value perturb}
\end{equation}
evaluated at $\omega=\overline{\omega}^{\prime}$. The numerator represents
the expected cost to the receiver; the denominator the slackening
of the incentive constraint (\ref{eq:easy_IC}); and the right hand
side follows because $\nicefrac{\phi'(\omega)}{\phi(\omega)}=-\omega$.
It is easy to see that this unit cost converges to $0$ as $\omega\rightarrow\infty$.
Hence, this punishment restores incentives at almost no cost to the
receiver. The cost of restoring incentives via surplus burning punishments
at the top is small because the likelihood of punishment in the tail
is very low under truth-telling; yet the relative deterrence effect
is extremely large because the punishment is far more likely after
a lie.\footnote{Formally, this vanishing unit cost relies on the log-concavity of
the Normal distribution: the identity $\nicefrac{\phi'(\omega)}{\phi(\omega)}=-\omega$
implies that the hazard-rate-like ratio in the denominator of (\ref{eq:unit value perturb})
grows without bound, driving the cost to zero. In a moral hazard context,
a similar intuition underlies Mirrlees's `unpleasant theorem' \citep{mirrlees1999moralhazard}:
punishment in the tails is effective because their relative likelihood
(following a deviation) is largest. However, our problem involves
no transfers, so punishments cannot all be loaded onto a single, vanishingly
unlikely event. Instead, they must be applied to a range of events,
leaving our solution bounded strictly away from the first best.}

The efficiency of punishing at the top is central to the optimality
of interval mechanisms among all simple ones. This can be seen by
examining the dual of problem (\ref{eq: P1}), where the simple envelope
condition (\ref{eq:easy_IC}) is used in place of (\ref{eq:env}).
After inspection of (\ref{eq:easy_IC}), it is easy to see that the
derivative of the appropriate Lagrangian with respect to $\sigma(\omega)$
is:
\begin{equation}
\omega+r-\alpha\left(-\left(\omega+b\right)\frac{\phi^{\prime}\left(\omega\right)}{\phi\left(\omega\right)}-1\right)=\omega+r-\alpha\left(\left(\omega+b\right)\omega-1\right),\label{eq:Lagrange_foc}
\end{equation}
where $\alpha>0$ is the associated multiplier for the constraint
(\ref{eq:easy_IC}). This first-order condition expresses the marginal
value of \emph{increasing} $\sigma(\omega)$ at state $\omega$, net
of any shadow cost for violating (\ref{eq:easy_IC}). Moreover, it
is easy to see that (\ref{eq:Lagrange_foc}) is quadratic and hence
positive on at most an intermediate interval of states. When $\omega$
is sufficiently small, increases in $\sigma(\omega)$ are both costly
to the receiver and tighten the sender's incentive constraint. Hence,
it is optimal to set $\sigma(\omega)=0$ there. As $\omega$ increases,
a tension arises between increasing the receiver's payoffs and providing
the sender with incentives: per (\ref{eq:Lagrange_foc}), this tension
is resolved precisely by punishing the sender at at the top, where
the unit cost (\ref{eq:unit value perturb}) to the receiver is smallest.\footnote{Notice, that equation (\ref{eq:unit value perturb}) implies that
the argument for punishing at the top should extend to other preferences
so long as the ratio of receiver-to-senders payoffs does not grow
faster than $\lvert\frac{\phi'(\omega)}{\phi(\omega)}\rvert$.}

\subsection{Optimality of Simple Mechanisms in the Large\label{subsec:Optimality-of-Simple}}

Given the example in Section \ref{sec:ternaryintuition}, it is not
at all obvious that it is without loss of optimality to focus on simple
mechanisms. However, we will see that optimality of simple mechanisms
arises as a consequence of the limiting incentive conditions expressed
by (\ref{eq:env}) and (\ref{eq:mon}).

Consider replacing an ICL mechanism $\sigma(\mathbf{h})$ with its
conditional mean $\bar{\sigma}\left(\omega\right)=\mathbb{E}\left[\sigma\left(\mathbf{h}\right)\mid\omega\right]$.
Clearly, doing so collapses the mechanism to a simple one, and by
the linearity of the receiver's expected utility in $\sigma$, leaves
her payoff unchanged. Hence, the only issue is whether the change
preserves incentives. For an optimal mechanism $\sigma$, this amounts
to checking that (\ref{eq:easy_IC}) holds for $\bar{\sigma}\left(\omega\right)$---because
(\ref{eq:mon}) is slack whenever $\sigma$ is informative, and the
uninformative mechanism is already simple.\footnote{Notice, if $\bar{\sigma}$ satisfies (\ref{eq:easy_IC}), it is feasible
in a relaxed problem in which (\ref{eq:mon}) is dropped. But the
solution to this relaxed problem is $\sigma^{*}$, which as we saw
in Section \ref{subsec:ICandPivot}, satisfies (\ref{eq:mon}). Hence,
if $\bar{\sigma}$ satisfies (\ref{eq:easy_IC}), then $\sigma^{*}(\omega)$
is a simple ICL mechanism with a weakly greater payoff than any ICL
$\sigma(\mathbf{h)}$.} We argue that (\ref{eq:easy_IC}) holds in two steps: first, (\ref{eq:env})
deters what we call \emph{state-shifting nudges}---combined perturbations
whose only aggregate effect is to shift the reported state $\tilde{\omega}$;
and second, because the limit sender is uninformed, the effect of
such a state-shifting nudge is identical whether it is applied through
$\sigma$ or through its state-conditional mean $\bar{\sigma}$.

\emph{Step 1: (\ref{eq:env}) Deters State-Shifting Nudges.} Recall
from Remark \ref{rem:ICL} that condition (\ref{eq:env}) is equivalent
to saying that an uninformed sender could not profit from nudging
the reported NEF $\tilde{\mathbf{h}}$ by an incremental amount $dh_{k}$,
in any direction $k$. These first-order conditions also deter the
sender from perturbing the empirical frequency distribution in \emph{any}
small way: that is, not only are nudges in each direction $k$ deterred,
but linear combinations of those nudges are also deterred. For instance,
he must not wish to increase the frequency of two types at the expense
of two others. In this way, incentive compatibility deters a large
set of complex deviations by the sender. Importantly, this implies
that (\ref{eq:env}) also deters simple state-shifting deviations,
in which the sender's combined nudge only has the effect of shifting
the (normalized) empirical mean of the distribution, $\tilde{\omega}=\mathbf{t}\cdot\tilde{\mathbf{h}}$.

We can construct state-shifting nudges via a linear combination of
deviations. In particular, the sender can do so by nudging each $h_{k}$
by the amount that would be expected following a $d\omega$ shift
of the state, that is, by $d\tilde{h}_{k}=\mathbb{E}\left[h_{k}\mid\omega=d\omega\right]$.
Importantly, this nudge does not change the conditional distribution
of the reported NEF $\tilde{\mathbf{h}}$ given the reported state
$\tilde{\omega}$, so its only effect is on $\tilde{\omega}$, which
shifts by $d\tilde{\omega}=\sum t_{k}d\tilde{h}_{k}$.\footnote{Formally, this is achieved by a nudge $d\mathbf{h}=\mathbf{\mathbf{\beta}}d\omega$
where direction $\beta$ is a $K\times1$ vector of regression coefficients
$\beta_{k}=\frac{\partial\mathbb{E}[h_{k}\mid\omega]}{\partial\omega}$,
so that $\beta_{k}d\omega$ represents the conditional expectation
of $h_{k}$ given an expected state of $d\omega$. Such a nudge implies
the induced distribution of $\tilde{\mathbf{h}}$ conditional on state
$\omega$ is equal to the distribution of the true NEF, $\mathbf{h}$,
conditional on state $\omega+d\omega$. We provide the calculation
in Online Appendix \ref{subsec:Verification-that-nudge}.}

\emph{Step 2: Essentially Uninformed Sender.} Since ICL mechanisms
automatically deter these state-shifting nudges, $\bar{\sigma}$ must
also deter them. From the perspective of an uninformed sender, a state-shifting
nudge implies that for each realization of the state, the probability
of acceptance shifts by $\mathbb{E}\left[\sigma\left(\mathbf{h}\right)\mid\omega+d\omega\right]-\mathbb{E}\left[\sigma\left(\mathbf{h}\right)\mid\omega\right]=\bar{\sigma}(\omega+d\omega)-\bar{\sigma}(\omega)$.
Since condition (\ref{eq:env}) implies this nudge is unprofitable,
then it must also be unprofitable under mechanism $\bar{\sigma}$
to nudge the distribution of $\tilde{\omega}$ upwards by $d\omega$.
Indeed, we show in Online Appendix \ref{subsec:ICL implies easy ICL}
that one can use this exercise to show that if $\sigma$ satisfies
(\ref{eq:env}), then $\bar{\sigma}\left(\omega\right)$ satisfies
(\ref{eq:easy_IC}).

Notice that the two-step argument above relied on the fact that the
sender is essentially uninformed in the limit, i.e., $\mathbb{E}\left[\sigma\left(\mathbf{h}\right)\mid\omega,t_{k}\right]=E\left[\sigma\left(\mathbf{h}\right)\mid\omega\right]$;
hence, from his perspective the state-shifting nudge described above
had expected impact $\bar{\sigma}(\omega+d\omega)-\bar{\sigma}(\omega)$
on acceptance in state $\omega$. By contrast, this equality does
not hold for finite $N$ as each sender has significant private information
about $\mathbf{h}$ conditional on $\omega$. For instance, in the
simple example where $N=2$, a sender can infer from his own signal
exactly what the other sender's signal must be for each realization
of $\omega$. As a result, incentive compatibility is violated by
the simple mechanism $\bar{\sigma}(\omega)=\mathbb{E}\left[\sigma^{\star}\left(\mathbf{h}\right)\mid\omega\right]$
that collapses the optimal mechanism described in Figure \ref{fig:A-fully-optimal}
down to a simple one. Notice, this collapse only changes the probability
of accept along the $\omega=0$ diagonal: it smoothes the probability
of acceptance yielding a constant $\bar{\sigma}(0)=\nicefrac{17}{21}$.
Now, $\bar{\sigma}(0)<\mathbb{E}\left[\sigma^{\star}\left(\mathbf{h}\right)\mid\omega=0,l\right]=\sigma^{\star}\!\bigl(l,h\bigr)$.
Clearly, this reduces the value of an $l$-report. Similarly, it increases
the value of an $m$-report as $\bar{\sigma}(0)>\sigma^{\star}\!\bigl(m,m\bigr)=\mathbb{E}\left[\sigma^{\star}\left(\mathbf{h}\right)\mid\omega=0,m\right]=\nicefrac{3}{7}$.
Because the sender has material private information and influence
when $N=2$, he knows that the effect of the smoothing is to reduce
the chances of acceptance when $\tilde{\omega}=0$ and he reports
$l$, and yet increase it when $\tilde{\omega}=0$ and he reports
$m$. Since type $l$ was indifferent between reporting $l$ and $m$
under $\sigma^{\star}$, he must strictly prefer to report $m$ under
$\bar{\sigma}$. Thus, the private information of the sender prevents
us from collapsing $\sigma^{\star}$ down to a simple mechanism.
\begin{rem}
\label{rem:taylor}As it is without loss to reduce mechanisms down
to simple ones in the limit, one might expect that any violation of
incentive compatibility caused by such a reduction for finite $N$,
should shrink as the population grows. Indeed, we show in Online Appendix
\ref{subsec:taylor} that this is the case.
\end{rem}

\subsection{Convergence Properties of Optimal Finite Mechanisms}

The uniqueness result in Theorem \ref{thm:The-optimum-is} implies
that any sequence of optimal mechanisms, $\sigma^{N,*}$, converge
in the appropriate sense. To see this, note that any sequence of optimal
outcome distributions $(\mathbf{h}^{N},a^{N,*})$ has a convergent
subsequence. This follows from Prokhorov\textquoteright s theorem;
see, for instance, \citet{billingsley2013convergence}. By the almost
everywhere uniqueness of $\sigma^{*}$, every convergent subsequence
converges in distribution to $(\mathbf{h},a^{*})$. But a sequence
converges if and only if every subsequence has a further subsequence
which converges to the same limit.\footnote{This argument is almost identical to the proof of Lemma 2.17 in \citet{guide2006infinite}.}
Thus:
\begin{cor}
Let $(\mathbf{h}^{N},a^{N,*})$ be the outcome induced by optimal
mechanism $\sigma^{N,*}$ for finite $N$, and $(\mathbf{h},a^{*})$
the outcome induced by $\sigma^{*}$. Then, $(\mathbf{h}^{N},a^{N,*})\overset{d}{\longrightarrow}(\mathbf{h},a^{*})$.
\end{cor}
This result implies that finding the optimal ICL mechanism was the
correct exercise for thinking about large economies, because $\sigma^{*}$
approximates the optimal mechanism for large finite populations.

\section{Heterogeneous Bias\label{sec:Heterogeneous-Bias-and}}

The main results of our paper are not tied to the common bias baseline.
To illustrate this, as well as how our ICL approach can be applied
to other settings, we consider a model in which senders' biases are
heterogeneous. Formally, the senders observe independent signals and
have payoffs given by
\[
a\cdot\left(\omega+b_{m}\right)
\]
where $b_{m}$ is the bias of a sender of class $m$ which is publicly
observed by the mechanism and $\omega$ is the same as before. In
this environment, an argument akin to that of Theorem \ref{thm:A-recommendation-mechanism}
implies that the ICL holds if and only if, for all $m$:
\begin{align}
\mathbb{E}\left[\sigma\left(\mathbf{h}_{1},\cdots,\mathbf{h}_{M}\right)\left(\left(\omega+b_{m}\right)\frac{h_{k,m}}{f_{k}}-t_{k}\right)\right] & =0,\label{eq:ICLhet-1}\\
\mathbb{E}\left[\sigma\left(\mathbf{h}_{1},\cdots,\mathbf{h}_{M}\right)\frac{h_{k,m}}{f_{k}}\right] & \text{is increasing in}\:k\nonumber 
\end{align}
where in the above $\mathbf{h}_{m}$ is the NEF vector among the senders
of class $m$.

In this setting, a simple mechanism is one in which the mechanism
$\sigma$ depends on the aggregate report of each group $\omega_{m}$
with
\[
\omega=\frac{\sum^{M}_{m=1}\sqrt{N_{m}}\omega_{m}}{\sqrt{N}}
\]
where $N_{m}$ is the number of senders of class $m$ and $N$ is
the total number of senders. In the Online Appendix \ref{sec:Implications-and-Extensions},
we show that an optimal mechanism is indeed simple in the sense mentioned
here. Moreover, it can be represented by a hyperbola in the space
of the average reports $\left(\omega_{1},\cdots,\omega_{M}\right)$
inside which accept is recommended while reject is again recommended
for extreme values outside of the hyperbola. The following example
illustrates this when there are two bias classes.

\begin{figure}[H]
\begin{centering}
\begin{tikzpicture}
\node[inner sep=0pt] (0,0)
    {\includegraphics[width=.45\textwidth]{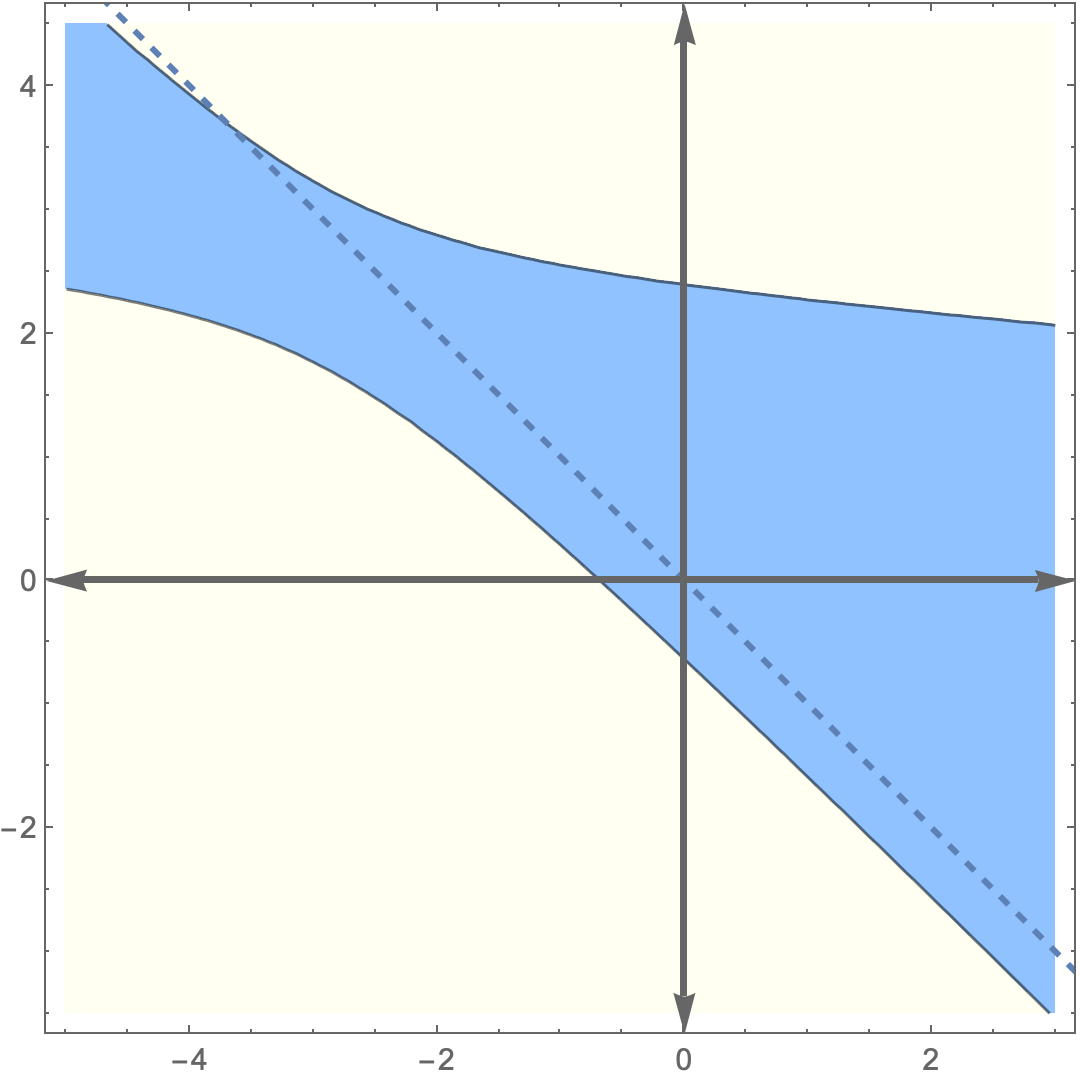}};
\draw[black] (4,-0.5) node[above] {$\omega_1$};
\draw[black] (0.5,3.8) node[right] {$\omega_2$};
\draw[black] (2,3) node[right] {$a=0$};
\draw[black] (-3,-2) node[right] {$a=0$};
\draw[black] (2,.5) node {$a=1$};
\draw[black] (-2.3,3.) node[right] {$\omega=0$};
\end{tikzpicture}
\par\end{centering}
\caption{Regions of action with heterogeneous bias\label{fig:Regions-of-action};
in the white region, $a=0$, while in the blue region, $a=1$. The
parameters are $\text{var}(s_{i})=1$, $b_{1}=0.1,b_{2}=0.3$, and
$r=0$.}
\end{figure}

\begin{example*}
\label{exa:hetbias} Suppose that there are two groups of senders
with biases $b_{1}=0.1,b_{2}=0.3$ and that the two groups of senders
are equal in size. In the large economy, the relevant statistics are
the average reports within each class rather than the full report
profile. Figure \ref{fig:Regions-of-action} depicts the optimal mechanism
in the space of average signal of each group $(\omega_{1},\omega_{2})$.
The acceptance region in this mechanism sits between two parabolas.
Similar to the optimal mechanism in Theorem \ref{thm:The-optimum-is},
the senders are punished with rejection in the unlikely event that
both group means are very high. However, there is a key difference
from the common bias case: the mediator listens more to the less biased
group--- higher values of $\omega$ are more likely to be rewarded
when $\omega_{1}$ is high.
\end{example*}
In the Online Appendix \ref{sec:Implications-and-Extensions}, we
provide the formal version of this model as well as one in which we
allow for a more general class of preferences.

\section{Related Literature\label{sec:Related-Literature}}

A large literature studies how information can be extracted by cross-checking
when agents have correlated information. \citet{cremer1988full} show
that full surplus extraction can be achieved via transfers when each
agent's information statistically identifies the others'. In multi-sender
cheap talk, \citet{krishna2001model}, \citet{battaglini2002multiple},
and \citet{meyer2019robustness} show how punishing disagreement can
support full revelation when senders\textquoteright{} information
is almost identical.\footnote{A notable exception to the result of \citet{battaglini2002multiple}
is provided in \citet{levy2007limits}. In their model, they show
full revelation may be impeded in multi-dimensional talk when the
states are correlated. In addition to the multiple-sender case, and
in contrast to \citet{Chakraborty2010}, they show full revelation
fails in a one sender version of their model. \citet{carroll2019strategic}
show that full information extraction is possible if a single dimension
of information can be verified.} \citet{battaglini2004policy} and \citet{gerardi2009aggregation}
extend this logic to imperfectly correlated information with large
populations. The common force in these papers is that one sender\textquoteright s
report can be compared with others\textquoteright . That is impossible
here, so we cannot implement the receiver-preferred allocation by
cross-checking. Indeed, our optimal mechanisms may punish agreement
rather than disagreement.

Similar to our paper, \citet{wolinsky2002eliciting} and \citet{kattwinkel2024optimal}
examine transfer-free mechanism design for a binary decision problem
with many senders who have common bias. Like us, the senders in \citet{wolinsky2002eliciting}
have \emph{unconditionally} independent signals. However, the signals
are binary and there is verifiable disclosure. He shows that there
is an optimal non-monotone mechanism in his setting, but it consists
of several acceptance intervals rather than one. \citet{kattwinkel2024optimal}
study a problem with binary \emph{conditionally} independent signals
and characterize the optimal mechanism for fixed $N$. Interestingly,
despite allowing for correlated signals, they show that interval mechanisms
are still optimal when the bias is large. While their setting features
partially correlated binary signals, one can show that the reason
for optimality in their setting is similar to that discussed in Section
\ref{sec:Optimality-in-Large} and \ref{subsec:punishtop}. By contrast,
they show that the sender-preferred mechanism can be optimal when
bias is small, whereas it is not optimal for any bias in the large-population
limit that we examine.\footnote{This is a feature of having an arbitrarily large population. As discussed
in Section \ref{sec:Model} and Online Appendix \ref{subsec:Optimality-of-Sender},
sender-preferred can be optimal when $N$ is finite.}

Beyond the differences in the optimal mechanisms, our paper has two
further offsets. First, by allowing for a richer type-space we showed
in Section \ref{sec:ternaryintuition} that it is typically with loss
to restrict attention to \foreignlanguage{american}{simple} mechanisms
in finite populations. However, we showed that as the population grows
large, it is without loss to use simple mechanisms that ignore higher
dimensions of the NEF. In their papers, this distinction does not
arise because the NEF is one-dimensional in the binary-signal case.
Second, we examine informational division. This implies that the conflict
of interest between the senders and receiver stays fixed as the population
grows. By contrast, information is additive in their papers, so the
probability of being in the disagreement region converges to zero
and the sender-preferred allocation converges to the receiver's first
best.

Non-monotonic decision rules can be optimal for other reasons. \citet{chwe2010anonymous}
finds that non-monotonic mechanisms help when the informed agents
have opposing preferences. In the context of protests, \citet{gui2025slippery}
show the government may not want to take an action when it is too
popular because this can encourage revolt by revealing that there
are large numbers of radicals in the population.

Our results can also be applied to voting (cf. Remark \ref{rem:voting}).
In contrast to our results, several papers show how information aggregation
by voters can be asymptotically efficient \citep{feddersen1997voting,mclennan1998consequences,gerardi2009aggregation,barelli2022full,bobkova2024optimality,chen2026communication}.
The seminal paper on this, \citet{feddersen1997voting}, examines
a model of voters with signals about candidate competence (the common
value state) and heterogeneous preferences over candidates. In equilibrium,
voters split into swing voters who vote based on their information,
and partisans who do not. As the population grows, the swing voters
become increasingly homogenous and their proportion shrinks to zero.
Even though the proportion shrinks, the absolute number of swing voters
increases and their collective information perfectly identifies the
state as $N\rightarrow\infty$. As a result, voting in their setting
replicates the perfect-information outcome.

A key difference between our paper and the rest of the literature
on information aggregation is that our limiting exercise captures
informational division rather than addition. In \citet{feddersen1997voting}
for instance, each voter adds to the aggregate information of the
population. This means that a vanishing fraction of voters can still
have almost perfect information collectively. By contrast, we keep
aggregate information constant by reducing signal precision as the
population grows. If information were similarly divided in their setting,
the vanishing fraction of swing voters would be of almost no informational
value.\footnote{Of course, if one were to change the nature of the asymptotic exercise,
then the actual equilibrium behavior would presumably be quite different.} We are not the first to identify barriers to efficient aggregation.
Indeed, \citet{feddersen1997voting} show aggregate uncertainty about
preferences is one such barrier.\footnote{In light of this issue, \citet{kawamura2013eliciting} investigates
the value of surveying strict subsets of the population.} \citet{downs1957economic} identifies costly information acquisition
as another, and \citet{martinelli2006would} studies its implications.
Even when information is costless, \citet{bobkova2024optimality}
shows all threshold voting rules except majority voting fail when
voters choose their information source.

Beyond the difference in results, we provide a new methodology for
studying aggregation by examining the large-population limit directly.
Much of the voting literature instead fixes a voting rule and studies
its asymptotic properties as they\emph{ add} agents with \emph{signals
of fixed precision} \citep{feddersen1997voting,martinelli2006would,gerardi2009aggregation,bobkova2024optimality}.
By contrast, our paper shows how one can use the infinite-population
limit to approximate incentive compatibility for large populations
via the notion of incentive compatibility in the large. This allows
us to examine the full mechanism design problem and thus the optimal
mechanism.

There is also a voting literature that examines the large-population
limit in settings without private information \citep{persson2000comparative,persson2002political}.
In these papers, people vote as if they are pivotal even though they
have no chance of being so. In the absence of private information,
this is innocuous, since it clearly captures incentive compatibility
for large finite populations. However, when there is private information,
voters condition their beliefs on the fact that they are pivotal \citep{austen1996information,feddersen1996swing,feddersen1998convicting,chwe2010anonymous,mcmurray2013aggregating}.
This makes it less obvious how the large-population limit should serve
as an approximation for large finite populations. We show how to adapt
these pivotality-style arguments at the limit.\footnote{\citet{matvejka2021electoral} is one of the few papers on voting
where there the population is infinite and agents are privately informed.
They avoid the problem by assuming that agents do not condition their
beliefs on their own pivotality. While such an approach is not consistent
with Bayesian information aggregation, this is not the focus of their
paper.}

Our work is also related to several other papers that examine the
limiting properties of mechanisms as the population grows. \citet{pesendorfer1997loser,pesendorfer2000efficiency,atakan2014auctions}
study when common value auctions aggregate information efficiently.
\citet{mclean2004informational} show mechanisms that use cross-checking
with side-payments can attain first-best for any population size,
but that the size of the payments goes to zero in the limit. \citet{battaglini2024organizing}
study collective action problems where agents hold private information
about their cost of contribution and optimal mechanisms to mitigate
free-riding. Apart from the obvious differences in subject matter,
our paper provides methods to characterize the asymptotics by studying
an unrestricted mechanism design problem at the limit.

Finally, there is a methodological connection to \citet{frick2026multidimensional}
and \citet{frick2026monitoring} which examine mechanisms as the quality
of the designer's information becomes asymptotically perfect. In the
limit, there is no private information and full surplus extraction
is possible. However, rather than solving for the optimal mechanism,
the authors show that certain fixed simple mechanisms converge quickly
to the first-best. By contrast, we characterize the asymptotic properties
of the optimal mechanism indirectly, by solving a non-trivial limit
problem. Our approaches are complimentary: we identify the point to
which mechanisms must converge if they are to do so optimally. Future
work could combine their methods and ours to identify simple mechanisms
with rapid convergence properties in environments where the limit
problem is non-trivial.

\section{Concluding Discussion \label{sec:Conclusion-and-Extensions}}

Our results show that a decision maker can benefit from information
being divided across multiple senders even when verification by cross-checking
is impossible. The optimal mechanism characterized in Theorem \ref{thm:The-optimum-is}
achieves this gain by punishing consensus in the direction of the
senders' bias.\footnote{In a related applied work on using network structure to detect coordinated
disinformation without direct content verification, \citet{CasillasFarboodiHashemiSaeediWilson2024}
recommend that social media platforms such as X use a mechanism in
which the platform gives warning when many suspicious accounts start
the same information to detect and deter disinformation dissemination.} However, despite the improvement, the optimal mechanism still falls
short of the first best.

Beyond the impossibility of cross-checking, this is because our model
features informational division. If information were additive instead,
conflicts of interest would disappear in the limit and first best
would be achievable even with uncorrelated signals.\footnote{In our setting, informational addition would be a state that is just
the sum of signals, $\omega=\sum^{N}_{i=1}s_{i}$. This would imply
that the probability of the state being in the disagreement region
would go to zero as $N\rightarrow\infty$.} This points to an open question: with correlated signals, can cross-checking
recover first best when the relevant asymptotic exercise still captures
division rather than addition? On one hand, keeping aggregate information
constant would make each individual essentially uninformed in the
limit, which is discouraging for cross-checking. On the other hand,
one can imagine information structures in which conditional correlations
remain strong---because groups of individuals observe essentially
the same signals as each other---even as these signals become uninformative
about the state. Thus, the potential for cross-checking in the limit
may be sensitive to the particulars of the information environment.

This raises a broader question: how should the distribution of information
in large populations be modeled more generally? So far, the literature
has taken what we describe as an additive approach, which implies
any fixed proportion of the population becomes perfectly informed
in the limit. By contrast, informational division implies a fixed
proportion's information is constant. A general framework that nests
these cases may help us better understand information aggregation
in large populations and open up new problems for investigation.

These considerations also suggest that new insights may be gained
by reexamining the literature on information aggregation through the
lens of informational division (or a general framework that nests
it). The papers discussed above span many economic environments, including
voting, common-value auctions, and collective action, and the notion
of incentive compatibility in the large can be applied in them all.
Hence, similar methods to those developed in Theorem \ref{thm:A-recommendation-mechanism}
offer an approach to analyzing division in these settings as well.
Likewise, these methods may also prove useful in online settings where
fake reviews, deep fakes, and bots create new problems for information
aggregation.

\bibliographystyle{ecta}
\bibliography{ocd}

\begin{thebibliography}{59}
\newcommand{\enquote}[1]{``#1''}
\expandafter\ifx\csname natexlab\endcsname\relax\def\natexlab#1{#1}\fi

\bibitem[\protect\citeauthoryear{Aliprantis and Border}{Aliprantis and
  Border}{2006}]{guide2006infinite}
\textsc{Aliprantis, C.~D. and K.~C. Border} (2006): \emph{Infinite dimensional
  analysis: A Hitchhiker's Guide}, Springer.

\bibitem[\protect\citeauthoryear{Ambrus, Azevedo, and Kamada}{Ambrus
  et~al.}{2013}]{ambrus_etal_2013}
\textsc{Ambrus, A., E.~M. Azevedo, and Y.~Kamada} (2013): \enquote{Hierarchical
  Cheap Talk,} \emph{Theoretical Economics}, 8, 233--261.

\bibitem[\protect\citeauthoryear{Antic, Chakraborty, and Harbaugh}{Antic
  et~al.}{2025}]{antic2025subversive}
\textsc{Antic, N., A.~Chakraborty, and R.~Harbaugh} (2025): \enquote{Subversive
  conversations,} \emph{Journal of Political Economy}, 133, 1621--1660.

\bibitem[\protect\citeauthoryear{Atakan and Ekmekci}{Atakan and
  Ekmekci}{2014}]{atakan2014auctions}
\textsc{Atakan, A.~E. and M.~Ekmekci} (2014): \enquote{Auctions, actions, and
  the failure of information aggregation,} \emph{American Economic Review},
  104, 2014--2048.

\bibitem[\protect\citeauthoryear{Aumann and Hart}{Aumann and
  Hart}{2003}]{Aumann2003}
\textsc{Aumann, R.~J. and S.~Hart} (2003): \enquote{{Long cheap talk},}
  \emph{Econometrica}, 71, 1619--1660.

\bibitem[\protect\citeauthoryear{Austen-Smith and Banks}{Austen-Smith and
  Banks}{1996}]{austen1996information}
\textsc{Austen-Smith, D. and J.~S. Banks} (1996): \enquote{Information
  aggregation, rationality, and the Condorcet jury theorem,} \emph{American
  Political Science Review}, 90, 34--45.

\bibitem[\protect\citeauthoryear{Ball}{Ball}{2024}]{ballscoring}
\textsc{Ball, I.} (2024): \enquote{Scoring Strategic Agents,} \emph{American
  Economic Journal: Microeconomics. Forthcoming}.

\bibitem[\protect\citeauthoryear{Barelli, Bhattacharya, and Siga}{Barelli
  et~al.}{2022}]{barelli2022full}
\textsc{Barelli, P., S.~Bhattacharya, and L.~Siga} (2022): \enquote{Full
  information equivalence in large elections,} \emph{Econometrica}, 90,
  2161--2185.

\bibitem[\protect\citeauthoryear{Battaglini}{Battaglini}{2002}]{battaglini2002multiple}
\textsc{Battaglini, M.} (2002): \enquote{Multiple referrals and
  multidimensional cheap talk,} \emph{Econometrica}, 70, 1379--1401.

\bibitem[\protect\citeauthoryear{Battaglini}{Battaglini}{2004}]{battaglini2004policy}
---\hspace{-.1pt}---\hspace{-.1pt}--- (2004): \enquote{Policy advice with
  imperfectly informed experts,} \emph{Advances in Theoretical Economics}, 4.

\bibitem[\protect\citeauthoryear{Battaglini and Palfrey}{Battaglini and
  Palfrey}{2024}]{battaglini2024organizing}
\textsc{Battaglini, M. and T.~R. Palfrey} (2024): \enquote{Organizing for
  Collective Action: Olson Revisited,} \emph{Journal of Political Economy},
  132, 2881--2936.

\bibitem[\protect\citeauthoryear{Best and Quigley}{Best and
  Quigley}{2024}]{best2024persuasion}
\textsc{Best, J. and D.~Quigley} (2024): \enquote{Persuasion for the long run,}
  \emph{Journal of Political Economy}, 132, 1740--1791.

\bibitem[\protect\citeauthoryear{Billingsley}{Billingsley}{1995}]{billingsley1995}
\textsc{Billingsley, P.} (1995): \emph{Probability and measure}, John Wiley \&
  Sons, 3. ed ed.

\bibitem[\protect\citeauthoryear{Billingsley}{Billingsley}{2013}]{billingsley2013convergence}
---\hspace{-.1pt}---\hspace{-.1pt}--- (2013): \emph{Convergence of probability
  measures}, John Wiley \& Sons.

\bibitem[\protect\citeauthoryear{Bobkova}{Bobkova}{2024}]{bobkova2024optimality}
\textsc{Bobkova, N.} (2024): \enquote{The Optimality of Majority Rule: An
  Information-Choice Perspective,} \emph{Available at SSRN 5018717}.

\bibitem[\protect\citeauthoryear{Carroll and Egorov}{Carroll and
  Egorov}{2019}]{carroll2019strategic}
\textsc{Carroll, G. and G.~Egorov} (2019): \enquote{Strategic communication
  with minimal verification,} \emph{Econometrica}, 87, 1867--1892.

\bibitem[\protect\citeauthoryear{Casillas, Farboodi, Hashemi, Saeedi, and
  Wilson}{Casillas et~al.}{2024}]{CasillasFarboodiHashemiSaeediWilson2024}
\textsc{Casillas, A., M.~Farboodi, L.~Hashemi, M.~Saeedi, and S.~Wilson}
  (2024): \enquote{{(Dis)Information Wars},} NBER Working Paper 32896, National
  Bureau of Economic Research.

\bibitem[\protect\citeauthoryear{Chakraborty and Harbaugh}{Chakraborty and
  Harbaugh}{2010}]{Chakraborty2010}
\textsc{Chakraborty, A. and R.~Harbaugh} (2010): \enquote{{Persuasion by cheap
  talk},} \emph{American Economic Review}, 100, 2361--2382.

\bibitem[\protect\citeauthoryear{Chen}{Chen}{2026}]{chen2026communication}
\textsc{Chen, K.} (2026): \enquote{Communication as Voting,} \emph{arXiv
  preprint arXiv:2505.14639}.

\bibitem[\protect\citeauthoryear{Chwe}{Chwe}{2010}]{chwe2010anonymous}
\textsc{Chwe, M. S.-Y.} (2010): \enquote{Anonymous procedures for Condorcet's
  model: Robustness, nonmonotonicity, and optimality,} \emph{Quarterly Journal
  of Political Science}, 5, 45--70.

\bibitem[\protect\citeauthoryear{Corrao and Dai}{Corrao and
  Dai}{2023}]{corrao2023bounds}
\textsc{Corrao, R. and Y.~Dai} (2023): \enquote{The Bounds of Mediated
  Communication,} \emph{arXiv preprint arXiv:2303.06244}.

\bibitem[\protect\citeauthoryear{Cover and Thomas}{Cover and
  Thomas}{2006}]{cover2006elements}
\textsc{Cover, T.~M. and J.~A. Thomas} (2006): \emph{Elements of information
  theory}, John Wiley \& Sons, 2. ed ed.

\bibitem[\protect\citeauthoryear{Crawford and Sobel}{Crawford and
  Sobel}{1982}]{Crawford1982}
\textsc{Crawford, V. and J.~Sobel} (1982): \enquote{{Strategic information
  transmission},} \emph{Econometrica}, 50, 1431--1451.

\bibitem[\protect\citeauthoryear{Cr{\'e}mer and McLean}{Cr{\'e}mer and
  McLean}{1988}]{cremer1988full}
\textsc{Cr{\'e}mer, J. and R.~P. McLean} (1988): \enquote{Full extraction of
  the surplus in Bayesian and dominant strategy auctions,} \emph{Econometrica},
  1247--1257.

\bibitem[\protect\citeauthoryear{Downs}{Downs}{1957}]{downs1957economic}
\textsc{Downs, A.} (1957): \enquote{An economic theory of political action in a
  democracy,} \emph{Journal of political economy}, 65, 135--150.

\bibitem[\protect\citeauthoryear{Feddersen and Pesendorfer}{Feddersen and
  Pesendorfer}{1997}]{feddersen1997voting}
\textsc{Feddersen, T. and W.~Pesendorfer} (1997): \enquote{Voting behavior and
  information aggregation in elections with private information,}
  \emph{Econometrica: Journal of the Econometric Society}, 1029--1058.

\bibitem[\protect\citeauthoryear{Feddersen and Pesendorfer}{Feddersen and
  Pesendorfer}{1998}]{feddersen1998convicting}
---\hspace{-.1pt}---\hspace{-.1pt}--- (1998): \enquote{Convicting the innocent:
  The inferiority of unanimous jury verdicts under strategic voting,}
  \emph{American Political Science Review}, 92, 23--35.

\bibitem[\protect\citeauthoryear{Feddersen and Pesendorfer}{Feddersen and
  Pesendorfer}{1996}]{feddersen1996swing}
\textsc{Feddersen, T.~J. and W.~Pesendorfer} (1996): \enquote{The swing voter's
  curse,} \emph{The American economic review}, 408--424.

\bibitem[\protect\citeauthoryear{Frick, Iijima, and Ishii}{Frick
  et~al.}{2026{\natexlab{a}}}]{frick2026monitoring}
\textsc{Frick, M., R.~Iijima, and Y.~Ishii} (2026{\natexlab{a}}):
  \enquote{Monitoring with rich data,} \emph{arXiv preprint arXiv:2312.16789}.

\bibitem[\protect\citeauthoryear{Frick, Iijima, and Ishii}{Frick
  et~al.}{2026{\natexlab{b}}}]{frick2026multidimensional}
---\hspace{-.1pt}---\hspace{-.1pt}--- (2026{\natexlab{b}}):
  \enquote{Multidimensional Screening with Precise Seller Information,}
  \emph{Econometrica}, 94, 35--70.

\bibitem[\protect\citeauthoryear{Gerardi, McLean, and Postlewaite}{Gerardi
  et~al.}{2009}]{gerardi2009aggregation}
\textsc{Gerardi, D., R.~McLean, and A.~Postlewaite} (2009):
  \enquote{Aggregation of expert opinions,} \emph{Games and Economic Behavior},
  65, 339--371.

\bibitem[\protect\citeauthoryear{Golosov, Skreta, Tsyvinski, and
  Wilson}{Golosov et~al.}{2014}]{golosov_etal_2014}
\textsc{Golosov, M., V.~Skreta, A.~Tsyvinski, and A.~Wilson} (2014):
  \enquote{Dynamic Strategic Information Transmission,} \emph{Journal of
  Economic Theory}, 151, 304--341.

\bibitem[\protect\citeauthoryear{Goltsman, H\"{o}rner, Pavlov, and
  Squintani}{Goltsman et~al.}{2009}]{Goltsman_etal_2009}
\textsc{Goltsman, M., J.~H\"{o}rner, G.~Pavlov, and F.~Squintani} (2009):
  \enquote{Mediation, arbitration and negotiation,} \emph{Journal of Economic
  Theory}, 144, 1397--1420.

\bibitem[\protect\citeauthoryear{Gui and Ma}{Gui and
  Ma}{2025}]{gui2025slippery}
\textsc{Gui, Z. and Z.~Ma} (2025): \enquote{Slippery Protests and Information
  Aggregation,} \emph{Available at SSRN 5315735}.

\bibitem[\protect\citeauthoryear{He, Sandomirskiy, and Tamuz}{He
  et~al.}{2026}]{he2026private}
\textsc{He, K., F.~Sandomirskiy, and O.~Tamuz} (2026): \enquote{Private private
  information,} \emph{Journal of Political Economy}, 134, 000--000.

\bibitem[\protect\citeauthoryear{Kattwinkel and Winter}{Kattwinkel and
  Winter}{2024}]{kattwinkel2024optimal}
\textsc{Kattwinkel, D. and A.~Winter} (2024): \enquote{Optimal Decision
  Mechanisms for Committees: Acquitting the Guilty,} \emph{arXiv preprint
  arXiv:2407.07293}.

\bibitem[\protect\citeauthoryear{Kawamura}{Kawamura}{2013}]{kawamura2013eliciting}
\textsc{Kawamura, K.} (2013): \enquote{Eliciting information from a large
  population,} \emph{Journal of Public Economics}, 103, 44--54.

\bibitem[\protect\citeauthoryear{Krishna and Morgan}{Krishna and
  Morgan}{2001}]{krishna2001model}
\textsc{Krishna, V. and J.~Morgan} (2001): \enquote{A model of expertise,}
  \emph{The Quarterly Journal of Economics}, 116, 747--775.

\bibitem[\protect\citeauthoryear{Krishna and Morgan}{Krishna and
  Morgan}{2004}]{krishna_morgan2004}
---\hspace{-.1pt}---\hspace{-.1pt}--- (2004): \enquote{The art of conversation:
  eliciting information from experts through multi-stage communication,}
  \emph{Journal of Economic Theory}, 117, 147--179.

\bibitem[\protect\citeauthoryear{Levy and Razin}{Levy and
  Razin}{2007}]{levy2007limits}
\textsc{Levy, G. and R.~Razin} (2007): \enquote{On the Limits of Communication
  in Multidimensional Cheap Talk: a Comment,} \emph{Econometrica}, 75,
  885--893.

\bibitem[\protect\citeauthoryear{Luenberger}{Luenberger}{1997}]{luenberger1997optimization}
\textsc{Luenberger, D.~G.} (1997): \emph{Optimization by vector space methods},
  John Wiley \& Sons.

\bibitem[\protect\citeauthoryear{Martinelli}{Martinelli}{2006}]{martinelli2006would}
\textsc{Martinelli, C.} (2006): \enquote{Would rational voters acquire costly
  information?} \emph{Journal of economic theory}, 129, 225--251.

\bibitem[\protect\citeauthoryear{Mat{\v{e}}jka and Tabellini}{Mat{\v{e}}jka and
  Tabellini}{2021}]{matvejka2021electoral}
\textsc{Mat{\v{e}}jka, F. and G.~Tabellini} (2021): \enquote{Electoral
  competition with rationally inattentive voters,} \emph{Journal of the
  European Economic Association}, 19, 1899--1935.

\bibitem[\protect\citeauthoryear{McLean and Postlewaite}{McLean and
  Postlewaite}{2004}]{mclean2004informational}
\textsc{McLean, R. and A.~Postlewaite} (2004): \enquote{Informational size and
  efficient auctions,} \emph{The Review of Economic Studies}, 71, 809--827.

\bibitem[\protect\citeauthoryear{McLennan}{McLennan}{1998}]{mclennan1998consequences}
\textsc{McLennan, A.} (1998): \enquote{Consequences of the Condorcet jury
  theorem for beneficial information aggregation by rational agents,}
  \emph{American Political Science Review}, 92, 413--418.

\bibitem[\protect\citeauthoryear{McMurray}{McMurray}{2013}]{mcmurray2013aggregating}
\textsc{McMurray, J.~C.} (2013): \enquote{Aggregating information by voting:
  The wisdom of the experts versus the wisdom of the masses,} \emph{Review of
  Economic Studies}, 80, 277--312.

\bibitem[\protect\citeauthoryear{Meyer, Moreno~de Barreda, and Nafziger}{Meyer
  et~al.}{2019}]{meyer2019robustness}
\textsc{Meyer, M., I.~Moreno~de Barreda, and J.~Nafziger} (2019):
  \enquote{Robustness of full revelation in multisender cheap talk,}
  \emph{Theoretical Economics}, 14, 1203--1235.

\bibitem[\protect\citeauthoryear{Migrow}{Migrow}{2021}]{migrow2021designing}
\textsc{Migrow, D.} (2021): \enquote{Designing communication hierarchies,}
  \emph{Journal of Economic Theory}, 198, 105349.

\bibitem[\protect\citeauthoryear{Mirrlees}{Mirrlees}{[1975]
  1999}]{mirrlees1999moralhazard}
\textsc{Mirrlees, J.~A.} ([1975] 1999): \enquote{The Theory of Moral Hazard and
  Unobservable Behaviour: Part I,} \emph{The Review of Economic Studies}, 66,
  3--21, paper completed 1975, published 1999.

\bibitem[\protect\citeauthoryear{Myerson}{Myerson}{1981}]{myerson1981optimal}
\textsc{Myerson, R.~B.} (1981): \enquote{Optimal auction design,}
  \emph{Mathematics of operations research}, 6, 58--73.

\bibitem[\protect\citeauthoryear{Myerson}{Myerson}{1982}]{myerson1982optimal}
---\hspace{-.1pt}---\hspace{-.1pt}--- (1982): \enquote{Optimal Coordination
  Mechanisms in Generalized Principal--Agent Problems,} \emph{Journal of
  Mathematical Economics}, 10, 67--81.

\bibitem[\protect\citeauthoryear{Persson, Roland, and Tabellini}{Persson
  et~al.}{2000}]{persson2000comparative}
\textsc{Persson, T., G.~Roland, and G.~Tabellini} (2000): \enquote{Comparative
  politics and public finance,} \emph{Journal of political Economy}, 108,
  1121--1161.

\bibitem[\protect\citeauthoryear{Persson and Tabellini}{Persson and
  Tabellini}{2002}]{persson2002political}
\textsc{Persson, T. and G.~Tabellini} (2002): \emph{Political economics:
  explaining economic policy}, MIT press.

\bibitem[\protect\citeauthoryear{Pesendorfer and Swinkels}{Pesendorfer and
  Swinkels}{1997}]{pesendorfer1997loser}
\textsc{Pesendorfer, W. and J.~M. Swinkels} (1997): \enquote{The loser's curse
  and information aggregation in common value auctions,} \emph{Econometrica:
  Journal of the Econometric Society}, 1247--1281.

\bibitem[\protect\citeauthoryear{Pesendorfer and Swinkels}{Pesendorfer and
  Swinkels}{2000}]{pesendorfer2000efficiency}
---\hspace{-.1pt}---\hspace{-.1pt}--- (2000): \enquote{Efficiency and
  information aggregation in auctions,} \emph{American Economic Review}, 90,
  499--525.

\bibitem[\protect\citeauthoryear{Salamanca}{Salamanca}{2021}]{salamanca2021value}
\textsc{Salamanca, A.} (2021): \enquote{The value of mediated communication,}
  \emph{Journal of Economic Theory}, 192, 105191.

\bibitem[\protect\citeauthoryear{Strack and Yang}{Strack and
  Yang}{2024}]{strack2024privacy}
\textsc{Strack, P. and K.~H. Yang} (2024): \enquote{Privacy-Preserving
  Signals,} \emph{Econometrica}, 92, 1907--1938.

\bibitem[\protect\citeauthoryear{Whitmeyer}{Whitmeyer}{2024}]{whitmeyer2024bayesian}
\textsc{Whitmeyer, M.} (2024): \enquote{Bayesian Elicitation,} \emph{Arizona
  State University Working Paper}.

\bibitem[\protect\citeauthoryear{Wolinsky}{Wolinsky}{2002}]{wolinsky2002eliciting}
\textsc{Wolinsky, A.} (2002): \enquote{Eliciting information from multiple
  experts,} \emph{Games and Economic Behavior}, 41, 141--160.

\end{thebibliography}

\appendix

\section{Proofs\label{sec:Proofs}}

\subsection{Proof of Theorem \ref{thm:A-recommendation-mechanism}}
\begin{proof}
Let $a^{N}\in\{0,1\}$ denote the recommended action induced by finite-$N$
mechanism $\sigma^{N}$. That is, $a^{N}$ is a random variable with
conditional distribution
\[
\Pr\left(a^{N}=1\mid\mathbf{h}^{N}\right)=\sigma^{N}\left(\mathbf{h}^{N}\right).
\]
For any integrable function $g$, the Law of Iterated Expectations
implies that
\begin{equation}
\mathbb{E}^{N}\!\left[\sigma^{N}\!\left(\mathbf{h}^{N}\right)g\!\left(\mathbf{h}^{N}\right)\right]=\mathbb{E}^{N}\!\left[a^{N}g\!\left(\mathbf{h}^{N}\right)\right].\label{eq:aN_id}
\end{equation}
In what follows, this observation will be useful for two reasons.
First, when $g$ is a continuous function of $\mathbf{h}^{N}$, the
function $a^{N}g(\mathbf{h}^{N})$ is also continuous in $\left(\mathbf{h}^{N},a^{N}\right)\in\mathbb{R}^{K}_{0}\times\mathbb{R}$.
Second, since $\{0,1\}$ is metrized by the discrete metric, the space
$\mathbb{R}^{K}_{0}\times\{0,1\}$ is metrizable. As will become clear,
these facts will aid convergence arguments.

\textbf{``Only If'' Direction. }Assume that $\sigma$ is ICL in
the sense of Definition~\ref{def:A-recommendation-mechanism}. Then,
there exists a sequence of incentive compatible mechanisms $\{\sigma^{N}\}_{N\ge1}$
satisfying (\ref{eq:freqIC}) such that 
\[
\left(\mathbf{h}^{N},a^{N}\right)\overset{d}{\longrightarrow}\left(\mathbf{h},a\right),
\]
where $\mathbf{h}\sim\mathcal{N}\left(\mathbf{0},\mathbf{\Sigma}\right)$
(recall Lemma \ref{lem:(Donsker.-CLT)-Let}) and $\Pr(a=1\mid\mathbf{h})=\sigma(\mathbf{h})$.
Similar to $a^{N}$ above, $a$ is the random variable induced by
the limit mechanism $\sigma$ and the random variable $\mathbf{h}$.

For any $k>l$, incentive compatibility requires that $\sigma^{N}$
satisfy the bounds (\ref{eq:finite_N_envelope_bounds}). Using (\ref{eq:derivative}),
(\ref{eq:allocation}) and (\ref{eq:aN_id}) to substitute terms,
(\ref{eq:finite_N_envelope_bounds}) can be equivalently expressed
\begin{align}
\mathbb{E}^{N}\!\left[a^{N}\left(1+\frac{h^{N}_{k}}{f_{k}\sqrt{N}}\right)\right] & \geq\frac{1}{t_{k}-t_{l}}\mathbb{E}^{N}\!\left[a^{N}\left(\mathbf{h}^{N}\cdot\mathbf{t}+b\right)\left(\frac{h^{N}_{k}}{f_{k}}-\frac{h^{N}_{l}}{f_{l}}\right)\right]\nonumber \\
 & \geq\mathbb{E}^{N}\!\left[a^{N}\left(1+\frac{h^{N}_{l}}{f_{l}\sqrt{N}}\right)\right].\label{eq:env_aN}
\end{align}

We now show that (\ref{eq:env}) is satisfied by passing (\ref{eq:env_aN})
to the limit. Each of the three integrands in (\ref{eq:env_aN}) is
continuous in the variables $(a^{N},\mathbf{h}^{N})$. Though these
integrands are clearly not bounded (so we cannot appeal to Portmanteau's
theorem directly), we show that they are \textit{uniformly integrable},
and as a result the desired convergence follows from Theorem 25.12
in \citet{billingsley1995} and the continuous mapping theorem. Recall
that a sequence of random variables $Y_{N}$ are said to be uniformly
integrable if they satisfy
\[
\lim_{M\rightarrow\infty}\sup_{N}\mathbb{E}^{N}\left[\left|Y_{N}\right|\mathbf{1}\left[\left|Y_{N}\right|>M\right]\right]=0
\]
We will do this only for $Y_{N}=a^{N}\left(\mathbf{h}^{N}\cdot\mathbf{t}+b\right)\frac{h^{N}_{k}}{f_{k}}$.
The argument for the other terms in (\ref{eq:env_aN}) are identical.
Note that
\[
\left|Y_{N}\right|\leq\frac{1}{f_{k}}\left|\mathbf{h}^{N}\cdot\mathbf{t}+b\right|\left\Vert \mathbf{h}^{N}\right\Vert \leq\frac{1}{f_{k}}\left(\left\Vert \mathbf{h}^{N}\right\Vert \left\Vert \mathbf{t}\right\Vert +b\right)\left\Vert \mathbf{h}^{N}\right\Vert \leq C(1+\left\Vert \mathbf{h}^{N}\right\Vert ^{2})
\]
where in the above we have used the Cauchy--Schwartz inequality,
the fact that $\left\Vert \mathbf{h}^{N}\right\Vert \leq1+\left\Vert \mathbf{h}^{N}\right\Vert ^{2}$
and $C=\left(\left\Vert t\right\Vert +b\right)/\min f_{k}$.

The above inequality implies that $\left|Y_{N}\right|>M\Rightarrow\sqrt{\frac{M-C}{C}}\leq\left\Vert \mathbf{h}^{N}\right\Vert $.
We can apply Hoeffding's inequality to $\mathbf{h}^{N}$ and thus
we have 
\[
\Pr\left(\lVert\mathbf{h}^{N}\rVert\geq z\right)\leq\Pr\left(\max_{k}\left|h^{N}_{k}\right|\geq\frac{z}{\sqrt{K}}\right)\leq\sum^{K}_{k=1}\Pr\left(\left|h^{N}_{k}\right|\geq\frac{z}{\sqrt{K}}\right)\leq2Ke^{-\frac{2z^{2}}{K}}.
\]
Hence, we can write 
\[
\mathbb{E}^{N}\left[\left|Y_{N}\right|\mathbf{1}\left[\left|Y_{N}\right|>M\right]\right]\leq\mathbb{E}^{N}\left[C(1+\left\Vert \mathbf{h}^{N}\right\Vert ^{2})\mathbf{1}\left[\left\Vert \mathbf{h}^{N}\right\Vert >\sqrt{\frac{M-C}{C}}\right]\right]
\]
if $\mu^{N}$ is the probability distribution of $\mathbf{h}^{N}$,
we can use the earlier version of Hoeffding's inequality and write
the above as
\begin{align}
\mathbb{E}^{N}\left[\left|Y_{N}\right|\mathbf{1}\left[\left|Y_{N}\right|>M\right]\right] & \leq\int_{\lVert\mathbf{h}^{N}\rVert\geq\sqrt{\frac{M-C}{C}}}C\left(1+\lVert\mathbf{h}^{N}\rVert^{2}\right)d\mu^{N}\nonumber \\
 & =\mu^{N}\left(\left\{ \mathbf{h}\mid\lVert\mathbf{h}\rVert\geq\sqrt{\frac{M-C}{C}}\right\} \right)M+\int^{\infty}_{M/C}\mu^{N}\left(\left\{ \mathbf{h}\mid\lVert\mathbf{h}\rVert^{2}+1\geq z\right\} \right)dz\nonumber \\
 & \leq2KMe^{-2\frac{M-C}{C}}+2K\int^{\infty}_{M/C}e^{-2(z-1)}dz=K\left(2M+1\right)e^{-2\frac{M-C}{C}}.\label{eq:hoeffding}
\end{align}
The RHS of (\ref{eq:hoeffding}) converges to 0 as $M\rightarrow\infty$.
This concludes the proof of asymptotic uniform integrability. Therefore,
we may repeatedly apply Theorem 25.12 of \citet{billingsley1995}
to the terms in (\ref{eq:env_aN}) to recover (\ref{eq:env}) as we
pass to the limit. Hence, the argument in the text is valid and (\ref{eq:env})
is established. The monotonicity constraint is established by considering
the outermost inequalities in (\ref{eq:env_aN}) and multiplying by
$\sqrt{N}$. The same convergence argument as above establishes the
claim.

\medskip{}
\textbf{``If'' Direction. }Assume that $\sigma$ satisfies (\ref{eq:env})
and (\ref{eq:mon}). We construct a sequence of finite-$N$ incentive
compatible mechanisms $\{\sigma^{N}\}$ whose induced joint laws converge
to the joint law generated by $\sigma$.

\textbf{\textit{Step 1: Constructing a sequence of mechanisms that
approximate $\sigma$. }}In this step, we find a sequence of mechanisms
$\tilde{\sigma}^{N}$ whose induced joint laws over $(\mathbf{h}^{N},a^{N})$
converge in distribution to $(\mathbf{h},a)$; in subsequent steps
we make small adjustments along this sequence to ensure incentive
compatibility is satisfied while preserving the appropriate convergence
in distribution to $(\mathbf{h},a)$.

To aid this task, it is convenient to construct a probability space
on which the $\mathbf{h}^{N}$ satisfy a strong convergence property.
Since $\mathbf{h}^{N}\overset{d}{\longrightarrow}\mathbf{h}$, we
may apply the Skorokhod representation theorem (\citet{billingsley1995},
Theorem 25.6) to establish the existence of a collection $\left\{ \left(\hat{\mathbf{h}}^{N}\right)^{\infty}_{N=1},\hat{\mathbf{h}}\right\} $
of random vectors on a common probability space such that $\hat{\mathbf{h}}^{N}\overset{d}{=}\mathbf{h}^{N}$
for each $N$, $\hat{\mathbf{h}}\overset{d}{=}\mathbf{h}$, and the
sequence $\left\{ \mathbf{\hat{h}}^{N}\right\} $ converges to $\mathbf{h}$
almost surely. For notational simplicity, we relabel $(\hat{\mathbf{h}}^{N},\hat{\mathbf{h}})$
as $(\mathbf{h}^{N},\mathbf{h})$ below.

On this space, we may also define appropriate action recommendations
as follows. Let $y\sim\mathrm{Unif}[0,1]$ be independent of $\mathbf{h}$
and define
\[
a:=\mathbf{1}\left\{ y\le\sigma\!\left(\mathbf{h}\right)\right\} .
\]
For each $N$, let $\tilde{\sigma}^{N}:\mathbb{R}^{K}_{0}\to[0,1]$
be a (measurable) version of the conditional probability
\[
\tilde{\sigma}^{N}\!\left(\mathbf{h}\right):=\Pr\!\left(a=1\mid\mathbf{h}^{N}=\mathbf{h}\right),
\]
that is, the conditional probability that $a=1$ given observation
of the random variable $\mathbf{h}^{N}$. Let $y^{N}\sim\mathrm{Unif}[0,1]$
be a sequence of independent uniform random variables and set\footnote{$y$ and the $\{y^{N}\}^{\infty}_{N=1}$ are independent of each other
and all other variables, including $\mathbf{h}$ and the $\left\{ \mathbf{h}^{M}\right\} ^{\infty}_{M=1}$
.} 
\[
\tilde{a}^{N}:=\mathbf{1}\left\{ y^{N}\le\tilde{\sigma}^{N}\!\left(\mathbf{h}^{N}\right)\right\} .
\]
By construction, the pair $(\mathbf{h}^{N},\tilde{a}^{N})$ has the
same distribution as $(\mathbf{h}^{N},a)$, and since $\mathbf{h}^{N}\xrightarrow[a.s.]{}\mathbf{h}$
, we have $(\mathbf{h}^{N},a)\overset{d}{\longrightarrow}(\mathbf{h},a)$,\footnote{To see this, observe that the almost sure convergence of $\mathbf{h}^{N}$
to $\mathbf{h}$ implies that for each open hypercube $H\subset\mathbb{R}^{K}_{0}$
and $a\in\{0,1\}$ (and hence for each open subset of $\mathbb{R}^{K}_{0}\times\{0,1\}$),
we have $\Pr(a,\mathbf{h}^{N}\in H)\rightarrow\Pr(a,\mathbf{h}\in H)$.} and therefore
\begin{equation}
\left(\mathbf{h}^{N},\tilde{a}^{N}\right)\overset{d}{\longrightarrow}\left(\mathbf{h},a\right).\label{eq:prelim_joint_conv}
\end{equation}
It remains to show that $\tilde{\sigma}^{N}$ can be adjusted so that
\textit{(i)} the resulting mechanisms satisfy the finite-$N$ incentive
constraints, and \textit{(ii)} the resulting sequence of distributions
over $(\mathbf{h}^{N},a^{N})$ also converge to the distribution over
$(\mathbf{h},a)$.

\textbf{\textit{Step 2: A convenient reframing of incentive compatibility.}}
For each $k\geq2$, define 
\begin{align*}
w_{k}\left(\mathbf{h}\right) & :=g^{k}_{1}(\mathbf{h})-\left(t_{k}-t_{k-1}\right),\\
w_{k,N}\left(\mathbf{h}\right) & :=g^{k}_{1}(\mathbf{h})-\left(t_{k}-t_{k-1}\right)g^{k}_{2}(\mathbf{h},N),
\end{align*}
where $g^{k}_{1}(\mathbf{h}^{N})=\left(\mathbf{h}^{N}\cdot\mathbf{t}+b\right)\left(\frac{h^{N}_{k}}{f_{k}}-\frac{h^{N}_{k-1}}{f_{k-1}}\right)$
and $g^{k}_{2}(\mathbf{h}^{N})=1+\frac{h^{N}_{k-1}}{f_{k-1}\sqrt{N}}$.
For adjacent types $k$ and $k-1$, the lower bound in constraint
(\ref{eq:finite_N_envelope_bounds}) can be written (see (\ref{eq:env_aN}))
in terms of $w_{k,N}$ as 
\begin{equation}
\mathbb{E}^{N}\!\left[\sigma^{N}(\mathbf{h}^{N})w_{k,N}(\mathbf{h}^{N})\right]\ge0,\label{eq:IC_no_downward_deviation}
\end{equation}
for all $k=2,\ldots,K$. If $\sigma^{N}$ satisfies (\ref{eq:IC_no_downward_deviation})
with equality, then the upper bound in (\ref{eq:finite_N_envelope_bounds})
is satisfied if and only if
\begin{equation}
\mathbb{E}^{N}\!\left[\sigma^{N}(\mathbf{h}^{N})(h^{N}_{k}/f_{k}-h^{N}_{k-1}/f_{k-1})\right]\ge0,\label{eq:IC_app_mon}
\end{equation}
for all $k=2,\ldots,K$.\footnote{Note that (\ref{eq:IC_app_mon}) is just constraint (\ref{eq:mon}).}
In the next step, we work with this characterization of incentive
compatibility to identify our sequence of incentive compatible mechanisms
$\sigma^{N}$ whose outcome distributions converge appropriately.

There is no reason that the mechanisms $\tilde{\sigma}^{N}$ identified
in step 1 should satisfy (\ref{eq:IC_no_downward_deviation}) (or
indeed, (\ref{eq:IC_app_mon})). We measure the extent to which (\ref{eq:IC_no_downward_deviation})
differs from $0$ for $\tilde{\sigma}^{N}$, by the difference

\[
\delta_{k,N}:=\mathbb{E}^{N}\!\left[\tilde{\sigma}^{N}\!\left(\mathbf{h}^{N}\right)w_{k,N}\!\left(\mathbf{h}^{N}\right)\right]=\mathbb{E}^{N}\!\left[\tilde{a}^{N}w_{k,N}\!\left(\mathbf{h}^{N}\right)\right],\qquad k=2,\ldots,K,
\]
Since the outcome distributions of $\tilde{\sigma}^{N}$ satisfy (\ref{eq:prelim_joint_conv}),
identical arguments to those developed in the `only if' part apply
so that the expectations of the $w_{k,N}$ converge, and hence
\[
\delta_{k,N}\rightarrow\mathbb{E}\!\left[a\,w_{k}\!\left(\mathbf{h}\right)\right]=\mathbb{E}\!\left[\sigma\!\left(\mathbf{h}\right)w_{k}\!\left(\mathbf{h}\right)\right]=0,
\]
where the last equality is simply a rewriting of ((\ref{eq:env})
).

\textbf{\textit{Step 3: Correcting the $\tilde{\sigma}^{N}$ to satisfy
incentive compatibility.}} We now adjust the $\tilde{\sigma}^{N}$
to develop a new sequence $\sigma^{N}$ which satisfies (\ref{eq:IC_no_downward_deviation})
with equality and (\ref{eq:IC_app_mon}) for all sufficiently large
$N$. This will establish existence of the desired limiting sequence.
To focus on the main substantive arguments, we first show how to construct
the $\sigma^{N}$ under two assumptions, whose purpose is to avoid
boundary issues: (i) there exists a $\varepsilon>0$ such that $\sigma$
and the $\tilde{\sigma}^{N}$ satisfy $\sigma(\mathbf{h}),\tilde{\sigma}^{N}(\mathbf{h})\in\left[\varepsilon,1-\varepsilon\right]$,
for all $\mathbf{h}$ and $N$, and (ii) $\sigma$ satisfies the monotonicity
constraints (\ref{eq:mon}) with strict inequality. After proving
the result for this case (see step 4 below), we describe how to extend
this argument to apply to any $\sigma$ which satisfies (\ref{eq:env})
and (\ref{eq:mon}) in step 5.

We identify values of $\mathbf{h}$ around which we may adjust $\tilde{\sigma}^{N}$
to restore (\ref{eq:IC_no_downward_deviation}) with equality, for
$k=2,\dots,K$, as follows. Define the map $W:\mathbb{R}^{K}_{0}\to\mathbb{R}^{K-1}$
by $W(\mathbf{h})=(w_{2}(\mathbf{h}),\ldots,w_{K}(\mathbf{h}))$.
A direct calculation shows that for any $\mathbf{v}\in\mathbb{R}^{K}_{0}$,
\[
Dw_{k}(0)\cdot\mathbf{v}=b\left(\frac{v_{k}}{f_{k}}-\frac{v_{k-1}}{f_{k-1}}\right),\qquad k=2,\ldots,K.
\]
The matrix $DW(0)$ is invertible. To see this, note that if $DW(0)\mathbf{v}=0$,
then $v_{k}/f_{k}=c$ for all $k$, and so $\sum_{k}v_{k}=c\sum_{k}f_{k}=c=0$
(recall $\sum_{k}v_{k}=0$ for $\mathbf{v}\in\mathbb{R}^{K}_{0}$),
which means $\mathbf{v}=0$. Since $\dim(\mathbb{R}^{K}_{0})=K-1$,
$DW(0)$ has full rank (i.e., is invertible). Thus, by the inverse
function theorem the image of $W$ contains an open set in $\mathbb{R}^{K-1}$.
Thus we may choose points $\bar{\mathbf{h}}^{2},\ldots,\bar{\mathbf{h}}^{K}\in\mathbb{R}^{K}_{0}$
whose corresponding vectors $W(\bar{\mathbf{h}}^{j})$, $j=2,\ldots,K$,
are linearly independent. These $\bar{\mathbf{h}}^{2},\ldots,\bar{\mathbf{h}}^{K}$
are the locations around which our perturbations to the $\tilde{\sigma}^{N}$
will be made.

We adjust $\tilde{\sigma}^{N}$ as follows. Letting $\mu$ denote
the measure on $\mathbb{R}^{K}_{0}$ corresponding to the distribution
$\mathbf{h}\sim N(0,\Sigma)$, choose disjoint open balls $B_{j}$
around the $\bar{\mathbf{h}}^{j}$ with $\mu(B_{j})>0$ and pick bounded,
continuous functions $\psi_{j}:\mathbb{R}^{K}_{0}\to[0,1]$ with $\mathrm{supp}(\psi_{j})\subset B_{j}$
and $\psi_{j}(\bar{\mathbf{h}}^{j})=1$.\footnote{That is, $\psi_{j}(\mathbf{h})=0$ for $\mathbf{h}\notin B_{j}$.}
Since the supports are disjoint, we have $\sum^{K}_{j=2}\psi_{j}(\mathbf{h})\le1$
for all $\mathbf{h}$. By appropriately scaling these functions, we
will be able to make the desired modifications to the $\tilde{\sigma}^{N}$.

To help identify the appropriate scalings, define the matrix $M\in\mathbb{R}^{(K-1)\times(K-1)}$
whose $(k-1,j-1)^{th}$ entry is 
\[
M_{k-1,j-1}:=\mathbb{E}\!\left[w_{k}\!\left(\mathbf{h}\right)\psi_{j}\!\left(\mathbf{h}\right)\right],\qquad k,j=2,\ldots,K.
\]
Since $w_{k}$ is continuous and the $B_{j}$ can be taken arbitrarily
small, $M$ can be made arbitrarily close to the matrix with columns
$W(\bar{\mathbf{h}}^{j})\mathbb{E}[\psi_{j}(\mathbf{h})]$ and therefore
made invertible.\footnote{Recall that the $W(\bar{\mathbf{h}}^{j})$ are linearly independent.
As is well known, scaling vectors by nonzero constants (in this case
the $\mathbb{E}[\psi_{j}(\mathbf{h})]$) preserves their linear independence.}

For each $N$, define the analogous matrix 
\[
(M_{N})_{k-1,j-1}:=\mathbb{E}^{N}\!\left[w_{k,N}\!\left(\mathbf{h}^{N}\right)\psi_{j}\!\left(\mathbf{h}^{N}\right)\right],\qquad k,j=2,\ldots,K.
\]
Because $\psi_{j}$ is bounded and continuous, the same convergence
argument used in the ``only if'' direction applies here and so $M_{N}\to M$
pointwise. Hence $M_{N}$ is also invertible for all sufficiently
large $N$.

Choose the scaling vector $\alpha^{N}:=-M^{-1}_{N}\delta_{N}$, where
$\delta_{N}$ is the $(K-1)\times1$ vector whose entries $\delta_{k-1,N}$,
$k=2,\dots,K$, are defined in step 2. Since $\delta_{N}\rightarrow0$
and $M_{N}\rightarrow M$---where $M$ is invertible, we have $\alpha^{N}\to0$.
We are now ready to define the corrected mechanism 
\begin{equation}
\sigma^{N}\left(\mathbf{h}\right):=\tilde{\sigma}^{N}\left(\mathbf{h}\right)+\sum^{K}_{j=2}\alpha^{N}_{j}\psi_{j}\left(\mathbf{h}\right).\label{eq:sigmaN_corrected}
\end{equation}
Because the supports of $\psi_{j}$ are disjoint, we have $|\sum_{j}\alpha^{N}_{j}\psi_{j}|\le\max_{j}|\alpha^{N}_{j}|$.
Since, by hypothesis, $\tilde{\sigma}^{N}\in[\varepsilon,1-\varepsilon]$
for all $N$, $\alpha^{N}\rightarrow0$ implies that there exists
an $N^{\prime}$ such that for all $N\geq N^{\prime}$, $\sigma^{N}(\mathbf{h})\in[0,1]$.
Moreover, by construction, 
\[
\mathbb{E}^{N}\!\left[\sigma^{N}\!\left(\mathbf{h}^{N}\right)w_{k,N}\!\left(\mathbf{h}^{N}\right)\right]=\delta_{k,N}+M_{(k-1,\cdot);N}\alpha^{N}=0,\qquad k=2,\ldots,K.
\]
where $M_{(k-1,\cdot);N}$ is the $k-1^{th}$ row of $M_{N}$. Thus,
the sequence $\{\sigma^{N}\}$ satisfies (\ref{eq:IC_no_downward_deviation})
with equality for all $N\geq N^{\prime}$.

We still need to show that the sequence $\{\sigma^{N}\}$ satisfies
(\ref{eq:IC_app_mon}) for $N$ sufficiently large. By hypothesis,
the limits 
\[
m_{k}:=\mathbb{E}\!\left[\sigma\!\left(\mathbf{h}\right)\left(\frac{h_{k}}{f_{k}}-\frac{h_{k-1}}{f_{k-1}}\right)\right],\qquad k=2,\ldots,K,
\]
are strictly positive. Applying the same convergence arguments used
in the ``only if'' direction, we have 
\[
\mathbb{E}^{N}\!\left[\tilde{\sigma}^{N}\!\left(\mathbf{h}^{N}\right)\left(\frac{h^{N}_{k}}{f_{k}}-\frac{h^{N}_{k-1}}{f_{k-1}}\right)\right]=\mathbb{E}^{N}\!\left[\tilde{a}^{N}\left(\frac{h^{N}_{k}}{f_{k}}-\frac{h^{N}_{k-1}}{f_{k-1}}\right)\right]\to m_{k}.
\]
Since $\alpha^{N}\rightarrow0$ and the $\psi_{j}$ are continuous
and bounded, the same conclusion extends to $\sigma^{N}$ and thus,
there exists an $N^{\prime\prime}$ such that $\sigma^{N}$ satisfies
(\ref{eq:IC_app_mon}) with strict inequality for all $N\geq N^{\prime\prime}$.
Taking $N\geq N^{\star}:=\max\{N^{\prime},N^{\prime\prime}\}$, the
subsequence $\{\sigma^{N}\}_{N\geq N^{\star}}$ satisfies both (\ref{eq:IC_no_downward_deviation})
(with equality) and (\ref{eq:IC_app_mon}) for all $N\geq N^{\star}$.
By the usual Spence-Mirrlees arguments, this means that, for $N\geq N^{\star}$,
$\sigma^{N}$ satisfies the full set (\ref{eq:finite_N_envelope_bounds})
(equivalently, (\ref{eq:env_aN})) of incentive constraints.

\textbf{\textit{Step 4: Convergence of the induced joint distributions.}}
It remains to show that the joint distributions $(\mathbf{h}^{N},a^{N})$
induced by the sequence $\{\sigma^{N}\}_{N\geq N^{\star}}$ converge
to that of $(\mathbf{h},a)$.\footnote{One can extend the sequence to all $N$ per Definition (\ref{def:A-recommendation-mechanism}),
by simply choosing $\sigma^{N}=\sigma^{S}$ for all $N<N^{\star}$.} To this end, first define the recommendation $a^{N}$ induced by
$\sigma^{N}$ as
\[
a^{N}:=\mathbf{1}\left\{ y^{N}\le\sigma^{N}\!\left(\mathbf{h}^{N}\right)\right\} ,
\]
where the $y^{N}$ are those defined in part 1. Then $\Pr(a^{N}\neq\tilde{a}^{N}\mid\mathbf{h}^{N})=|\sigma^{N}(\mathbf{h}^{N})-\tilde{\sigma}^{N}(\mathbf{h}^{N})|$.
Hence for any bounded continuous function $g:\mathbb{R}^{K}_{0}\times\{0,1\}\rightarrow\mathbb{R}$
where $\sup|g(\mathbf{h},a)|:=G<\infty$, 
\[
\left|\mathbb{E}^{N}\!\left[g\left(\mathbf{h}^{N},a^{N}\right)\right]-\mathbb{E}^{N}\!\left[g\left(\mathbf{h}^{N},\tilde{a}^{N}\right)\right]\right|\le2G\,\mathbb{E}^{N}\!\left[\left|\sigma^{N}\!\left(\mathbf{h}^{N}\right)-\tilde{\sigma}^{N}\!\left(\mathbf{h}^{N}\right)\right|\right].
\]
The right-hand side converges to zero because $|\sigma^{N}-\tilde{\sigma}^{N}|\le\max_{j}|\alpha^{N}_{j}|$
and $\alpha^{N}\to0$ (see equation (\ref{eq:sigmaN_corrected})).
Combining this with (\ref{eq:prelim_joint_conv}) yields 
\[
\left(\mathbf{h}^{N},a^{N}\right)\overset{d}{\longrightarrow}\left(\mathbf{h},a\right).
\]
which is exactly ICL. This proves the result for all $\sigma$ which
satisfy \textit{(i)} and \textit{(ii)} described in step 3.

\textbf{\textit{Step 5: extension to any $\sigma$ satisfying (\ref{eq:env})-(\ref{eq:mon}).
}}Finally, we explain how to adapt the arguments from steps 3 and
4 for all $\sigma$ which satisfy (\ref{eq:env}) -(\ref{eq:mon}).
Take any such $\sigma$, and construct the nearby mechanism 
\[
\sigma^{\prime}_{\varepsilon,\varphi}(\mathbf{h})=\left(1-2\varepsilon-\varphi\right)\sigma(\mathbf{h})+\varepsilon+\varphi\sigma^{S}(\mathbf{h}),
\]
where the scalars $\varepsilon$ and $\varphi$ satisfy $\varepsilon,\varphi>0$
and $2\varepsilon+\phi<1$. $\sigma^{\prime}_{\varepsilon,\varphi}$
is a mixture over three mechanisms: $\sigma$, the sender-preferred
mechanism $\sigma^{S}$, and $\underline{\sigma}=\frac{1}{2}$, which
randomizes between $a\in\{0,1\}$ independently of $\mathbf{h}$.
The mechanisms $\sigma^{S}$ and $\underline{\sigma}$ are trivially
incentive compatible for all $N$, and so unsurprisingly $\sigma^{\prime}_{\varepsilon,\varphi}$
satisfies (\ref{eq:env}) and (\ref{eq:mon}). Moreover, $\sigma^{S}$
satisfies (\ref{eq:mon}) with strict inequality, and therefore so
too does $\sigma^{\prime}_{\varepsilon,\varphi}$. Let $(\mathbf{h},a^{\prime})$
be the joint distribution induced by $\sigma^{\prime}_{\varepsilon,\varphi}$
in the obvious way.

Similarly, replace the approximating sequence $\tilde{\sigma}^{N}$
defined in step 1 with
\[
\tilde{\sigma}^{\prime,N}_{\varepsilon,\varphi}=\left(1-2\varepsilon-\varphi\right)\tilde{\sigma}^{N}(\mathbf{h})+\varepsilon+\varphi\sigma^{S}(\mathbf{h}).
\]
Notice that the $\tilde{\sigma}^{\prime,N}_{\varepsilon,\varphi}$
satisfy $\tilde{\sigma}^{\prime,N}_{\varepsilon,\varphi}(\mathbf{h})\in[\varepsilon,1-\varepsilon]$
for all $\mathbf{h}\in\mathbb{R}^{K}_{0}$. Using essentially the
same arguments in step 1, it is easy to see that the $(\mathbf{h}^{N},\tilde{a}^{\prime,N})$
induced by $\tilde{\sigma}^{\prime,N}_{\varepsilon,\varphi}$ converge
in distribution to the $(\mathbf{h},a^{\prime}_{\varepsilon,\varphi})$
induced by $\sigma^{\prime}$. Moreover, defining the natural analogue
of the differences $\delta_{N}$ from step 2 by
\[
\delta^{\prime}_{k,N}:=\mathbb{E}^{N}\!\left[\tilde{\sigma}^{\prime,N}_{\varepsilon,\varphi}\!\left(\mathbf{h}^{N}\right)w_{k,N}\!\left(\mathbf{h}^{N}\right)\right]=\mathbb{E}^{N}\!\left[\tilde{a}^{\prime,N}_{\varepsilon,\varphi}w_{k,N}\!\left(\mathbf{h}^{N}\right)\right],\qquad k=2,\ldots,K,
\]
it is easily verified that we again have $\delta^{\prime}_{k,N}\rightarrow\mathbb{E}\!\left[\sigma^{\prime}\left(\mathbf{h}\right)w_{k}\!\left(\mathbf{h}\right)\right]=0$.\footnote{Note, we suppress the dependence of $\delta^{\prime}_{k,N}$ on $\varepsilon$,
$\varphi$ for notational ease.} By step 3, this implies that, for each $\varepsilon,\varphi$, there
is a sequence $\alpha^{N}_{\varepsilon,\varphi}\rightarrow0$ which
makes (\ref{eq:IC_no_downward_deviation}) hold with equality for
$\sigma^{\prime,N}_{\varepsilon,\varphi}$ (defined in the obvious
way, per (\ref{eq:sigmaN_corrected})). Since $\sigma^{\prime}_{\varepsilon,\varphi}$
satisfies \textit{(i)} and \textit{(ii)} from step 3, the $\sigma^{\prime,N}_{\varepsilon,\varphi}$
furthermore satisfy $\sigma^{\prime,N}_{\varepsilon,\varphi}(\mathbf{h})\in[0,1]$,
for all $\mathbf{h}$, and (\ref{eq:IC_app_mon}), for all $N$ greater
than a threshold $N^{\star}_{\varepsilon,\varphi}$.

Finally, we identify a sequence $\sigma^{N}$ that approximates $\sigma$
via a diagonal argument. Specifically, set sequences $\varepsilon_{m}=\varphi_{m}=\frac{1}{2^{m+1}}$,
$m=1,2,\dots$, and for each $m$ choose a $N^{m}\geq N^{\star}_{\varepsilon_{m},\varphi_{m}}$
such that $\alpha^{N}_{m}=\alpha^{N}_{\varepsilon_{m},\varphi_{m}}$
further satisfies $\max_{k}\left|(\alpha^{N}_{m})_{k}\right|<\frac{1}{2^{m+1}}$
for all $N\geq N^{m}$, where $(\alpha^{N}_{m})_{k}$ is the $k^{th}$
entry of $\alpha^{N}_{m}$. Furthermore, for each $N\geq N^{1}$,
let $\underline{m}(N)=\max\{N^{m}:N^{m}\leq N\}$, write $\sigma^{\prime,N}_{\varepsilon_{m},\varphi_{m}}=\sigma^{\prime,N}_{m}$,
and define the sequence
\[
\sigma^{N}=\begin{cases}
\sigma^{\prime,N}_{\underline{m}(N)}, & \text{ for }N\geq N^{1}\\
\sigma^{S}, & \text{ otherwise}.
\end{cases}
\]

For $N<N^{1}$, $\sigma^{N}$ is trivially incentive compatible. Thereafter,
by construction, the sequence $\sigma^{N}=\sigma^{\prime,N}_{\underline{m}(N)}=\tilde{\sigma}^{\prime,N}_{\underline{m}(N)}+\sum^{K}_{j=2}(\alpha^{N}_{\underline{m}(N)})_{j}\psi_{j}$,
satisfies (\ref{eq:IC_no_downward_deviation}) (with equality), (\ref{eq:IC_app_mon}),
and $\sigma^{N}\in[0,1]$ for each $N$. Moreover, since $\varepsilon_{\underline{m}(N)},\varphi_{\underline{m}(N)},\alpha^{N}_{\underline{m}(N)}\rightarrow0$,
essentially the same convergence argument in step 4 applies, so that
the $(\mathbf{h}^{N},a^{N})$ induced by the $\sigma^{N}$ convergence
in distribution to $(\mathbf{h},a)$. This concludes the proof.
\end{proof}

\subsection{Proof of Theorem \ref{thm:The-optimum-is}}
\begin{proof}
Note that any $\sigma\left(\mathbf{h}\right)$ that is ICL satisfies
\begin{equation}
\mathbb{E}\left[\sigma\left(\mathbf{h}\right)\left(\left(\mathbf{h}\cdot\mathbf{t}+b\right)\sum_{k}h_{k}t_{k}-\sum_{k}f_{k}t^{2}_{k}\right)\right]=0\Rightarrow\mathbb{E}\left[\sigma\left(\mathbf{h}\right)\left(\left(\omega\left(\mathbf{h}\right)+b\right)\omega\left(\mathbf{h}\right)-\text{Var}\left(s\right)\right)\right]=0.\label{eq: ICLr}
\end{equation}
Now, consider the relaxed optimization where the ICL requirement,
(\ref{eq:env}), is replaced with the above. Optimality implies that
any optimal mechanism $\sigma^{*}$ should satisfy
\[
\sigma^{*}\left(\mathbf{h}\right)=1\Leftrightarrow\omega\left(\mathbf{h}\right)+r-\alpha\left(\left(\omega\left(\mathbf{h}\right)+b\right)\omega\left(\mathbf{h}\right)-\text{Var}\left(s\right)\right)\geq0
\]
for some Lagrange multiplier associated with the relaxed constraint
(\ref{eq: ICLr}). This condition implies that $\sigma^{*}\left(\mathbf{h}\right)$
depends only on $\omega\left(\mathbf{h}\right)$.

Note that if $\alpha>0$ , the above condition defines two cutoffs
for $\underline{\omega}<\overline{\omega}$ that satisfy the following
quadratic equation: 
\begin{equation}
\omega+r+\alpha\text{Var}\left(s\right)-\alpha\left(\omega+b\right)\omega=0.\label{eq:optimal_large_interval}
\end{equation}
For the above to have two roots, we need to have $\left(1-\alpha b\right)^{2}+4\alpha\left(r+\alpha\text{Var}\left(s\right)\right)\geq0$.
As we will show in the Online Appendix, Section \ref{subsec:Existence-multi},
when $b>r>\frac{b-\sqrt{b^{2}+4}}{2}$, for any $\alpha>0$ , this
is the case. Moreover, it has to be that $-b<\underline{\omega}<0<-\underline{\omega}<\overline{\omega}$.
Finally, there is a unique $\alpha$ such that the relaxed version
of ICL, (\ref{eq: ICLr}), holds.

This implies that the optimal mechanism in the relaxed problem must
satisfy $\sigma^{*}\left(\mathbf{h}\right)=1$ when $\omega=\mathbf{h}\cdot\mathbf{t}\in\left[\underline{\omega},\overline{\omega}\right]$,
and $\sigma^{*}\left(\mathbf{h}\right)=0$, otherwise. Note that when
$\sigma\left(\mathbf{h}\right)$ is only a function of the sample
mean, basic properties of the normal distribution implies that $\mathbb{E}\left[h_{k}|\omega\right]=\frac{f_{k}t_{k}}{\text{Var}\left(s\right)}\omega$.
We can use this to show that
\begin{align*}
\mathbb{E}\left[\sigma^{*}\left(\mathbf{h}\right)\left(\left(\omega\left(\mathbf{h}\right)+b\right)\frac{h_{k}}{f_{k}}-t_{k}\right)\right] & =\mathbb{E}\left[\mathbb{E}\left[\sigma^{*}\left(\mathbf{h}\right)\left(\left(\omega\left(\mathbf{h}\right)+b\right)\frac{h_{k}}{f_{k}}-t_{k}\right)|\mathbf{\omega}\left(\mathbf{h}\right)=\omega\right]\right]\\
 & =\mathbb{E}\left[\sigma^{*}\left(\mathbf{h}\right)\left(\left(\omega\left(\mathbf{h}\right)+b\right)\mathbb{E}\left[\frac{h_{k}}{f_{k}}|\mathbf{\omega}\left(\mathbf{h}\right)=\omega\right]-t_{k}\right)\right]\\
 & =\mathbb{E}\left[\sigma^{*}\left(\mathbf{h}\right)\left(\left(\omega\left(\mathbf{h}\right)+b\right)\frac{t_{k}\omega\left(\mathbf{h}\right)}{\text{Var}\left(s\right)}-t_{k}\right)\right]\\
 & =\frac{t_{k}}{\text{Var}\left(s\right)}\mathbb{E}\left[\sigma^{*}\left(\mathbf{h}\right)\left(\left(\omega\left(\mathbf{h}\right)+b\right)\omega\left(\mathbf{h}\right)-\text{Var}\left(s\right)\right)\right]=0,
\end{align*}
where we have used the law of iterated expectations and the fact that
$\sigma^{*}\left(\mathbf{h}\right)$ depends only on $\omega\left(\mathbf{h}\right)$.
Similarly, we can write
\[
\mathbb{E}\left[\sigma^{*}\left(\mathbf{h}\right)\frac{h_{k}}{f_{k}}\right]=\mathbb{E}\left[\sigma^{*}\left(\mathbf{h}\right)\mathbb{E}\left[\frac{h_{k}}{f_{k}}|\omega\right]\right]=\frac{t_{k}}{\text{Var}\left(s_{i}\right)}\mathbb{E}\left[\sigma^{*}\left(\mathbf{h}\right)\omega\right].
\]
Since $\overline{\omega}+\underline{\omega}>0$, $\mathbb{E}\sigma^{*}\left(\mathbf{h}\right)\omega\left(\mathbf{h}\right)>0$
and thus the above expression is increasing in $k$. Therefore, when
$r>\frac{b-\sqrt{b^{2}+4}}{2}$, the solution of the relaxed problem
is also the optimal mechanism.

In the Online Appendix, Section \ref{subsec:Existence-multi}, we
show that $\alpha<0$ is not a possibility. Moreover, whenever $r\leq\frac{b-\sqrt{b^{2}+4}}{2}$,
we must have $\underline{\omega}=\overline{\omega}$, and thus the
solution of the relaxed problem as well as the optimal mechanism is
$\sigma^{*}\equiv0$.
\end{proof}

\section{Three-type Example\label{subsec:Simple-Example-Appendix}}

This subsection verifies: (i) the simple mechanism $\sigma^{\dagger}$
in (\ref{eq:ternary_sdpmech}) is optimal within the class of truthful
obedient simple mechanisms; and (ii) the mechanism $\sigma^{\star}$
in (\ref{eq:ternary_sdpmech}) solves \ref{eq:P}. As a preliminary,
we establish some useful conditions for reference. For $s_{i}\in\{l,m,h\}$,
the interim truth-telling utility of sender $s_{i}$ is

\[
U(s_{i}):=\E\!\left[\sigma(\mathbf{s})\bigl(\omega(\mathbf{s})+b\bigr)\mid s_{i}\right]=\frac{1}{3}\sum_{s_{-i}\in\{l,m,h\}}\sigma(s_{i},s_{-i})\Bigl(\frac{s_{i}+s_{-i}}{\sqrt{2}}+3\Bigr).
\]
The probability of acceptance given report $\tilde{s}$ is:

\[
\E\!\left[\sigma(\mathbf{s})\mid\tilde{s}\right]=\frac{1}{3}\sum_{s_{-i}\in\{l,m,h\}}\sigma(\tilde{s},s_{-i}).
\]
Finally, the incentive compatibility condition (\ref{eq:general_IC})
in this example is equivalent to

\begin{equation}
U(s_{i})-U(\tilde{s})\;\ge\;\frac{s_{i}-\tilde{s}}{\sqrt{2}}\,\E\!\left[\sigma(\mathbf{s})\mid\tilde{s}\right]\qquad\forall\,s_{i},\hat{s}\in\{l,m,h\}.\label{eq:ternary_no_masquerade}
\end{equation}
To see how we get this formulation of incentive compatibility, notice
that if a sender with signal $s_{i}$ reports $\tilde{s}$, then $\omega(s_{i},s_{-i})=\omega(\tilde{s},s_{-i})+\frac{s_{i}-\tilde{s}}{\sqrt{2}}$
for every $s_{-i}$, so his interim payoff from the lie $\tilde{s}$
is just $U(\tilde{s})+\frac{s_{i}-\tilde{s}}{\sqrt{2}}\E\!\left[\sigma(\mathbf{s})\mid\tilde{s}\right]$.

\subsection{The Optimal Simple Mechanism $\sigma^{\dagger}(\omega)$ \label{subsec:Simple-mechanism-Three}}

To show $\sigma^{\dagger}(\omega)$ is optimal, we first consider
a relaxed problem in which we only have to satisfy the two incentive
constraints for the adjacent upward deviations $l$ to $m$ and $m$
to $h$. We then verify that incentive compatibility holds globally
for the solution to the relaxed problem. Applying (\ref{eq:ternary_no_masquerade}),
and rearranging, these incentive constraints are: 
\begin{align}
\sigma(-2)+\sigma(-1)+\sigma(0) & \ge3\sigma(1),\tag{IC-\ensuremath{l\to m}}\label{eq:IC_lm_SD}\\
2\sigma(-1)+\sigma(0)+\sigma(1) & \ge4\sigma(2).\tag{IC-\ensuremath{m\to h}}\label{eq:IC_mh_SD}
\end{align}
The receiver's expected payoff is 
\begin{eqnarray}
\E\!\left[\omega\,\sigma(\omega)\right] & = & \frac{1}{9}\Bigl(-2\sigma(-2)-2\sigma(-1)+2\sigma(1)+2\sigma(2)\Bigr).\label{eq:rpayoff3}
\end{eqnarray}
As $\omega=0$ does not enter the objective, we may set $\sigma(0)=1$
without loss (it only relaxes (\ref{eq:IC_lm_SD})--(\ref{eq:IC_mh_SD})).
With $\sigma(0)=1$, constraints (\ref{eq:IC_lm_SD})--(\ref{eq:IC_mh_SD})
imply 
\[
\sigma(1)\le\frac{1+\sigma(-2)+\sigma(-1)}{3},\qquad\sigma(2)\le\frac{1+\sigma(1)+2\sigma(-1)}{4}.
\]
Since the objective is increasing in $\sigma(1)$ and $\sigma(2)$,
these constraints bind at an optimum. Substituting them into the objective
yields (up to the constant factor $1/9$) 
\[
-2\sigma(-2)-2\sigma(-1)+2\sigma(1)+2\sigma(2)=\frac{4}{3}-\frac{7}{6}\sigma(-2)-\frac{1}{6}\sigma(-1),
\]
which is strictly decreasing in $\sigma(-2)$ and $\sigma(-1)$. Therefore
$\sigma(-2)=\sigma(-1)=0$ at an optimum for this relaxed problem,
and then $\sigma(1)=\sigma(2)=1/3$. This is exactly $\sigma$ in
(\ref{eq:ternary_sdpmech}).

Finally, this mechanism is incentive compatible for all deviations.
First, $\E\!\left[\omega\,\sigma(\mathbf{s})\right]$ is increasing
in reports and incentive compatibility binds for upward adjacent lies,
so incentive compatibility is slack for downward adjacent lies. Second,
a single-crossing property implies local incentive compatibility is
sufficient for global.

\subsection{The Mechanism $\sigma^{\star}$ is Optimal \label{subsec:optimal-three}}

Recall, it is without loss to restrict attention to symmetric mechanisms
with $\sigma(s_{1},s_{2})=\sigma(s_{2},s_{1})$. Under symmetry, it
is therefore without loss to use the simple notation $\sigma(\omega)$
for all $\omega(\mathbf{s})\neq0$. Let $\sigma(-2):=\sigma(l,l)$,
$\sigma(-1):=\sigma(l,m)=\sigma(m,l)$, $\sigma(1):=\sigma(m,h)=\sigma(h,m)$,
and $\sigma(2):=\sigma(h,h)$. However, where $\omega(\mathbf{s})=0$
we separate out the profiles into $\sigma(l,h)=\sigma(h,l)$ and $\sigma(m,m)$,
which need not coincide. Notice, that this implies the receiver's
expected payoff is still given by equation (\ref{eq:rpayoff3}). The
only entries that matter directly for the receiver payoffs are $\sigma(-2),\sigma(-1),\sigma(1),\sigma(2)$;
the $\omega=0$ entries $\sigma(l,h)$ and $\sigma(m,m)$ matter only
through the incentive constraints.

We first establish an upper bound via the relaxed problem where we
only satisfy feasibility and the two adjacent upward deviations $l$
to $m$ and $m$ to $h$; then, we show the upper-bound is tight.
Applying (\ref{eq:ternary_no_masquerade}), and rearranging, these
incentive constraints are: 
\begin{align}
6\sigma(m,m)\le3\sigma(-2)+3\sigma(-1)+9\sigma(l,h)-9\sigma(1), & \tag{IC\ensuremath{^{\star}l\to m}}\label{eq:IC_lm_unres}\\[4pt]
6\sigma(m,m)\ge4\sigma(l,h)+8\sigma(2)-4\sigma(-1)-2\sigma(1). & \tag{IC\ensuremath{^{\star}m\to h}}\label{eq:IC_mh_unres}
\end{align}
Combining these inequalities yields

\begin{equation}
7\sigma(1)+8\sigma(2)\le3\sigma(-2)+7\sigma(-1)+5\sigma(l,h).\label{eq:dual_bound_ternary}
\end{equation}
Using (\ref{eq:dual_bound_ternary}), 
\begin{align*}
\sigma(1)+\sigma(2)-\sigma(-1)-\sigma(-2) & =\frac{1}{7}\Bigl(7\sigma(1)+7\sigma(2)-7\sigma(-1)-7\sigma(-2)\Bigr)\\
 & =\frac{1}{7}\Bigl((7\sigma(1)+8\sigma(2))-\sigma(2)-7\sigma(-1)-7\sigma(-2)\Bigr)\\
 & \le\frac{1}{7}\Bigl(3\sigma(-2)+7\sigma(-1)+5\sigma(l,h)-\sigma(2)-7\sigma(-1)-7\sigma(-2)\Bigr)\\
 & =\frac{1}{7}\Bigl(5\sigma(l,h)-4\sigma(-2)-\sigma(2)\Bigr)\\
 & \le\frac{5}{7},
\end{align*}
where the last inequality uses feasibility: $0\le\sigma(l,h)\le1$
and $\sigma(-2)\ge0$, $\sigma(2)\ge0$. In combination with equation
(\ref{eq:rpayoff3}), this implies $\E\!\left[\omega\,\sigma(\mathbf{s})\right]\le\frac{2}{9}\cdot\frac{5}{7}=\frac{10}{63}$.

To see this upper bound is tight observe: 1) the payoff of $\sigma^{\star}$
is $\E\!\left[\omega\,\sigma^{\star}(\mathbf{s})\right]=\frac{2}{9}\frac{5}{7}=\frac{10}{63}$;
2) incentive compatibility constraints for upward adjacent lies are
tight; 3) as $\E\!\left[\sigma^{\star}(\mathbf{\tilde{s}})\mid\tilde{s_{i}}\right]$
is weakly increasing in reports, incentive compatibility must be slack
for downward adjacent lies; and 4) a single-crossing property implies
local incentive compatibility is sufficient for global incentive compatibility.\newpage{}

\section*{Online Appendix}

\section{Supporting Materials}

\subsection{\label{subsec:remarkproofICL}Remark \ref{rem:ICL}}

To see the alternative interpretation (\ref{eq:env}) of in Remark
(\ref{rem:ICL}), note by Lemma (\ref{lem:(Donsker.-CLT)-Let}) that
such an uninformed sender believes $\mathbf{h}$ follows distribution
$N(\boldsymbol{0},\Sigma)$. Suppose this sender can shift the mean
of $\mathbf{h}$ by an amount $\tilde{\beta}$, so that the reported
NEF $\tilde{\mathbf{h}}$ will follow distribution $N(\tilde{\beta},\Sigma)$.
For this nudged distribution to correspond to a NEF, its entries must
sum to $0$. Thus, we consider nudges which satisfy $\tilde{\beta}_{l}=-\sum_{j\neq l}\tilde{\beta}_{j}$.

We show that (\ref{eq:env}) corresponds to the claimed set of first
order conditions. To do so, note that if this uninformed sender chooses
nudge $\tilde{\beta}$, his expected utility in a mechanism $\sigma$
is $U\left(\tilde{\beta}\right)=\mathbb{E}\left[(\boldsymbol{h}\cdot\boldsymbol{t}+b)\sigma(\mathbf{h}+\tilde{\beta})\right].$
Using the change of variables $h=\tilde{\boldsymbol{h}}-\tilde{\beta}$
and plugging in the constraints $h_{l}=-\sum_{j\neq l}h_{j}$ and
$\tilde{\beta}_{l}=-\sum_{j\neq l}\tilde{\beta}_{j}$, this can be
re-expressed
\begin{equation}
U\left(\tilde{\mathbf{\beta}}_{-l}\right)=\mathbb{E}\left[(\tilde{\boldsymbol{h}}_{-l}\cdot\hat{\boldsymbol{t}}+b)\sigma(\tilde{\mathbf{h}}_{-l})\right]-\left(\tilde{\mathbf{\beta}}_{-l}\cdot\hat{\mathbf{t}}\right)\mathbb{E}\left[\sigma\left(\tilde{\boldsymbol{h}}_{-l}\right)\right],\label{eq:uninformed_sender_payoff}
\end{equation}
where $\mathbf{h}_{-l}$, $\tilde{\boldsymbol{h}}_{-l}$ and $\tilde{\mathbf{\beta}}_{-l}$
are $(K-1)\times1$ sub-vectors of $\mathbf{h}$, $\tilde{\mathbf{h}}$
and $\tilde{\mathbf{\beta}}$ whose $l^{th}$ row has been removed,
and $\hat{\mathbf{t}}=t_{-l}-t_{l}\mathbf{1}_{-l}$, where $t_{-l}$
and $\mathbf{1}_{-l}$ are the corresponding $(K-1)\times1$ sub-vectors
of $\mathbf{t}$ and $\mathbf{1}$.\foreignlanguage{american}{ $\mathbf{h}_{-l}$
is distributed according to $\mathcal{N}(\boldsymbol{0},\hat{\Sigma})$,
where $\hat{\Sigma}$ is the $(K-1)\times(K-1)$ matrix formed by
deletion of the $l^{th}$ row and column from $\Sigma$. Moreover,
covariance matrix $\hat{\Sigma}$ is invertible, with entry $\alpha_{kj}=\frac{1}{f_{k}}\mathbf{1}(k=j)+\frac{1}{f_{l}}$
in the $k^{th}$ row, and $j^{th}$ column of $\hat{\Sigma}$.}\footnote{\selectlanguage{american}%
Indeed, direct calculation easily verifies that $\hat{\Sigma}\cdot\hat{\Sigma}^{-1}=\hat{\Sigma}^{-1}\hat{\Sigma}=I$.\selectlanguage{english}%
}\foreignlanguage{american}{ We are now ready to differentiate (\ref{eq:uninformed_sender_payoff})
with respect to $\tilde{\beta}_{k}$. First consider the term
\[
\mathbb{E}\left[(\tilde{\boldsymbol{h}}_{-l}\cdot\hat{\boldsymbol{t}}+b)\sigma(\tilde{\mathbf{h}}_{-l})\right]=C\intop_{\mathbb{R}^{K}}(\tilde{\boldsymbol{h}}_{-l}\cdot\hat{\boldsymbol{t}}+b)\sigma(\tilde{\boldsymbol{h}}_{-l})e^{-\frac{1}{2}\left(\tilde{\boldsymbol{h}}_{-l}-\tilde{\mathbf{\beta}}_{-l}\right)^{\top}\hat{\Sigma}^{-1}\left(\tilde{\boldsymbol{h}}_{-l}-\tilde{\mathbf{\beta}}_{-l}\right)}d\tilde{\boldsymbol{h}}_{-l},
\]
where $C=\frac{1}{\left(2\pi\right)^{\frac{K-1}{2}}\text{det}(\hat{\Sigma})^{\frac{1}{2}}}$
is a coefficient of the joint Normal distribution. Differentiation
with respect to $\tilde{\beta}_{k}$, evaluated at $\tilde{\mathbf{\beta}}_{-l}=\boldsymbol{0}$,
yields
\begin{eqnarray*}
 & C\intop_{\mathbb{R}^{K}}(\tilde{\boldsymbol{h}}_{-l}\cdot\hat{\boldsymbol{t}}+b)\sigma(\tilde{\boldsymbol{h}})e^{-\frac{1}{2}\tilde{\boldsymbol{h}}_{-l}\hat{\Sigma}^{-1}\tilde{\boldsymbol{h}}_{-l}}\times\boldsymbol{e}^{\top}_{k}\hat{\Sigma}^{-1}\tilde{\boldsymbol{h}}_{-l}d\tilde{\boldsymbol{h}}_{-l}\\
= & \mathbb{E}\left[(\boldsymbol{h}_{-K}\cdot\hat{\boldsymbol{t}}+b)\sigma(\boldsymbol{h})\left(\frac{h_{k}}{f_{k}}-\frac{h_{l}}{f_{l}}\right)\right]
\end{eqnarray*}
where $\mathbf{e}_{k}$ is a basis vector for the row in which $\tilde{\beta}_{k}$
enters $\tilde{\mathbf{\beta}}$. The first line follows after differentiation
of the quadratic form in $\tilde{\beta}_{k}$, and the second from
$\boldsymbol{e}^{\top}_{k}\hat{\Sigma}^{-1}\tilde{\boldsymbol{h}}_{-l}=\frac{h_{k}}{f_{k}}-\frac{h_{l}}{f_{l}}$.}\footnote{\selectlanguage{american}%
Observe that $\boldsymbol{e}^{\top}_{k}\hat{\Sigma}^{-1}\tilde{\boldsymbol{h}}_{-l}=\sum_{j\neq l}\alpha_{kj}\tilde{h}_{j}=\frac{h_{k}}{f_{k}}+\frac{1}{f_{l}}\sum_{j\neq l}h_{j}=\frac{h_{k}}{f_{k}}-\frac{h_{l}}{f_{l}}$.\selectlanguage{english}%
}\foreignlanguage{american}{ As for the second term in (\ref{eq:uninformed_sender_payoff}),
$-\left(\tilde{\mathbf{\beta}}_{-l}\cdot\hat{\boldsymbol{t}}\right)\mathbb{E}\left[\sigma(\tilde{\boldsymbol{h}})\right]$,
a trivial application of the product rule for differentiation shows
that this differentiates to
\[
-(t_{k}-t_{l})\mathbb{E}[\sigma(\boldsymbol{h})].
\]
Thus, the first order condition at $d\mathbf{h}_{-l}=0$ is 
\[
\mathbb{E}\left[(\boldsymbol{h}_{-l}\cdot\hat{\boldsymbol{t}}+b)\sigma(\boldsymbol{h})\left(\frac{h_{k}}{f_{k}}-\frac{h_{l}}{f_{l}}\right)\right]-(t_{k}-t_{l})\mathbb{E}[\sigma(\boldsymbol{h})].
\]
To get (\ref{eq:uninformed_sender_payoff}), integrate out $t_{l}$
as described in section (\ref{sec:The-Value-of_mediator}).}

\subsection{Simple Mechanisms in the Large Economy Limit \label{subsec:Simple-Mechanisms-appendix}}

\subsubsection{\label{subsec:ICL implies easy ICL}If $\sigma(\mathbf{h})$ satisfies
\ref{eq:env}, then $\bar{\sigma}(\omega)$ satisfies \ref{eq:easy_IC}}

Consider a marginal shift in the reported frequency distribution so
that \foreignlanguage{american}{$\tilde{\boldsymbol{h}}=\boldsymbol{h}+\boldsymbol{\beta}d\omega$,
where }$\boldsymbol{\beta}$ is a vector of regression coefficients
$\beta_{k}=\frac{\partial\mathbb{E}[h_{k}\mid\omega]}{\partial\omega}=\frac{\text{cov}(h_{k},\omega)}{\text{var}(\omega)}=\frac{t_{k}f_{k}}{var(\omega)}$.\foreignlanguage{american}{
After some algebra, we can directly evaluate the following $\beta_{k}$-weighted
sum as:}

\selectlanguage{american}%
\begin{eqnarray*}
\sum^{K}_{k=1}\beta_{k}\mathbb{E}\left[\left(\omega+b\right)\sigma\left(\mathbf{h}\right)\frac{h_{k}}{f_{k}}\right] & = & \mathbb{E}\left[\left(\omega+b\right)\sigma\left(\mathbf{h}\right)\sum^{K}_{k=1}\frac{t_{k}f_{k}}{\text{var}\left(\omega\right)}\frac{h_{k}}{f_{k}}\right]\\
 & = & \frac{1}{\text{var}\left(\omega\right)}\mathbb{E}\left[\left(\omega+b\right)\omega\sigma\left(\mathbf{h}\right)\right]\\
 & = & -\intop\left(\omega+b\right)\,\mathbb{E}\left[\sigma\left(\mathbf{h}\right)\mid\omega\right]\phi^{\prime}\left(\frac{\omega}{\text{var}\left(\omega\right)}\right)d\omega.
\end{eqnarray*}
where the final line follows after noting that the Normal density
$\phi$ satisfies $\phi^{\prime}\left(\frac{\omega}{\text{var}(\omega)}\right)=-\frac{\omega}{\text{var}\left(\omega\right)}\phi\left(\frac{\omega}{\text{var}\left(\omega\right)}\right)$
and uses the Law of Iterated Expectations. On the other hand, evaluating
the same sum using (\ref{eq:env}) yields
\begin{eqnarray*}
\sum^{K}_{k=1}\beta_{k}\mathbb{E}\left[\left(\omega+b\right)\sigma\left(\mathbf{h}\right)\frac{h_{k}}{f_{k}}\right] & = & \sum^{K}_{k=1}\beta_{k}t_{k}\mathbb{E}\left[\sigma\left(\mathbf{h}\right)\right]\\
 & = & \mathbb{E}\left[\mathbb{E}\left[\sigma\left(\mathbf{h}\right)\mid\omega\right]\right],
\end{eqnarray*}
where the second line uses $\sum^{K}_{k=1}t_{k}\beta_{k}=1$ and the
Law of Iterated Expectations. Equating the two derived expressions,
recalling that $\mathbb{E}\left[\sigma(\mathbf{h})\mid\omega\right]$
and comparing to the state-dependent version (\ref{eq:easy_IC}) of
the envelope condition, establishes the claim.

\subsubsection{Verification that nudge in direction $\beta$ shifts $\tilde{\omega}$
but preserves conditional distributions $\tilde{\mathbf{h}}\mid\tilde{\omega}$\label{subsec:Verification-that-nudge}}

\selectlanguage{english}%
Though the derivation (\ref{subsec:ICL implies easy ICL}) already
proves that $\bar{\sigma}$ satisfies (\ref{eq:easy_IC}), described
an underlying intuition based on the statistical properties of the
considered nudge. Here, we briefly verify the described effects of
a nudge in the direction $\tilde{\mathbf{\beta}}=\mathbf{\beta}dw$.

\selectlanguage{american}%
$\mathbf{h}$ and $\omega$ are jointly normal (Lemma (\ref{lem:(Donsker.-CLT)-Let})),
and the conditional distribution of $\mathbf{h}\mid\omega$ is well-known
to be $\mathcal{N}(\boldsymbol{\beta}\omega,\Sigma-\boldsymbol{\beta}\boldsymbol{\beta}^{\top}\text{Var}(\omega))$.
Consider the random variables $\tilde{\boldsymbol{h}}=\boldsymbol{h}+\boldsymbol{\beta}d\omega$,
which distorts the reported frequencies in the direction of vector
$\tilde{\mathbf{\beta}}=\boldsymbol{\beta}d\omega$, and $\tilde{\omega}=\tilde{\boldsymbol{h}}\cdot\boldsymbol{t}$.
Observe that $\sum t_{k}\beta_{k}=1$, and therefore $\tilde{\omega}=\omega+d\omega$.

Now consider the conditional distribution of $\tilde{\boldsymbol{h}}$
given $\tilde{\omega}$. In the move from $\boldsymbol{h}$ to $\tilde{\boldsymbol{h}}$,
and $\omega$ to $\tilde{\omega}$, the means of these random variables
have changed in the obvious fashion, while the covariances are unaffected.
The variables are still jointly Normal, and the conditional distribution
is 
\[
\tilde{\boldsymbol{h}}\mid\tilde{\omega}\sim\mathcal{N}\left(\boldsymbol{\beta}d\omega+\boldsymbol{\beta}\left(\tilde{\omega}-d\omega\right),\Sigma-\boldsymbol{\beta}\boldsymbol{\beta}^{\top}\text{Var}(\omega)\right)=\mathcal{N}\left(\boldsymbol{\beta}\tilde{\omega},\Sigma-\boldsymbol{\beta}\boldsymbol{\beta}^{\top}\text{Var}(\omega)\right).
\]

Hence, given a reported state $\tilde{\omega}$, the joint distribution
over $\tilde{h}$ is exactly the same as it would have been if the
agent did not distort the mechanism. Of course, since the conditional
distributions are unchanged, this implies that the expectations of
any function of $\boldsymbol{h}$ (such as $\sigma$) are also unchanged.
\selectlanguage{english}%

\subsubsection{Remark \ref{rem:taylor}\label{subsec:taylor}}

Consider a mechanism $\sigma^{N}$ that is incentive compatible. Then,
consider an alternative mechanism $\bar{\sigma}^{N}\left(\hat{\omega}\right)=\mathbb{E}^{N}\left[\sigma^{N}\left(\mathbf{h}\right)|\omega=\omega\left(\mathbf{h}\right)\right]$
that replaces the probability of accept for each realization $\mathbf{h}$
with the probability of accept conditional on its associated state
value $\omega(\mathbf{h})$. Since the receiver only cares about $\omega$,
this mechanism provides her with the same payoffs. However, this modification
of the mechanism may no longer satisfy sender incentive compatibility.
Nonetheless, it has a notable property, specifically, that the effect
of an individual sender's information, $s_{i}$, on the mechanism
or the payoffs is an order of magnitude lower than it would be if
we were simply controlling for $\omega$.

To see this, consider the payoff of sender $i$, whose signal realization
$s_{i}$ takes a value $t_{k}\in S$ when he reports this signal truthfully.
This is given by
\[
U_{k}=\mathbb{E}^{N}\left[\sigma^{N}\left(\mathbf{h}\right)\left(\omega+b\right)|s_{i}=t_{k}\right].
\]
Using the law of iterated expectations, we can write the above as
\[
U_{k}=\mathbb{E}^{N}\left[\mathbb{E}^{N}\left[\sigma^{N}\left(\mathbf{h}\right)|\omega,s_{i}=t_{k}\right]\left(\omega+b\right)|s_{i}=t_{k}\right].
\]
We can also write
\begin{align*}
\mathbb{E}^{N}\left[\sigma^{N}\left(\mathbf{h}\right)|\omega,s_{i}=t_{k}\right] & =\frac{\sum_{\mathbf{h}:\mathbf{h}\cdot\mathbf{t}=\omega}\text{Pr}^{N}\left(\mathbf{h}|s_{i}=t_{k}\right)\sigma^{N}\left(\mathbf{h}\right)}{\sum_{\mathbf{h}:\mathbf{h}\cdot\mathbf{t}=\omega}\text{Pr}^{N}\left(\mathbf{h}|s_{i}=t_{k}\right)}\\
 & =\frac{\sum_{\mathbf{h}:\mathbf{h}\cdot\mathbf{t}=\omega}\left(\frac{h_{k}}{f_{k}\sqrt{N}}+1\right)\text{Pr}^{N}\left(\mathbf{h}\right)\sigma^{N}\left(\mathbf{h}\right)}{\sum_{\mathbf{h}:\mathbf{h}\cdot\mathbf{t}=\omega}\left(\frac{h_{k}}{f_{k}\sqrt{N}}+1\right)\text{Pr}^{N}\left(\mathbf{h}\right)}.
\end{align*}
We can view the above as a function of $\frac{1}{\sqrt{N}}$ (without
considering the effect of changes of $N$ on the probabilities, $\text{Pr}^{N}$),
and a Taylor approximation around $1/\sqrt{N}=0$ implies that
\begin{align*}
\mathbb{E}^{N}\left[\sigma^{N}\left(\mathbf{h}\right)|\omega,s_{i}=t_{k}\right] & =\bar{\sigma}^{N}\left(\omega\right)+\frac{1}{\sqrt{N}}\text{cov}^{N}\left(\frac{h_{k}}{f_{k}},\sigma^{N}\left(\mathbf{h}\right)|\omega\right)+O\left(\frac{1}{N}\right).
\end{align*}
The above states that controlling for $\omega$, the additional information
contained in $s_{i}=t_{k}$ affects the expected probability of recommending
$a=1$ only by an order of $1/\sqrt{N}$, and thus this impact shrinks
at that rate as $N$ converges to infinity. A similar property holds
for the payoff of the senders of a given type under truth-telling:
\begin{align*}
U_{k} & =\mathbb{E}^{N}\left[\mathbb{E}^{N}\left[\sigma^{N}\left(\mathbf{h}\right)|\omega,s_{i}=t_{k}\right]\left(\omega+b\right)|s_{i}=t_{k}\right]\\
 & =\mathbb{E}^{N}\left[\left(\bar{\sigma}^{N}\left(\omega\right)+\frac{\text{cov}^{N}\left(\frac{h_{k}}{f_{k}},\sigma^{N}\left(\mathbf{h}\right)|\omega\right)+O\left(1/\sqrt{N}\right)}{\sqrt{N}}\right)\left(\omega+b\right)|s_{i}=t_{k}\right].\\
 & =\mathbb{E}^{N}\left[\left(\frac{h_{k}}{f_{k}\sqrt{N}}+1\right)\bar{\sigma}^{N}\left(\omega\right)\left(\omega+b\right)\right]+O\left(1/N\right)
\end{align*}
In arriving at the last expression, we have used the typical property
of the multinomial distribution in (\ref{eq:conditional_h_prob}).
Recall that we can cast the incentive compatibility using its envelope
form and write
\[
U_{k}-U_{l}\geq\frac{t_{k}-t_{l}}{\sqrt{N}}\mathbb{E}^{N}\left[\sigma^{N}\left(\mathbf{h}\right)|s_{i}=t_{l}\right].
\]
The right-hand side represents the extra surplus that a sender of
type $k$ is able to guarantee by pretending to be of type $l$. Therefore,
it determines the marginal information rent captured by that sender.
Applying the aforementioned logic to this incentive compatibility
has a few implications as $N$ becomes large. First, the right-hand
side of the above can be replaced by $\frac{t_{k}-t_{l}}{\sqrt{N}}\mathbb{E}^{N}\left[\bar{\sigma}^{N}\left(\omega\right)\right]+O\left(1/N\right)$.
Second, the left-hand side can be replaced by $\mathbb{E}^{N}\left[\left(\frac{h_{k}}{f_{k}\sqrt{N}}-\frac{h_{l}}{f_{l}\sqrt{N}}\right)\bar{\sigma}^{N}\left(\omega\right)\left(\omega+b\right)\right]+O\left(1/N\right)$.
In other words, the conditional mean of $\sigma^{N}$ is the main
determinant of incentive provision up to a first order, and all of
its higher moments---which fully determine the distribution of $\sigma^{N}$---affect
the incentives with a lower order of magnitude. Since as we take limit,
only the first-order effects become relevant, we can simply replace
$\sigma^{N}$ with $\bar{\sigma}^{N},$ and the resulting limit (as
$N\rightarrow\infty$) will satisfy the conditions in Theorem \ref{thm:A-recommendation-mechanism}.

\section{Extensions\label{sec:Implications-and-Extensions}}

In this section, we extend our analysis in two directions. In Section
\ref{subsec:Heterogeneous-Bias}, we study optimal mechanisms in the
large when bias is heterogeneous across senders. In Section \ref{subsec:General-Preferences},
we study the problem under more general preferences.

\subsection{Heterogeneous Bias\label{subsec:Heterogeneous-Bias}}

Here, we consider an extension where the senders' bias is heterogeneous
but observable. We show that the general structure of the optimal
mechanisms in the large does not change.

Formally, there are $M$ bias classes, where a sender in class $m\in\left\{ 1,\cdots,M\right\} $
has bias $b_{m}$. There are $N_{m}$ senders in bias class $m$,
and $N=\sum_{m}N_{m}$. The relative size of class $m$ is $\nu_{m}=N_{m}/N,$
with $\nu_{1}+\cdots+\nu_{M}=1$. We assume that the senders' signals
are independent and distributed according to a discrete distribution.
That is, for a sender $i$ of bias class $m$, $s_{i,m}\in S_{m}=\left\{ t_{1,m}<\cdots<t_{K,m}\right\} $,
with $f_{k,m}=\Pr\left(s_{i,m}=t_{k,m}\right)$ such that
\[
\sum^{K}_{k=1}f_{k,m}t_{k,m}=0,\sum^{K}_{k=1}f_{k,m}t^{2}_{k,m}=\text{var}\left(s_{i,m}\right)=\eta.
\]
Moreover, the payoff of the sender is given by
\[
\left(\frac{\sum^{M}_{m=1}\sum^{N_{m}}_{i=1}s_{i,m}}{\sqrt{N}}+b_{m}\right)a=\left(\omega+b_{m}\right)a,
\]
and the payoff of the receiver is $\left(\omega+r\right)a$. As in
the main model, we can define $\mathbf{h}_{m}\in\mathbb{R}^{K}_{0}$
as the deviations of the sample distribution of signals among the
senders of type $m$ from the true distribution $\mathbf{f}_{m}$
multiplied by $\sqrt{N_{m}}$. By the central limit theorem, $\mathbf{h}_{m}\rightarrow_{d}\mathcal{N}\left(\mathbf{0},\mathbf{\Sigma}_{m}\right)$,
where the $k,k'$ element of $\Sigma_{m}$ is $f_{k,m}\left(\mathbf{1}\left[k=k'\right]-f_{k',m}\right)$.
We will refer to $\omega_{m}=\mathbf{h}_{m}\cdot\mathbf{t}_{m}$ as
the mean of group $m$. Again, we focus on symmetric mechanisms, which
are of the form $\sigma\left(\mathbf{h}_{1},\cdots,\mathbf{h}_{M}\right)\in\left[0,1\right]$.
We can write 
\[
\omega=\frac{\sum^{M}_{m=1}\mathbf{h}_{m}\cdot\mathbf{t}_{m}\sqrt{N_{m}}}{\sqrt{N}}=\sum^{M}_{m=1}\omega_{m}\sqrt{\nu_{m}}.
\]
The central limit theorem implies that $\omega_{m}\rightarrow\mathcal{N}\left(0,\eta\right)$,
and by assumption, $\omega_{m}$'s are independent across groups.
In this environment, an argument akin to that of Theorem \ref{thm:A-recommendation-mechanism}
implies that the ICL holds if and only if, for all $m$:
\begin{align}
\mathbb{E}\left[\sigma\left(\mathbf{h}_{1},\cdots,\mathbf{h}_{M}\right)\left(\left(\omega+b_{m}\right)\frac{h_{k,m}}{f_{k,m}}-t_{k,m}\right)\right] & =0,\label{eq:ICLhet}\\
\mathbb{E}\left[\sigma\left(\mathbf{h}_{1},\cdots,\mathbf{h}_{M}\right)\frac{h_{k,m}}{f_{k,m}}\right] & \text{is increasing in}\:k.\nonumber 
\end{align}

With this result in hand, the rest of the characterization follows
that of Theorem \ref{thm:The-optimum-is}. The following proposition
states the main result for this extension:
\begin{prop}
\label{prop:The-receiver-optimal} The receiver-optimal ICL mechanism
is a function of the sample mean for each class $\omega_{m}=\mathbf{h}_{m}\cdot\mathbf{b}_{m}$.
Moreover, two vectors, $\lambda=\left(\lambda_{1},\cdots,\lambda_{M}\right)\in\mathbb{R}^{M}$
and $\zeta=\left(\zeta_{1},\cdots,\zeta_{M}\right)\in\mathbb{R}^{M}$,
exist such that the optimal mechanism $\sigma^{*}$ satisfies
\begin{align}
\sigma^{*}\left(\omega_{1},\cdots,\omega_{M}\right) & =1\Leftrightarrow\omega+r+\sum_{m}\lambda_{m}\left[\omega\omega_{m}+b_{m}\omega_{m}-\eta\right]+\sum_{m}\zeta_{m}\omega_{m}\geq0,\label{eq: heterogeneous}\\
 & \mathbb{E}\left[\sigma^{*}\cdot\left(\left(\omega+b_{m}\right)\omega_{m}-\eta\right)\right]=0,\label{eq:aggIChet}\\
 & \zeta_{m}\mathbb{E}\left[\sigma^{*}\cdot\omega_{m}\right]=0,\zeta_{m}\geq0,\mathbb{E}\left[\sigma^{*}\cdot\omega_{m}\right]\geq0.\label{eq: compS}
\end{align}
\end{prop}
In Proposition 2, the $\lambda_{m}$'s are the Lagrange multipliers
associated with the aggregated version of the incentive compatibility,
(\ref{eq:ICLhet}), in each bias class. Additionally, the $\zeta_{m}$'s
are the Lagrange multipliers on the monotonicity constraints, which
become $\mathbb{E}\left[\sigma^{*}\cdot\omega_{m}\right]\geq0$. Thus,
(\ref{eq: compS}) is the complementary slackness associated with
this constraint. The proof of Proposition \ref{prop:The-receiver-optimal}
closely follows that of Theorems \ref{thm:A-recommendation-mechanism}
and \ref{thm:The-optimum-is}, and is relegated to the Appendix.

The variables $\lambda,\zeta$ that determine the region for which
$\sigma=1$ can be found by solving the system of equations defined
by the incentive constraints, (\ref{eq:aggIChet}) and (\ref{eq: compS}).
Their existence is guaranteed by the existence of the solution of
the mechanism design problem, as we show in the proof of Proposition
\ref{prop:The-receiver-optimal}.

\subsection{General Preferences\label{subsec:General-Preferences}}

In this section, we show how our main results on optimal aggregation
in the large extend to more general preferences.

As before, there are $N$ senders, and each has type $s_{i}$ independently
drawn from $\left\{ t_{1},\cdots,t_{K}\right\} $ with probability
$f_{k}$. Again, let $h_{k}$ be the $k$-NEF. Focusing on symmetric
mechanisms, we can assume that a mechanism is a function of NEF. Hence,
we let the payoffs of the senders and of the receiver satisfy
\begin{equation}
u_{R}\left(a,\mathbf{h}\right)=\begin{cases}
u_{r}\left(\mathbf{h}\right) & a=1\\
0 & a=0
\end{cases},u_{S}\left(a,\mathbf{h}\right)=\begin{cases}
u_{s}\left(\mathbf{h}\right) & a=1\\
0 & a=0
\end{cases}.\label{eq: genpayoffs}
\end{equation}
We make the following assumption on $u_{s},u_{r}$:
\begin{assumption}
\label{assu:The-payoff-functions}The payoff functions $u_{r},u_{s}:\mathbb{R}^{K}_{0}\rightarrow\mathbb{R}$
satisfy the following properties:
\begin{enumerate}
\item They are continuous and differentiable.
\item The marginal value of $h_{k}$ for the sender, $\frac{\partial u_{s}\left(\mathbf{h}\right)}{\partial h_{k}}$,
is higher for higher values of $k$.
\item There exists $p>1$ such that $u_{s}\left(\mathbf{h}\right)=O\left(\left\Vert \mathbf{h}\right\Vert ^{p}_{K}\right)$
and $\left\Vert \nabla u_{s}\left(\mathbf{h}\right)\right\Vert _{K}=O\left(\left\Vert \mathbf{h}\right\Vert ^{p}_{K}\right)$.
\item The functions $\left\{ u_{s}\left(\mathbf{h}\right)\frac{h_{k}}{f_{k}}-\frac{\partial u_{s}\left(\mathbf{h}\right)}{\partial h_{k}}\right\} ^{K}_{k=1}$
are linearly independent.
\end{enumerate}
\end{assumption}
In our model so far, the payoff functions are linear and thus satisfy
the aforementioned assumption. The monotonicity assumption on the
senders' marginal value of an additional type $k$ sender is akin
to the standard single crossing assumption used in mechanism design.
The last parts of the assumption are technical ones that allow us
to show the equivalent of Theorem \ref{thm:A-recommendation-mechanism}
in this general setting.

The following proposition is the equivalent of Theorem \ref{thm:A-recommendation-mechanism}
for arbitrary payoffs:
\begin{prop}
\label{prop:A-recommendation-mechanism-gen}Suppose payoffs are given
by (\ref{eq: genpayoffs}) and satisfy Assumption \ref{assu:The-payoff-functions}.
A recommendation mechanism $\sigma\left(\mathbf{h}\right)$ is incentive
compatible in the large if and only if there exists $U$ such that
it satisfies
\begin{align}
\mathbb{E}\left[\sigma\left(\mathbf{h}\right)\left(u_{s}\left(\mathbf{h}\right)\frac{h_{k}}{f_{k}}-\frac{\partial u_{s}\left(\mathbf{h}\right)}{\partial h_{k}}\right)\right] & =U,\forall k\label{eq: ICLgen}\\
\mathbb{E}\left[\sigma\left(\mathbf{h}\right)\left(\frac{h_{k}}{f_{k}}-\frac{h_{l}}{f_{l}}\right)\left(\frac{\partial u_{s}\left(\mathbf{h}\right)}{\partial h_{k}}-\frac{\partial u_{s}\left(\mathbf{h}\right)}{\partial h_{l}}\right)\right] & \geq\mathbb{E}\left[\sigma\left(\mathbf{h}\right)\left(\frac{\partial^{2}u_{s}\left(\mathbf{h}\right)}{\partial h^{2}_{k}}+\frac{\partial^{2}u_{s}\left(\mathbf{h}\right)}{\partial h^{2}_{l}}-2\frac{\partial^{2}u_{s}\left(\mathbf{h}\right)}{\partial h_{l}\partial h_{k}}\right)\right]\label{eq: monGen}
\end{align}
for all $k>l$.
\end{prop}
Proposition \ref{prop:A-recommendation-mechanism-gen} describes how
incentive compatibility in the large is affected by having arbitrary
payoffs. The first equality is the equivalent of the standard envelope
condition as in (\ref{eq:env}). The second inequality is the standard
monotonicity condition. Notably, with general payoffs, the second
derivative or curvature of the payoff function is also relevant to
the standard monotonicity condition.

The complications arising from arbitrary payoffs imply that a general
characterization of optimal mechanisms is not so straightforward.
However, once the assumption of monotonicity is not binding, the characterization
of the resulting optimal mechanism is simple:
\begin{prop}
Let $\sigma^{*}:\mathbb{R}^{K}_{0}\rightarrow\left\{ 0,1\right\} $
be a mechanism that is ICL, i.e., satisfies the conditions in Proposition
\ref{prop:A-recommendation-mechanism-gen}. If $\left\{ \lambda_{k}\right\} ^{K}_{k=1}$
exist such that 
\begin{equation}
\sigma^{*}\left(\mathbf{h}\right)=1\Leftrightarrow u_{r}\left(\mathbf{h}\right)+\sum^{K}_{k=1}\lambda_{k}\left(\frac{h_{k}}{f_{k}}u_{s}\left(\mathbf{h}\right)-\frac{\partial u_{s}}{\partial h_{k}}\left(\mathbf{h}\right)\right)\geq0;\label{eq: Optimal}
\end{equation}
then $\sigma^{*}$ is an optimal mechanism.
\end{prop}
This condition is equivalent to the interval condition in the case
of our initial payoffs.

A particularly tractable specification is when $u_{r}$ and $u_{s}$
are linear and satisfy the specification in Proposition \ref{prop:Suppose-that-}.
Notably, this model is different from our baseline example, because
the senders and the receiver evaluate different signal realizations
differentially. For example, in the context of voting, the receiver's
preferences can represent how an uninformed voter cares about different
issues, while the senders' preferences represent the interests of
the differentially informed political elite in terms of the same issues.
In this case, optimal aggregation can be described by the following
Proposition:
\begin{prop}
\label{prop:Suppose-that-}Suppose that $u_{r}\left(\mathbf{h}\right)=\mathbf{t}_{r}\cdot\mathbf{h}$
and that $u_{s}\left(\mathbf{h}\right)=\mathbf{t}_{s}\cdot\mathbf{h}$
with $\mathbf{t}_{s},\mathbf{t}_{r}\in\mathbb{R}^{K}$ such that 
\begin{align*}
\sum_{k}f_{k}t_{s,k}= & \sum_{k}f_{k}t_{r,k}=0,\\
\sum_{k}f_{k}t^{2}_{s,k}= & \sum_{k}f_{k}t^{2}_{r,k}=1,\sum_{k}f_{k}t_{r,k}t_{s,k}=\gamma
\end{align*}
and $t_{s,k},t_{r,k}$ are increasing in $k$. Then, the optimal mechanism
$\sigma^{*}$ is a function of $\omega_{s}=\mathbf{t}_{s}\cdot\mathbf{h},\omega_{r}=\mathbf{t}_{r}\cdot\mathbf{h}$
and satisfies
\begin{align*}
\sigma^{*}\left(\omega_{r},\omega_{s}\right)=1\Leftrightarrow\omega_{r} & +\left(\lambda_{r}\omega_{r}+\lambda_{s}\omega_{s}\right)\omega_{s}-\gamma\lambda_{r}-\lambda_{s}\\
 & +\omega_{r}\sum_{k}\left(t_{r,k}-\gamma t_{s,k}\right)\left(\zeta_{k}-\zeta_{k+1}\right)\\
 & +\omega_{s}\sum_{k}\left(t_{s,k}-\gamma t_{r,k}\right)\left(\zeta_{k}-\zeta_{k+1}\right)\geq0
\end{align*}
for some $\alpha_{r},\alpha_{s}$ and $\zeta_{k}\geq0$ with $\zeta_{1}=\zeta_{K+1}=0$
such that
\begin{align}
\mathbb{E}\left[\sigma^{*}\cdot\left(\omega^{2}_{s}-1\right)\right] & =0\label{eq: AggICL}\\
\mathbb{E}\left[\sigma^{*}\cdot\left(\omega_{s}\omega_{r}-\gamma\right)\right] & =0\label{eq:AggICL2}
\end{align}
and 
\begin{equation}
\zeta_{k}\left[\left(t_{s,k}-t_{s,k-1}\right)\mathbb{E}\sigma^{*}\cdot\left(\omega_{s}-\gamma\omega_{r}\right)+\left(t_{r,k}-t_{r,k-1}\right)\mathbb{E}\sigma^{*}\cdot\left(\omega_{r}-\gamma\omega_{s}\right)\right]=0.\label{eq: CSgen}
\end{equation}
\end{prop}
Proposition \ref{prop:Suppose-that-} illustrates that the optimal
mechanism takes a tractable form and is only a function of the ex
post payoffs of the senders and of the receiver. Moreover, (\ref{eq: CSgen})
is the usual complementary slackness associated with the monotonicity
constraint (\ref{eq: monGen}). Hence, much in line with standard
models of mechanism design, Proposition \ref{prop:Suppose-that-}
illustrates a procedure for finding the optimal mechanism. That is,
one can conjecture that the monotonicity constraints are slack, i.e.,
$\zeta_{k}=0$, and find $\lambda_{r},\lambda_{s}$ so that ICL constraints
(\ref{eq: AggICL}) and (\ref{eq:AggICL2}) are satisfied. If the
resulting mechanism $\sigma^{*}$ satisfies the monotonicity constraint,
it is the optimum. Otherwise, a procedure akin to ironing is required
to find which monotonicity constraints are binding.

The next example illustrates this result.
\begin{example}
\label{exa:Consider-an-example}Consider an example in which $K=3$
and $u_{r}\left(\mathbf{h}\right)=\rho\left(-2h_{1}-h_{2}+3h_{3}\right),u_{s}\left(\mathbf{h}\right)=\rho\left(-\sqrt{7}h_{1}+\sqrt{7}h_{3}\right),$
where $\rho=\sqrt{3/14}$ and $f_{1}=f_{2}=f_{3}=1/3$. In this case,
$\gamma=\frac{5}{2\sqrt{7}}$. Using Proposition \ref{prop:Suppose-that-},
we can numerically find the values of $\lambda_{r},\lambda_{s}$ that
satisfy the conditions in Proposition \ref{prop:Suppose-that-}. Figure
\ref{fig:Optimal-Mechanism-in} depicts the set of $\omega_{s},\omega_{r}$'s
for which $a=1$. In this example, the agreement regions are the positive
and negative quadrants, while the disagreement regions are the remaining
regions. The optimal mechanism creates value for the receiver by recommending
$a=1$ when $\omega_{r}>0$ and $\omega_{s}<0$---the disagreement
quadrant preferred by the receiver. Surplus burning is occurring in
the positive and negative quadrants. In the positive quadrant, high
reports of $\omega_{s}$ and low reports of $\omega_{r}$ are punished
by recommending $a=0$. In the negative quadrant, surplus burning
is happening by recommending $a=1$ when both the senders and the
receiver prefer $a=0$.
\begin{figure}[h]
\begin{centering}
\begin{tikzpicture}
\node[inner sep=0pt] (0,0)
    {\includegraphics[width=.45\textwidth]{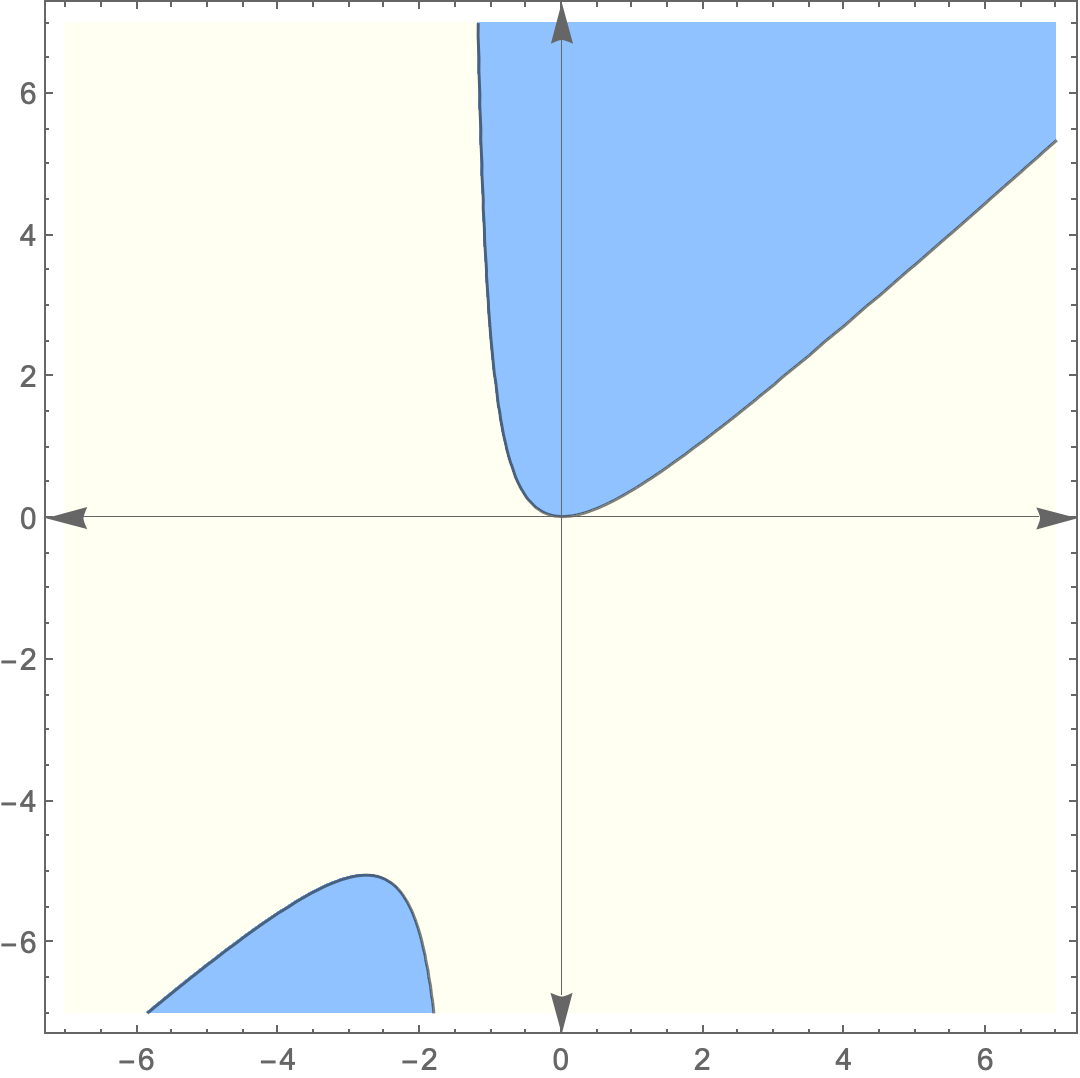}};
\draw[black] (4,0) node[above] {$\omega_s$};
\draw[black] (0,3.8) node[right] {$\omega_r$};
\draw[black] (1.2,2) node[right] {$a=1$};

\draw[black] (-2,-1.5) node {$a=0$};
\end{tikzpicture}
\par\end{centering}
\caption{{\small Optimal mechanism in Example \ref{exa:Consider-an-example}.
The sender's payoff $\omega_{s}$ is on the x-axis, and the receiver's
payoff $\omega_{r}$ is on the y-axis. \label{fig:Optimal-Mechanism-in}}}
\end{figure}
\end{example}

\subsection{Optimality of Sender Preferred for N>1 \label{subsec:Optimality-of-Sender}}

Let each sender's signal be distributed according to a common cumulative
distribution function $F\left(s\right)$, with density $f\left(s\right)$
on an interval $S=\left[\underline{s},\overline{s}\right]\subset\mathbb{R}$,
that satisfies $\int_{S}sf\left(s\right)ds=0$. We use $\mathbf{s}=\left(s_{1},s_{2},...,s_{N}\right)$
to denote the senders' type profile. Additionally, with a slight abuse
of notation, we let $F(\mathbf{s})=\prod_{j}F\left(s_{j}\right)$
and $F_{-i}\left(\mathbf{s}_{-i}\right)=\prod_{j\neq i}F\left(s_{j}\right)$
be the distribution functions of $\mathbf{s}$ and $\mathbf{s}_{-i},$
respectively, with the obvious corresponding density functions.
\begin{prop}
\label{prop:The-bias-}Let $f\left(\cdot\right)$ be a $C^{1}$ function
with a finite derivative and full support, and define $\ell=\inf_{s_{i}\in S}f'\left(s_{i}\right)\left(\overline{s}-s_{i}\right)/f\left(s_{i}\right)$.
Then, for any sender and receiver payoff parameters, $(b,r)$, there
exists a positive finite value $\underline{N}(b,r,\ell,\overline{s})$
such that the sender-preferred mechanism $\sigma^{S}\left(\mathbf{s}\right)=\mathbf{1}\left[\omega\left(\mathbf{s}\right)+b\geq0\right]$
is a solution to (\ref{eq:P})whenever
\[
N\leq\underline{N}(b,r,\ell,\overline{s}).
\]
\end{prop}
Proposition \ref{prop:The-bias-} states that if the number of senders
is low enough, then the best incentive compatible mechanism from the
receiver's perspective is the same as that of the senders. It can
be shown that the finite bound on the number of senders for the optimality
of the sender-preferred mechanism, $\underline{N}(b,r,\ell,\overline{s})$,
is decreasing in the bias of the sender, $b-r>0$. Of course, for
large levels of bias, even the uninformative allocation may be preferable
to $\sigma^{S}$. Perhaps also unsurprising, when $\sigma^{S}$ tends
to the receiver first best, i.e., $b-r\rightarrow0$, then this bound
tends to infinity.\footnote{The bound is decreasing in $0\geq\ell>-\infty$. The value of $\ell$
is the upper bound on the negative curvature of $F$. Thus, for highly
negative values of $\ell$, there may be a very large probability
mass in the disagreement region relative to the probability of being
in the agreement region. By considering the $N=1$ case, one can see
how this could easily imply that the cost of disagreement is just
too high relative to the value of accepting when all parties agree.} As we discuss in more detail later, one way to interpret $N\leq\underline{N}(b,r,\ell,\overline{s})$
is as a condition under which unrestricted private communication among
the senders is indeed the best outcome from the receiver's perspective.
Conversely, it identifies a necessary condition for the commitment
of the mediator to be able to create value for the receiver by exploiting
the informational division between the senders.
\begin{proof}
We prove the claim in two steps: First, we show that one can use standard
Lagrangian techniques to characterize the solution of problem (\ref{eq:P}).
Second, we construct the Lagrange multipliers -- members of an appropriate
dual space to be precisely defined below -- so that the sender-preferred
allocation, $\sigma^{*}_{S}\left(\mathbf{s}\right)$, maximizes the
appropriate Lagrangian.

We begin by rewriting the problem (\ref{eq:P}) as follows:
\[
\max_{\sigma:S^{N}\rightarrow\left[0,1\right],\underline{v}}\int_{S^{N}}\left(\omega\left(\mathbf{s}\right)+r\right)\sigma\left(\mathbf{s}\right)f\left(\mathbf{s}\right)d\mathbf{s}
\]
subject to

\begin{align}
\mathbb{E}\left[\left(\omega\left(\mathbf{s}\right)+b\right)\sigma\left(\mathbf{s}\right)|s_{i}\right] & =\underline{v}+\frac{1}{\sqrt{N}}\int^{s_{i}}_{-1}\mathbb{E}\left[\sigma\left(\mathbf{s}\right)|\tilde{s}_{i}\right]d\tilde{s}_{i},\forall s_{i}\in S,i\in\left\{ 1,\cdots,N\right\} ,\label{eq: C1}\\
\mathbb{E}\left[\sigma\left(\mathbf{s}\right)|s_{i}\right] & \leq\mathbb{E}\left[\sigma\left(\mathbf{s}\right)|s_{i}'\right],\forall s_{i}\leq s_{i}'\label{eq: Mon}\\
0\leq & \sigma\left(\mathbf{s}\right)\leq1,\mathbf{s}\in S^{N}.\label{eq: C2}
\end{align}
The first constraint is the integral form of the envelope version
of incentive compatibility (\ref{eq:general_IC}), while the second
constraint is the standard monotonicity of the interim allocations
$\mathbb{E}\left[\sigma\left(\mathbf{s}\right)|s_{i}\right]$. Finally,
the last condition is the upper and lower bounds on $\sigma\left(\mathbf{s}\right)$.\footnote{That incentive compatibility is equivalent to the envelope condition
(\ref{eq: C1}), and the monotonicity of $\mathbb{E}\left[\sigma\left(\mathbf{s}\right)|s_{i}\right]$
is standard. See, for example, \citet{myerson1981optimal}.} We can thus frame the problem in a convenient way for establishing
weak duality or standard sufficiency conditions for optimality---see
Theorem 1 in Section 8.4 in \citet{luenberger1997optimization}.

More specifically, let $x=\left(\sigma,\underline{v}\right)\in X=L_{2}\left(S^{N}\right)\times\mathbb{R}$
and $Z=L_{2}\left(S\right)^{3N}\times L_{2}\left(S^{N}\right)^{2}$.
The above program is a linear programming problem whose constraint
set is of the form $G\left(x\right)\leq0$ for some $G:X\rightarrow Z$,
where $\leq$ is defined below. The function $G\left(x\right)$ is
given by the quintuple $\left(G_{1}\left(x\right),\cdots,G_{5}\left(x\right)\right)$.
In this quintuple, $G_{1},G_{2}:X\rightarrow L_{2}\left(S\right)^{N}$
and are given by
\begin{align*}
G_{1,i}\left(x\right)\left(s_{i}\right)=\mathbb{E}\left[\left(\omega\left(\mathbf{s}\right)+b\right)\sigma\left(\mathbf{s}\right)|s_{i}\right]-\underline{v}-\frac{1}{\sqrt{N}}\int^{s_{i}}_{-1}\mathbb{E}\left[\sigma\left(\mathbf{s}\right)|\tilde{s}_{i}\right]d\tilde{s}_{i} & \leq0\\
G_{2,i}\left(x\right)\left(s_{i}\right)=-G_{1,i}\left(x\right)\left(s_{i}\right) & \leq0.
\end{align*}
That is, instead of using an equality constraint in (\ref{eq: C1}),
we use two inequality constraints that have the same implication;
$G_{3,i}:X\rightarrow L_{2}\left(S\right)$ is given by $G_{3,i}\left(x\right)\left(s_{i}\right)=-\mathbb{E}\left[\sigma\left(\mathbf{s}\right)|s_{i}\right]$
(associated with (\ref{eq: C2})), and $G_{4},G_{5}:X\rightarrow L_{2}\left(S^{N}\right)$
are simply $G_{4}\left(x\right)\left(\mathbf{s}\right)=-\sigma\left(\mathbf{s}\right)$
and $G_{5}\left(x\right)\left(\mathbf{s}\right)=1-\sigma\left(\mathbf{s}\right)$
(associated with (\ref{eq: Mon})). The relation $\leq$ is associated
with the cone of members of $Z$ whose first, second, fourth, and
fifth elements are nonnegative while its third element is a non-decreasing
function.

Given this formulation, we can use Theorem 1 in Section 8.4 in \citet{luenberger1997optimization}
(switched to maximization), which states that if there exist $x_{0}\in X$
and $z^{*}_{0}\in Z^{*}$ such that the Lagrangian $L\left(x,z^{*}\right)=w\left(x\right)+\left\langle z^{*},G\left(x\right)\right\rangle $
has a saddle point at $x_{0},z^{*}_{0}$, 
\[
L\left(x,z^{*}_{0}\right)\leq L\left(x_{0},z^{*}_{0}\right)\leq L\left(x_{0},z^{*}\right),\forall x\in X,z^{*}\geq0.
\]
Then $x_{0}\in\arg\max_{x\in X,G\left(x\right)\leq0}w\left(x\right)$.
In our setting, $G$ is as defined before, while $w\left(x\right)=\mathbb{E}\left[\sigma\left(\mathbf{s}\right)\omega\left(\mathbf{s}\right)\right]$.
If $x_{0}$ satisfies $\left\langle z^{*}_{0},G\left(x_{0}\right)\right\rangle =0$,
then the second inequality is satisfied by the definition of the dual
cone in $Z^{*}$. Thus, we have to construct the Lagrange multiplier
$z^{*}_{0}$ such that $\left\langle z^{*}_{0},G\left(x_{0}\right)\right\rangle =0,$
and the first of the aforementioned inequalities holds.

Using the Riesz representation theorem---see Theorem 14.12 in \citet{guide2006infinite}---we
can write the Lagrangian as

\begin{align*}
L\left(x,z^{*}\right)= & \int\left(\omega\left(\mathbf{s}\right)+r\right)\sigma\left(\mathbf{s}\right)f\left(\mathbf{s}\right)d\mathbf{s}\\
 & +\sum_{i}\int_{S}G_{1,i}\left(x\right)\left(s_{i}\right)d\Lambda^{1}_{i}\left(s_{i}\right)+\sum_{i}\int_{S}G_{2,i}\left(x\right)\left(s_{i}\right)d\Lambda^{2}_{i}\left(s_{i}\right)\\
 & +\sum_{i}\int_{S}G_{3,i}\left(s_{i}\right)d\Omega_{i}\left(s_{i}\right)+\int\sigma\left(\mathbf{s}\right)d\underline{\zeta}\left(\mathbf{s}\right)+\int\left(1-\sigma\left(\mathbf{s}\right)\right)d\overline{\zeta}.
\end{align*}
Here, $\Lambda^{1}_{i},\Lambda^{2}_{i}$'s are positive Borel measures
over $S$, and $\underline{\zeta},\overline{\zeta}$ are positive
Borel measures over $S^{N}$, since $z^{*}_{0}$ should be a member
of the dual cone of $\geq$ in $Z^{*}$. Additionally, $\Omega_{i}$
is a signed measure that satisfies $\Omega_{i}\left(\left[s,\overline{s}\right]\right)\geq0,\forall s\in S$.\footnote{Recall that the dual cone of increasing functions is the set of signed
Borel measures $\Omega$ so that $\int f\left(s_{i}\right)d\Omega\geq0$
for all increasing $f$ . This coincides with the set of measures
that satisfy $\Omega\left(\left[s,\overline{s}\right]\right)\geq0$.} Let $\Delta\Lambda_{i}=\Lambda^{1}_{i}-\Lambda^{2}_{i}$. We can
use integration by parts to write the Lagrangian as
\begin{align*}
L\left(x,z^{*}\right)= & \int\left(\omega\left(\mathbf{s}\right)+r\right)\sigma\left(\mathbf{s}\right)f\left(\mathbf{s}\right)d\mathbf{s}\\
 & +\sum_{i}\int_{S}\int_{S^{N-1}}\left[\left(\omega\left(s_{i};s_{-i}\right)+b\right)\sigma\left(s_{i};s_{-i}\right)f_{-i}\left(s_{-i}\right)ds_{-i}-\underline{v}\right]d\Delta\Lambda_{i}\\
 & -\frac{1}{\sqrt{N}}\sum_{i}\int_{S}\int_{S^{N-1}}\sigma\left(s_{i};s_{-i}\right)f_{-i}\left(s_{-i}\right)ds_{-i}\Delta\Lambda_{i}\left(\left[s_{i},\overline{s}\right]\right)ds_{i}\\
 & +\int\sigma\left(\mathbf{s}\right)d\underline{\zeta}\left(\mathbf{s}\right)+\int\left(1-\sigma\left(\mathbf{s}\right)\right)d\overline{\zeta}\left(\mathbf{s}\right)+\sum_{i}\int_{S}\int_{S^{N-1}}\sigma\left(s_{i};s_{-i}\right)f_{-i}\left(s_{-i}\right)ds_{-i}d\Omega_{i}\left(s_{i}\right).
\end{align*}
We are now ready to identify the relevant Lagrange multipliers and
verify that---under the upper bound on $N$---they support the sender-preferred
allocation $x_{0}=\left(\sigma^{S}\left(\mathbf{s}\right),v^{S}\left(\underline{s}\right)\right)$
as a saddle point of $L$. To that end, set
\begin{align*}
\Delta\Lambda_{i}\left(\left[\underline{s},s_{i}\right]\right) & =\begin{cases}
0 & s_{i}=\underline{s}\\
f\left(s_{i}\right)\frac{b-r}{b+\sqrt{N}\overline{s}}\left(\overline{s}-s_{i}\right) & s_{i}>\underline{s}
\end{cases}\\
\Omega_{i}\left(\left[\underline{s},s_{i}\right]\right) & =0.
\end{align*}
By the Jordan decomposition theorem, any signed measure can be written
as the difference of two positive measures. Therefore, consider two
positive measures $\Lambda^{1}_{i},\Lambda^{2}_{i}$ such that $\Delta\Lambda_{i}=\Lambda^{1}_{i}-\Lambda^{2}_{i}$.
Given the aforementioned Lagrangian, let us define the ``first order
condition'' as follows:\footnote{Formally, $\lambda\left(\mathbf{s}\right)$ is the Fréchet derivative
of $L\left(x,z\right)$ evaluated at $x^{*}$ and in the direction
of the Dirac delta function at $\mathbf{s}$.}
\begin{align*}
\forall i,s_{i}>\underline{s}:\lambda\left(\mathbf{s}\right)= & \left(\omega\left(\mathbf{s}\right)+r\right)f\left(\mathbf{s}\right)\\
 & +\sum^{N}_{i=1}\left(\omega\left(\mathbf{s}\right)+b\right)\frac{f\left(\mathbf{s}\right)}{f\left(s_{i}\right)}\frac{b-r}{b+\sqrt{N}\overline{s}}\frac{d}{ds_{i}}\left(f\left(s_{i}\right)\left(\overline{s}-s_{i}\right)\right)\\
 & +\frac{1}{\sqrt{N}}\sum^{N}_{i=1}f\left(\mathbf{s}\right)\frac{b-r}{b+\sqrt{N}\overline{s}}\left(\overline{s}-s_{i}\right).
\end{align*}
Moreover, we can write
\begin{align*}
\mathbf{s},\exists i,s_{i}=\underline{s}:\frac{\lambda\left(\mathbf{s}\right)}{\left(\omega\left(\mathbf{s}\right)+b\right)f\left(\mathbf{s}\right)}= & \frac{b-r}{b+\sqrt{N}\overline{s}}\left(\overline{s}-\underline{s}\right)\sum^{N}_{j=1}\mathbf{1}\left[s_{j}=\underline{s}\right].
\end{align*}
This holds since $\Lambda_{i}$ has a mass point at $\underline{s},$
and when calculating the Fréchet--Gateaux derivative of the Lagrangian,
the effect of changes in the direction of\textbf{ }the Dirac's delta
at $\mathbf{s}$, only mass points are relevant.

In order to show optimality of sender-preferred allocation, it is
sufficient to show that 
\[
\lambda\left(\mathbf{s}\right)\geq0\iff\omega\left(\mathbf{s}\right)+b\geq0.
\]
When $\forall i,s_{i}>\underline{s}$, we have
\begin{align*}
s_{i}>\underline{s}:\lambda\left(\mathbf{s}\right)= & \left(\omega\left(\mathbf{s}\right)+r\right)f\left(\mathbf{s}\right)\\
 & +\frac{b-r}{b+\sqrt{N}\overline{s}}\sum^{N}_{i=1}\left(\omega\left(\mathbf{s}\right)+b\right)f\left(\mathbf{s}\right)\left(-1+\frac{f'\left(s_{i}\right)\left(\overline{s}-s_{i}\right)}{f\left(s_{i}\right)}\right)\\
 & +\frac{1}{\sqrt{N}}\sum^{N}_{i=1}f\left(\mathbf{s}\right)\frac{b-r}{b+\sqrt{N}\overline{s}}\left(\overline{s}-s_{i}\right)\\
= & \left(\omega\left(\mathbf{s}\right)+b\right)f\left(\mathbf{s}\right)+\left(r-b\right)f\left(\mathbf{s}\right)\\
 & +\left(\omega\left(\mathbf{s}\right)+b\right)f\left(\mathbf{s}\right)\frac{b-r}{b+\sqrt{N}\overline{s}}\sum^{N}_{i=1}\left(-1+\frac{f'\left(s_{i}\right)\left(\overline{s}-s_{i}\right)}{f\left(s_{i}\right)}\right)\\
 & +f\left(\mathbf{s}\right)\frac{b-r}{b+\sqrt{N}\overline{s}}\sum^{N}_{i=1}\frac{\left(\overline{s}-s_{i}\right)}{\sqrt{N}}\\
= & \left(\omega\left(\mathbf{s}\right)+b\right)f\left(\mathbf{s}\right)\left[1+\frac{b-r}{b+\sqrt{N}\overline{s}}\sum^{N}_{i=1}\left(-1+\frac{f'\left(s_{i}\right)\left(\overline{s}-s_{i}\right)}{f\left(s_{i}\right)}\right)\right]\\
 & +\left(r-b\right)f\left(\mathbf{s}\right)+f\left(\mathbf{s}\right)\frac{b-r}{b+\sqrt{N}\overline{s}}\left(\sqrt{N}\overline{s}-\omega\left(\mathbf{s}\right)\right)\\
= & \left[1-\frac{b-r}{b+\sqrt{N}\overline{s}}+\frac{b-r}{b+\sqrt{N}\overline{s}}\times\sum^{N}_{i=1}\left(-1+\frac{f'\left(s_{i}\right)\left(\overline{s}-s_{i}\right)}{f\left(s_{i}\right)}\right)\right]\times\\
 & \left(\omega\left(\mathbf{s}\right)+b\right)f\left(\mathbf{s}\right)+\left(r-b\right)f\left(\mathbf{s}\right)+f\left(\mathbf{s}\right)\frac{b-r}{b+\sqrt{N}\overline{s}}\left(\sqrt{N}\overline{s}+b\right)\\
= & \frac{b-r}{b+\sqrt{N}\overline{s}}\sum^{N}_{i=1}\left(-1+\frac{f'\left(s_{i}\right)\left(\overline{s}-s_{i}\right)}{f\left(s_{i}\right)}+\frac{1}{N}\frac{r+\sqrt{N}\overline{s}}{b-r}\right)\times.\\
 & \left(\omega\left(\mathbf{s}\right)+b\right)f\left(\mathbf{s}\right)
\end{align*}
This implies that if the term in the last brackets is always positive,
$\lambda\left(\mathbf{s}\right)\geq0$ if and only $\omega\left(\mathbf{s}\right)+b\geq0$.
Note that the term in the brackets is always positive if and only
if
\[
\frac{1}{N}\frac{r+\sqrt{N}\overline{s}}{b-r}\geq1-\inf_{s_{i}\in S}\frac{f'\left(s_{i}\right)\left(\overline{s}-s_{i}\right)}{f\left(s_{i}\right)}=1-\ell\rightarrow r+\sqrt{N}\overline{s}\geq N\left(b-r\right)\left(1-\ell\right)
\]
The above is a quadratic equation in $\sqrt{N}$ and since $1-\ell>0$,
it should hold when $\sqrt{N}$ is below its higher root. The squared
value of this higher root is given by$\underline{N}(b,r,\ell,\overline{s})=\left(\frac{\sqrt{\overline{s}^{2}+4r\left(b-r\right)\left(1-\ell\right)}+\overline{s}}{2\left(b-r\right)\left(1-\ell\right)}\right)^{2}$which
gives us the (sufficient) bound on $N$.

When for some $i$, $s_{i}=\underline{s}$, then from the calculation
of $L$, it is clear that $\lambda\left(\mathbf{s}\right)\geq0$ if
and only if $\omega\left(\mathbf{s}\right)+b\geq0$. This concludes
the proof.
\end{proof}

\subsection{Other Proofs}

\subsubsection{Existence of the Multiplier in Theorem \ref{thm:The-optimum-is}\label{subsec:Existence-multi}}
\begin{proof}
Here, we show that when $r>\frac{b-\sqrt{b^{2}+4\text{Var}\left(s\right)}}{2}$
there is a unique $\alpha$ for which 
\begin{align*}
\int_{D_{\alpha}}\left(1-\frac{\omega\left(\omega+b\right)}{\text{Var}\left(s\right)}\right)\phi\left(\frac{\omega}{\sqrt{\text{Var}\left(s\right)}}\right)d\omega & =0\\
\left\{ \omega|\psi\left(\omega,\alpha\right)=\omega+r+\alpha\text{Var}\left(s\right)-\alpha\left(\omega+b\right)\omega\geq0\right\}  & =D_{\alpha}
\end{align*}
Moreover $\alpha\in\left(0,1/b\right)$ which implies that $D_{\alpha}=\left[\underline{\omega},\overline{\omega}\right]$
where 
\[
-b<\underline{\omega}<-r<-\underline{\omega}<\overline{\omega}.
\]
Additionally, we show that if $r\leq\frac{b-\sqrt{b^{2}+4}}{2}$,
then whenever the above holds, $\underline{\omega}=\overline{\omega}$.
We do so via a series of claims.

\textbf{Claim 1. }\textit{For any value of $\alpha$, $\int_{D_{\alpha}}\left(1-\frac{\omega\left(\omega+b\right)}{\text{Var}\left(s\right)}\right)\phi\left(\frac{\omega}{\sqrt{\text{Var}\left(s\right)}}\right)d\omega$
is increasing in $\alpha$.}

To prove this claim, note that if $\alpha<0$, then either $\psi\left(\omega,\alpha\right)$
is always positive or at most two values of $\underline{\omega},\overline{\omega}$
exist such that $\psi\left(\underline{\omega},\alpha\right)=\psi\left(\overline{\omega},\alpha\right)=0$
in which case $D_{\alpha}=\mathbb{R}\backslash\left(\underline{\omega},\overline{\omega}\right)$.
On the other hand, if $\alpha>0$, either $D_{\alpha}=\left[\underline{\omega},\overline{\omega}\right]$
with $\psi\left(\underline{\omega},\alpha\right)=\psi\left(\overline{\omega},\alpha\right)=0$
or $D_{\alpha}=\emptyset$. Evidently, if $\psi\left(\omega,\alpha\right)$
does not have a zero, then the integral is 0.

Note that when the zeros exist, they are given by
\begin{align}
\underline{\omega} & =\frac{1-b\alpha}{2\alpha}-\frac{1}{2\alpha}\sqrt{\left(1-b\alpha\right)^{2}+4r\alpha+4\alpha^{2}\text{Var}\left(s\right)}=\frac{1-\alpha b-\sqrt{\Delta}}{2\alpha}\label{eq: omegl}\\
\overline{\omega} & =\frac{1-\alpha b}{2\alpha}+\frac{1}{2\alpha}\sqrt{\left(1-b\alpha\right)^{2}+4r\alpha+4\alpha^{2}\text{Var}\left(s\right)}=\frac{1-\alpha b+\sqrt{\Delta}}{2\alpha}\label{eq:omegah}
\end{align}
We thus have that
\begin{align*}
\frac{d\overline{\omega}}{d\alpha} & =-\frac{\psi_{\alpha}\left(\overline{\omega},\alpha\right)}{\psi_{\omega}\left(\overline{\omega},\alpha\right)}=-\frac{\text{Var}\left(s\right)-\overline{\omega}\left(\overline{\omega}+b\right)}{1-\alpha\left(2\overline{\omega}+b\right)}=-\frac{-\frac{\overline{\omega}+r}{\alpha}}{-\sqrt{\Delta}}=-\frac{\overline{\omega}+r}{\alpha\sqrt{\Delta}}\\
\frac{d\underline{\omega}}{d\alpha} & =-\frac{\psi_{\alpha}\left(\underline{\omega},\alpha\right)}{\psi_{\omega}\left(\underline{\omega},\alpha\right)}=-\frac{\text{Var}\left(s\right)-\underline{\omega}\left(\underline{\omega}+b\right)}{1-\alpha\left(2\underline{\omega}+b\right)}=-\frac{-\frac{\underline{\omega}+r}{\alpha}}{\sqrt{\Delta}}=\frac{\underline{\omega}+r}{\alpha\sqrt{\Delta}}
\end{align*}
When $\alpha>0$, $\overline{\omega}\geq\underline{\omega}$ and we
have
\begin{align*}
\frac{d}{d\alpha}\int_{D_{\alpha}}\left[1-\frac{\omega\left(\omega+b\right)}{\text{Var}\left(s\right)}\right]\phi\left(\frac{\omega}{\sqrt{\text{Var}\left(s\right)}}\right)d\omega & =\\
\frac{d}{d\alpha}\int^{\overline{\omega}}_{\underline{\omega}}\left[1-\frac{\omega\left(\omega+b\right)}{\text{Var}\left(s\right)}\right]\phi\left(\frac{\omega}{\sqrt{\text{Var}\left(s\right)}}\right)d\omega & =\\
\phi\left(\frac{\overline{\omega}}{\sqrt{\text{Var}\left(s\right)}}\right)\frac{d\overline{\omega}}{d\alpha}\left[1-\frac{\overline{\omega}\left(\overline{\omega}+b\right)}{\text{Var}\left(s\right)}\right]-\phi\left(\frac{\underline{\omega}}{\sqrt{\text{Var}\left(s\right)}}\right)\frac{d\underline{\omega}}{d\alpha}\left[1-\frac{\underline{\omega}\left(\underline{\omega}+b\right)}{\text{Var}\left(s\right)}\right] & =\\
\phi\left(\frac{\overline{\omega}}{\sqrt{\text{Var}\left(s\right)}}\right)\left(-\frac{\overline{\omega}+r}{\alpha\sqrt{\Delta}}\right)\left(-\frac{\overline{\omega}+r}{\alpha\text{Var}\left(s\right)}\right)-\phi\left(\frac{\underline{\omega}}{\sqrt{\text{Var}\left(s\right)}}\right)\left(\frac{\underline{\omega}+r}{\alpha\sqrt{\Delta}}\right)\left(-\frac{\underline{\omega}+r}{\alpha\text{Var}\left(s\right)}\right) & =\\
\frac{1}{\sqrt{\Delta}\text{Var}\left(s\right)}\left[\phi\left(\frac{\overline{\omega}}{\sqrt{\text{Var}\left(s\right)}}\right)\left(\frac{\overline{\omega}+r}{\alpha}\right)^{2}+\phi\left(\frac{\underline{\omega}}{\sqrt{\text{Var}\left(s\right)}}\right)\left(\frac{\underline{\omega}+r}{\alpha}\right)^{2}\right] & >0
\end{align*}
where the above holds since both terms are squares multiplied by a
density. Similarly when $\alpha<0$, $\overline{\omega}<\underline{\omega}$
and thus
\begin{align*}
\int_{D_{\alpha}}\left[1-\frac{\omega\left(\omega+b\right)}{\text{Var}\left(s\right)}\right]\phi\left(\frac{\omega}{\sqrt{\text{Var}\left(s\right)}}\right)d\omega & =\\
\int_{\mathbb{R}\backslash\left(\overline{\omega},\underline{\omega}\right)}\left[1-\frac{\omega\left(\omega+b\right)}{\text{Var}\left(s\right)}\right]\phi\left(\frac{\omega}{\sqrt{\text{Var}\left(s\right)}}\right)d\omega & =\\
\int^{\overline{\omega}}_{\underline{\omega}}\left[1-\frac{\omega\left(\omega+b\right)}{\text{Var}\left(s\right)}\right]\phi\left(\frac{\omega}{\sqrt{\text{Var}\left(s\right)}}\right)d\omega
\end{align*}
where in the above we have used the fact that the integral over the
entire real line is 0. Then a logic similar to the case of $\alpha>0$
implies that the above is increasing.

\textbf{Claim 2. }\textit{For all values $\alpha<0$, $\int_{D_{\alpha}}\left[1-\frac{\omega\left(\omega+b\right)}{\text{Var}\left(s\right)}\right]\phi\left(\frac{\omega}{\sqrt{\text{Var}\left(s\right)}}\right)d\omega<0$.}

To show this, it is sufficient to show that it holds as $\alpha\rightarrow0$
from below. Note that as $\alpha\nearrow0$, $\underline{\omega}\rightarrow-r$
and $\overline{\omega}\rightarrow-\infty$. Hence,
\begin{align*}
\int_{D_{\alpha}}\left[1-\frac{\omega\left(\omega+b\right)}{\text{Var}\left(s\right)}\right]\phi\left(\frac{\omega}{\sqrt{\text{Var}\left(s\right)}}\right)d\omega & =\\
\int^{-\infty}_{-r}\left[1-\frac{\omega\left(\omega+b\right)}{\text{Var}\left(s\right)}\right]\phi\left(\frac{\omega}{\sqrt{\text{Var}\left(s\right)}}\right)d\omega & =\\
-\int^{-r}_{-\infty}\left[1-\frac{\omega\left(\omega+b\right)}{\text{Var}\left(s\right)}\right]\frac{e^{-\frac{\omega^{2}}{2\text{Var}\left(s\right)}}}{\sqrt{2\pi}}d\omega & =-\left.\left(\omega+b\right)\frac{e^{-\frac{\omega^{2}}{2\text{Var}\left(s\right)}}}{\sqrt{2\pi}}\right|^{-r}_{-\infty}=-\left(b-r\right)\frac{e^{-\frac{r^{2}}{2\text{Var}\left(s\right)}}}{\sqrt{2\pi}}<0
\end{align*}
which proves the claim.

\textbf{Claim 3. }\textit{If $r>\frac{b-\sqrt{b^{2}+4\text{Var}\left(s\right)}}{2}$,
then $\exists!\alpha\in\left(0,1/b\right)$ such that $\int_{D_{\alpha}}\left[1-\frac{\omega\left(\omega+b\right)}{\text{Var}\left(s\right)}\right]\phi\left(\omega/\sqrt{\text{Var}\left(s\right)}\right)d\omega=0$.
At this value of $\alpha$, $-b<\underline{\omega}<-r<\overline{\omega}$
and $0<\underline{\omega}+\overline{\omega}$.}

Note that as $\alpha\searrow0$, $\underline{\omega}\rightarrow-r$
and $\overline{\omega}\rightarrow\infty$. Hence and as above, $\int_{D_{\alpha}}\left[1-\frac{\omega\left(\omega+b\right)}{\text{Var}\left(s\right)}\right]\phi\left(\omega/\sqrt{\text{Var}\left(s\right)}\right)d\omega<0$
holds for $\alpha=0$. In contrast, at $\alpha=1/b$, 
\begin{align*}
\underline{\omega} & =-\frac{b}{2}\sqrt{4r/b+4\text{Var}\left(s\right)/b^{2}}=-\sqrt{rb+\text{Var}\left(s\right)}\\
\overline{\omega} & =\sqrt{rb+\text{Var}\left(s\right)}
\end{align*}
At these values 
\begin{align*}
\int_{D_{\alpha}}\left[1-\frac{\omega\left(\omega+b\right)}{\text{Var}\left(s\right)}\right]\phi\left(\omega/\sqrt{\text{Var}\left(s\right)}\right)d\omega= & \int^{\sqrt{rb+\text{Var}\left(s\right)}}_{-\sqrt{rb+\text{Var}\left(s\right)}}\left[1-\frac{\omega\left(\omega+b\right)}{\text{Var}\left(s\right)}\right]\phi\left(\omega/\sqrt{\text{Var}\left(s\right)}\right)d\omega\\
= & \left(\sqrt{rb+\text{Var}\left(s\right)}+b\right)\frac{e^{-\frac{rb+\text{Var}\left(s\right)}{2\text{Var}\left(s\right)}}}{\sqrt{2\pi}}\\
 & -\left(b-\sqrt{rb+\text{Var}\left(s\right)}\right)\frac{e^{-\frac{rb+\text{Var}\left(s\right)}{2\text{Var}\left(s\right)}}}{\sqrt{2\pi}}\\
= & 2\sqrt{rb+\text{Var}\left(s\right)}\frac{e^{-\frac{rb+\text{Var}\left(s\right)}{2\text{Var}\left(s\right)}}}{\sqrt{2\pi}}>0
\end{align*}
Hence, there exists a unique $\alpha\in\left(0,1/b\right)$ for which
$\int_{D_{\alpha}}\left[1-\frac{\omega\left(\omega+b\right)}{\text{Var}\left(s\right)}\right]\phi\left(\omega/\sqrt{\text{Var}\left(s\right)}\right)d\omega=0$.
Since $\underline{\omega},\overline{\omega}$ satisfy (\ref{eq: omegl})
and (\ref{eq:omegah}), their sum is $\frac{1-\alpha b}{\alpha}>0$.
Moreover, since $\Delta=\left(1-b\alpha\right)^{2}+4r\alpha+4\alpha^{2}\text{Var}\left(s\right)$,
we can sign $\underline{\omega},\overline{\omega}$ by signing $\Delta$.
The expression for $\Delta$ is a quadratic function of $\alpha$
whose discriminant is $\left(4r-2b\right)^{2}-4\left(b^{2}+4\text{Var}\left(s\right)\right)=16\left(r^{2}-rb-\text{Var}\left(s\right)\right)$.
When $b>r>\frac{b-\sqrt{b^{2}+4\text{Var}\left(s\right)}}{2}$, this
expression is always negative which means that for all values of $\alpha$,
$\Delta>0$. Since $\alpha<1/b$, we must have that $\overline{\omega}>0>\underline{\omega}$.
Finally, $\psi\left(-r,\alpha\right)=\alpha\left(\text{Var}\left(s\right)+rb-r^{2}\right)$.
When $b>r>\frac{b-\sqrt{b^{2}+4\text{Var}\left(s\right)}}{2}$, this
expression is positive which means that $\underline{\omega}<-r<\overline{\omega}$.

\textbf{Claim 4. }\textit{If $r\leq\frac{b-\sqrt{b^{2}+4\text{Var}\left(s\right)}}{2}$,
then $\exists\alpha>0$ such that $\int_{D_{\alpha}}\left[1-\frac{\omega\left(\omega+b\right)}{\text{Var}\left(s\right)}\right]\phi\left(\omega/\sqrt{\text{Var}\left(s\right)}\right)d\omega=0$.
For all such values of $\alpha$, $\underline{\omega}=\overline{\omega}$.}

The proof of this claim follows from the same argument as before together
with the fact that when $r<\frac{b-\sqrt{b^{2}+4\text{Var}\left(s\right)}}{2}$,
there is an $1/b>\alpha>0$ for which $\Delta=\left(1-b\alpha\right)^{2}+4r\alpha+4\alpha^{2}\text{Var}\left(s\right)=0$.
To see this, note that the values of $\alpha$ for which this holds
are given by
\[
\alpha=\frac{b-2r\pm2\sqrt{r^{2}-rb-\text{Var}\left(s\right)}}{b^{2}+4\text{Var}\left(s\right)}
\]
Since $r\leq\frac{b-\sqrt{b^{2}+4\text{Var}\left(s\right)}}{2}$,
we must have that $2r\leq b-\sqrt{b^{2}+4\text{Var}\left(s\right)}$
which implies that the lower root above is positive. Since as we have
argued before, $\int_{D_{\alpha}}\left[1-\frac{\omega\left(\omega+b\right)}{\text{Var}\left(s\right)}\right]\phi\left(\omega/\sqrt{\text{Var}\left(s\right)}\right)d\omega$
is strictly increasing in $\alpha$ whenever $\alpha>0$ and $\underline{\omega},\overline{\omega}$
are real and at the above values of $\alpha$, $\underline{\omega}=\overline{\omega}$
and $\int_{D_{\alpha}}\left[1-\frac{\omega\left(\omega+b\right)}{\text{Var}\left(s\right)}\right]\phi\left(\omega/\sqrt{\text{Var}\left(s\right)}\right)d\omega=0$,
this establishes the claim.
\end{proof}

\subsection{Extensions: Proofs\label{sec:Extensions:-Proofs}}

\paragraph{Heterogeneous Biases: Proof of Proposition \ref{prop:The-receiver-optimal}}
\begin{proof}
The fact that ICL is equivalent to \ref{prop:The-receiver-optimal}
follows a similar existence proof as of that of Theorem \ref{thm:A-recommendation-mechanism}.
Such an extension is possible because the number of bias types and
signal realizations is finite. Hence, we can apply the same technique
-- consider a linear operator for all signal realization in each
bias group and show the existence of a mechanism for any finite number
of senders.

Now, consider the problem of choosing $\sigma\left(\mathbf{h}_{1},\cdots,\mathbf{h}_{M}\right)$
to maximize the receiver's payoff $\mathbb{E}\left[\sigma\left(\mathbf{h}_{1},\cdots,\mathbf{h}_{M}\right)\left(\omega+r\right)\right]$
subject to the ICL requirements:
\begin{align*}
\mathbb{E}\left[\sigma\left(\mathbf{h}_{1},\cdots,\mathbf{h}_{M}\right)\left(\frac{h_{k,m}}{f_{k,m}}\left(\omega+b_{m}\right)-t_{k,m}\right)\right] & =0\\
\mathbb{E}\left[\sigma\left(\mathbf{h}_{1},\cdots,\mathbf{h}_{M}\right)\frac{h_{k,m}}{f_{k,m}}\right] & \geq\mathbb{E}\left[\sigma\left(\mathbf{h}_{1},\cdots,\mathbf{h}_{M}\right)\frac{h_{k-1,m}}{f_{k-1,m}}\right],k>1
\end{align*}
By multiplying the top condition by $f_{k,m}t_{k,m}$ and summing
over $k$ for a fixed $m$ and using the fact that $\sum_{k}f_{k,m}t^{2}_{k,m}=\eta$,
we arrive at 
\begin{equation}
\mathbb{E}\left[\sigma\left(\mathbf{h}_{1},\cdots,\mathbf{h}_{M}\right)\left(\omega_{m}\left(\omega+b_{m}\right)-\eta\right)\right]=0\label{eq: IChetagg}
\end{equation}
Moreover, since $t_{k,m}$ and $\mathbb{E}\left[\sigma\frac{h_{k,m}}{f_{k,m}}\right]$
are both increasing in $k$, their covariance has to be positive.
In other words,
\[
\sum_{k}f_{k,m}t_{k,m}\mathbb{E}\left[\sigma\frac{h_{k,m}}{f_{k,m}}\right]\geq\sum_{k}f_{k,m}t_{k,m}\sum_{k}f_{k,m}\mathbb{E}\left[\sigma\frac{h_{k,m}}{f_{k,m}}\right]
\]
The left hand side of the above is $\mathbb{E}\left[\sigma\omega_{m}\right]$
while the right hand side is 0. Hence, we must have that for all $m$,
$\mathbb{E}\left[\sigma\omega_{m}\right]\geq0$.

In other words, if a mechanism is ICL, then it must satisfy (\ref{eq: IChetagg})
and $\mathbb{E}\left[\sigma\omega_{m}\right]\geq0$. One can thus
focus on the relaxed problem of maximizing $\mathbb{E}\left[\sigma\cdot\left(\omega+r\right)\right]$
subject to (\ref{eq: IChetagg}) and $\mathbb{E}\left[\sigma\cdot\omega_{m}\right]\geq0$.
In this relaxed problem, all constraint only depend on $\left(\mathbf{h}_{1},\cdots,\mathbf{h}_{M}\right)$
via $\omega_{1},\cdots,\omega_{M}$ which implies that the solution
should only depend on $\omega_{1},\cdots,\omega_{M}$. Additionally,
$\sigma^{*}$ is a solution to this relaxed problem if and only if
multipliers $\lambda_{1},\cdots,\lambda_{M}$ associated with (\ref{eq: IChetagg})
and $\zeta_{m}\geq0$ associated with $\mathbb{E}\left[\sigma\cdot\omega_{m}\right]\geq0$
exists that satisfy the conditions provided in statement of the proposition.

It therefore remains to be shown that if $\sigma^{*}$ solves the
relaxed problem, then it is indeed ICL. To see this, Since $\mathbf{h}_{m}\sim\mathcal{N}\left(0,\Sigma_{m}\right)$
and $h_{k,m}$'s are independent across different bias groups, we
must have that
\[
\mathbb{E}\left[h_{k,m}|\omega_{1},\cdots,\omega_{M}\right]=\mathbb{E}\left[h_{k,m}|\omega_{m}\right]=\frac{\mathbb{E}\left[h_{k,m}\omega_{m}\right]}{\mathbb{E}\left[\omega^{2}_{m}\right]}\omega_{m}=\frac{f_{k,m}t_{k,m}}{\eta}\omega_{m}
\]
Therefore,
\begin{align*}
\mathbb{E}\left[\sigma^{*}\cdot\left(\frac{h_{k,m}}{f_{k,m}}\left(\omega+b_{m}\right)-t_{k,m}\right)\right] & =\mathbb{E}\left[\mathbb{E}\left[\sigma^{*}\cdot\frac{h_{k,m}}{f_{k,m}}\left(\omega+b_{m}\right)|\omega_{1},\cdots,\omega_{M}\right]-\sigma^{*}\cdot t_{k,m}\right]\\
 & =\mathbb{E}\left[\sigma^{*}\cdot\left(\omega+b_{m}\right)\mathbb{E}\left[\frac{h_{k,m}}{f_{k,m}}|\omega_{1},\cdots,\omega_{M}\right]-\sigma^{*}\cdot t_{k,m}\right]\\
 & =\mathbb{E}\left[\sigma^{*}\cdot\left(\omega+b_{m}\right)\frac{t_{k,m}}{\eta}\omega_{m}-\sigma^{*}\cdot t_{k,m}\right]\\
 & =\frac{t_{k,m}}{\eta}\mathbb{E}\left[\sigma^{*}\cdot\left(\omega+b_{m}\right)\omega_{m}-\eta\sigma^{*}\right]=0
\end{align*}
where in the above we have used the law of iterated expectations and
that $\sigma^{*}$ is the solution to the relaxed problem. Similarly,
\begin{align*}
\mathbb{E}\left[\sigma^{*}\cdot\frac{h_{k,m}}{f_{k,m}}\right] & =\mathbb{E}\left[\sigma^{*}\cdot\mathbb{E}\left[\frac{h_{k,m}}{f_{k,m}}|\omega_{1},\cdots,\omega_{M}\right]\right]\\
 & =\frac{t_{k,m}}{\eta}\mathbb{E}\left[\sigma^{*}\cdot\omega_{m}\right]
\end{align*}
Since $\sigma^{*}$ is the solution to the relaxed problem, it must
satisfy $\mathbb{E}\left[\sigma^{*}\cdot\omega_{m}\right]$ which
implies that the above is increasing in $k$. Therefore, $\sigma^{*}$
is ICL.
\end{proof}

\subsubsection{General Preferences: Proof of Proposition \ref{prop:A-recommendation-mechanism-gen}}
\begin{proof}
Similar to the first part of Theorem \ref{thm:A-recommendation-mechanism},
we can write
\begin{equation}
U^{N}_{l,k}=\sum_{\mathbf{s}_{-}\in S^{N-1}}\text{Pr}_{N-1}\left(\mathbf{s}_{-}\right)\sigma^{N}\left(\mathbf{h}^{N}\left(\mathbf{s}_{-}+\mathbf{e}_{l}\right)\right)u_{s}\left(\mathbf{h}^{N}\left(\mathbf{s}_{-}+\mathbf{e}_{k}\right)\right)\label{eq: Ulk-1}
\end{equation}
where in the above $\mathbf{e}_{k}$ is a $K$--dimensional vector
whose elements are 0 except for its $k$-th element which is 1

Consider a vector of realizations $\mathbf{s}_{-}\in S^{N-1}$ for
which $\mathbf{h}^{N}\left(\mathbf{s}_{-}+\mathbf{e}_{k}\right)=\mathbf{h}$.
In this case, the count of $s_{i}$'s which are of type $m$ is given
by $n_{m}=\sqrt{N}h_{m}+Nf_{m}$ with $\sqrt{N}h_{k}+Nf_{k}\geq1$.
Then probability of this occurring -- using the multi--nomial distribution
-- is given by
\begin{align*}
\text{Pr}_{N-1}\left(\mathbf{s}_{-}\right) & =\left(\begin{array}{c}
N-1\\
n_{1},\cdots,n_{k}-1,\cdots,n_{K}
\end{array}\right)f^{n_{1}}_{1}\cdots f^{n_{k}-1}_{k}\cdots f^{n_{K}}_{K}\\
 & =\left(\begin{array}{c}
N\\
n_{1},\cdots,n_{K}
\end{array}\right)f^{n_{1}}_{1}\cdots f^{n_{K}}_{K}\frac{n_{k}}{f_{k}N}\\
 & =\text{Pr}_{N}\left(\mathbf{s}_{-}+\mathbf{e}_{k}\right)\frac{h_{k}\sqrt{N}+Nf_{k}}{f_{k}N}
\end{align*}
Using this adjustment of the probabilities, we can write
\begin{align*}
U^{N}_{l,k}= & \sum_{\mathbf{s}_{-}\in S^{N-1}}\text{Pr}_{N-1}\left(\mathbf{s}_{-}\right)\sigma^{N}\left(\mathbf{h}^{N}\left(\mathbf{s}_{-}+\mathbf{e}_{l}\right)\right)u_{s}\left(\mathbf{h}^{N}\left(\mathbf{s}_{-}+\mathbf{e}_{k}\right)\right)\\
= & \sum_{\mathbf{s}_{-}\in S^{N-1}}\text{Pr}_{N-1}\left(\mathbf{s}_{-}\right)\sigma^{N}\left(\mathbf{h}^{N}\left(\mathbf{s}_{-}+\mathbf{e}_{l}\right)\right)u_{s}\left(\mathbf{h}^{N}\left(\mathbf{s}_{-}+\mathbf{e}_{l}\right)\right)+\\
 & \sum_{\mathbf{s}_{-}\in S^{N-1}}\text{Pr}_{N-1}\left(\mathbf{s}_{-}\right)\sigma^{N}\left(\mathbf{h}^{N}\left(\mathbf{s}_{-}+\mathbf{e}_{l}\right)\right)\left[u_{s}\left(\mathbf{h}^{N}\left(\mathbf{s}_{-}+\mathbf{e}_{k}\right)\right)-u_{s}\left(\mathbf{h}^{N}\left(\mathbf{s}_{-}+\mathbf{e}_{l}\right)\right)\right]\\
 & =U^{N}_{l,l}+\sum_{\mathbf{s}_{-}\in S^{N-1}}\text{Pr}_{N-1}\left(\mathbf{s}_{-}\right)\sigma^{N}\left(\mathbf{h}^{N}\left(\mathbf{s}_{-}+\mathbf{e}_{l}\right)\right)\left[u_{s}\left(\mathbf{h}^{N}\left(\mathbf{s}_{-}+\mathbf{e}_{k}\right)\right)-u_{s}\left(\mathbf{h}^{N}\left(\mathbf{s}_{-}+\mathbf{e}_{l}\right)\right)\right]
\end{align*}
Let $\mu^{N}\left(\mathbf{h}\right)$ be the probability of the adjusted
frequencies being equal to $\mathbf{h}$. If $\mathbf{h}=\mathbf{h}^{N}\left(\mathbf{s}\right)$,
then it must be that $\mu^{N}\left(\mathbf{h}\right)=\text{Pr}_{N}\left(\mathbf{s}\right)$.
Let us also define $H^{N}\subset\mathbb{R}^{K}_{0}$ to be the support
of $\mu^{N}$. We can thus write the above as 
\begin{align*}
U^{N}_{l,k}= & U^{N}_{l,l}+\sum_{\mathbf{s}_{-}\in S^{N-1}}\text{Pr}_{N-1}\left(\mathbf{s}_{-}\right)\sigma^{N}\left(\mathbf{h}^{N}\left(\mathbf{s}_{-}+\mathbf{e}_{l}\right)\right)\left[u_{s}\left(\mathbf{h}^{N}\left(\mathbf{s}_{-}+\mathbf{e}_{k}\right)\right)-u_{s}\left(\mathbf{h}^{N}\left(\mathbf{s}_{-}+\mathbf{e}_{l}\right)\right)\right]\\
= & U^{N}_{l,l}+\sum_{\mathbf{s}_{-}\in S^{N-1}}\text{Pr}_{N}\left(\mathbf{s}_{-}+\mathbf{e}_{l}\right)\frac{h^{N}_{l}\left(\mathbf{s}_{-}+\mathbf{e}_{l}\right)\sqrt{N}+Nf_{l}}{f_{l}N}\sigma^{N}\left(\mathbf{h}^{N}\left(\mathbf{s}_{-}+\mathbf{e}_{l}\right)\right)\times\\
 & \qquad\qquad\qquad\qquad\qquad\left[u_{s}\left(\mathbf{h}^{N}\left(\mathbf{s}_{-}+\mathbf{e}_{k}\right)\right)-u_{s}\left(\mathbf{h}^{N}\left(\mathbf{s}_{-}+\mathbf{e}_{l}\right)\right)\right]\\
= & U^{N}_{l,l}+\sum_{\mathbf{h}\in H^{N},h_{l}\geq\frac{1-f_{l}\sqrt{N}}{N}}\mu^{N}\left(\mathbf{h}\right)\sigma^{N}\left(\mathbf{h}\right)\frac{h_{l}\sqrt{N}+Nf_{l}}{f_{l}N}\left[u_{s}\left(\mathbf{h}+\frac{\mathbf{e}_{k}-\mathbf{e}_{l}}{\sqrt{N}}\right)-u_{s}\left(\mathbf{h}\right)\right]\\
= & U^{N}_{l,l}+\sum_{\mathbf{h}\in H^{N},h_{l}\geq\frac{1-f_{l}\sqrt{N}}{N}}\mu^{N}\left(\mathbf{h}\right)\sigma^{N}\left(\mathbf{h}\right)\frac{h_{l}\sqrt{N}+Nf_{l}}{f_{l}N}\left[u_{s}\left(\mathbf{h}+\frac{\mathbf{e}_{k}-\mathbf{e}_{l}}{\sqrt{N}}\right)-u_{s}\left(\mathbf{h}\right)\right]+\\
 & \sum_{\mathbf{h}\in H^{N},h_{l}=-f_{l}\sqrt{N}}\mu^{N}\left(\mathbf{h}\right)\sigma^{N}\left(\mathbf{h}\right)\frac{h_{l}\sqrt{N}+Nf_{l}}{f_{l}N}\left[u_{s}\left(\mathbf{h}+\frac{\mathbf{e}_{k}-\mathbf{e}_{l}}{\sqrt{N}}\right)-u_{s}\left(\mathbf{h}\right)\right]
\end{align*}
If we use $\mathbb{E}^{N}$ to refer to the expectation with respect
to $\mu^{N}$, we can write the incentive constraints as
\begin{align}
U^{N}_{k,k}-U^{N}_{l,l}\geq\mathbb{E}^{N}\left[\sigma^{N}\left(\mathbf{h}\right)\frac{h_{l}\sqrt{N}+f_{l}N}{f_{l}N}\left(u_{s}\left(\mathbf{h}+\frac{\mathbf{e}_{k}-\mathbf{e}_{l}}{\sqrt{N}}\right)-u_{s}\left(\mathbf{h}\right)\right)\right]\label{eq: ICgen}
\end{align}
As $N$ tends to infinity, $\sqrt{N}\left(u_{s}\left(\mathbf{h}+\frac{\mathbf{e}_{k}-\mathbf{e}_{l}}{\sqrt{N}}\right)-u_{s}\left(\mathbf{h}\right)\right)$
converges to $\frac{\partial u_{s}}{\partial h_{k}}-\frac{\partial u_{s}}{\partial h_{l}}$.
Similar to the specific case of section \ref{sec:The-Value-of_mediator},
we have
\begin{align*}
\sqrt{N}\left(U^{N}_{k,k}-U^{N}_{l,l}\right) & =\sqrt{N}\mathbb{E}^{N}\left[\sigma\left(\mathbf{h}\right)\left(\frac{h_{k}\sqrt{N}+f_{k}N}{f_{k}N}-\frac{h_{l}\sqrt{N}+f_{l}N}{f_{l}N}\right)u_{s}\left(\mathbf{h}\right)\right]\\
 & =\mathbb{E}^{N}\left[\sigma\left(\mathbf{h}\right)\left(\frac{h_{k}}{f_{k}}-\frac{h_{l}}{f_{l}}\right)u_{s}\left(\mathbf{h}\right)\right]
\end{align*}
Now taking a limit in (\ref{eq: ICgen}) gives us
\[
\mathbb{E}\left[\sigma\left(\mathbf{h}\right)u_{s}\left(\mathbf{h}\right)\left(\frac{h_{k}}{f_{k}}-\frac{h_{l}}{f_{l}}\right)\right]\geq\mathbb{E}\left[\sigma\left(\mathbf{h}\right)\left(\frac{\partial u_{s}\left(\mathbf{h}\right)}{\partial h_{k}}-\frac{\partial u_{s}\left(\mathbf{h}\right)}{\partial h_{l}}\right)\right]
\]
Since this has to hold for all $k,l$, it should hold with equality
and thus
\[
\mathbb{E}\left[\sigma\left(\mathbf{h}\right)\left(u_{s}\left(\mathbf{h}\right)\frac{h_{k}}{f_{k}}-\frac{\partial u_{s}\left(\mathbf{h}\right)}{\partial h_{k}}\right)\right]=\overline{U}
\]
Similar to before, we also need that when $k>l$,
\begin{align*}
 & \mathbb{E}^{N}\left[\sigma^{N}\left(\mathbf{h}\right)\left(\frac{h_{k}}{f_{k}\sqrt{N}}+1\right)\left(u_{s}\left(\mathbf{h}\right)-u_{s}\left(\mathbf{h}+\frac{\mathbf{e}_{l}-\mathbf{e}_{k}}{\sqrt{N}}\right)\right)\right]\geq\\
 & \qquad\qquad\mathbb{E}^{N}\left[\sigma^{N}\left(\mathbf{h}\right)\left(\frac{h_{l}}{f_{l}\sqrt{N}}+1\right)\left(u_{s}\left(\mathbf{h}+\frac{\mathbf{e}_{k}-\mathbf{e}_{l}}{\sqrt{N}}\right)-u_{s}\left(\mathbf{h}\right)\right)\right]
\end{align*}
Using Taylor's formula, we can write
\begin{align*}
u_{s}\left(\mathbf{h}+\frac{\mathbf{e}_{l}-\mathbf{e}_{k}}{\sqrt{N}}\right)-u_{s}\left(\mathbf{h}\right) & =\frac{\mathbf{e}_{l}-\mathbf{e}_{k}}{\sqrt{N}}\nabla u_{s}\left(\mathbf{h}\right)^{\text{tr}}+\frac{1}{2}\frac{\mathbf{e}_{l}-\mathbf{e}_{k}}{\sqrt{N}}\nabla^{2}u_{s}\left(\mathbf{h}\right)\frac{\mathbf{e}^{\text{tr}}_{l}-\mathbf{e}^{\text{tr}}_{k}}{\sqrt{N}}+O\left(\frac{1}{N\sqrt{N}}\right)\\
u_{s}\left(\mathbf{h}+\frac{\mathbf{e}_{k}-\mathbf{e}_{l}}{\sqrt{N}}\right)-u_{s}\left(\mathbf{h}\right) & =\frac{\mathbf{e}_{k}-\mathbf{e}_{l}}{\sqrt{N}}\nabla u_{s}\left(\mathbf{h}\right)^{\text{tr}}+\frac{1}{2}\frac{\mathbf{e}_{k}-\mathbf{e}_{l}}{\sqrt{N}}\nabla^{2}u_{s}\left(\mathbf{h}\right)\frac{\mathbf{e}^{\text{tr}}_{k}-\mathbf{e}^{\text{tr}}_{l}}{\sqrt{N}}+O\left(\frac{1}{N\sqrt{N}}\right)
\end{align*}
Thus the above inequality becomes
\begin{align*}
\mathbb{E}^{N}\left[\sigma^{N}\left(\mathbf{h}\right)\left(\frac{h_{k}}{f_{k}\sqrt{N}}+1\right)\left(\frac{\mathbf{e}_{k}-\mathbf{e}_{l}}{\sqrt{N}}\nabla u_{s}\left(\mathbf{h}\right)^{\text{tr}}-\frac{1}{2}\frac{\mathbf{e}_{k}-\mathbf{e}_{l}}{\sqrt{N}}\nabla^{2}u_{s}\left(\mathbf{h}\right)\frac{\mathbf{e}^{\text{tr}}_{k}-\mathbf{e}^{\text{tr}}_{l}}{\sqrt{N}}+O\left(\frac{1}{N\sqrt{N}}\right)\right)\right] & \geq\\
\mathbb{E}^{N}\left[\sigma^{N}\left(\mathbf{h}\right)\left(\frac{h_{l}}{f_{l}\sqrt{N}}+1\right)\left(\frac{\mathbf{e}_{k}-\mathbf{e}_{l}}{\sqrt{N}}\nabla u_{s}\left(\mathbf{h}\right)^{\text{tr}}+\frac{1}{2}\frac{\mathbf{e}_{k}-\mathbf{e}_{l}}{\sqrt{N}}\nabla^{2}u_{s}\left(\mathbf{h}\right)\frac{\mathbf{e}^{\text{tr}}_{k}-\mathbf{e}^{\text{tr}}_{l}}{\sqrt{N}}+O\left(\frac{1}{N\sqrt{N}}\right)\right)\right]
\end{align*}
We thus have
\begin{align*}
\frac{1}{N}\mathbb{E}^{N}\left[\sigma^{N}\left(\mathbf{h}\right)\left(\frac{h_{k}}{f_{k}}\left(\mathbf{e}_{k}-\mathbf{e}_{l}\right)\nabla u_{s}\left(\mathbf{h}\right)^{\text{tr}}-\frac{1}{2}\left(\mathbf{e}_{k}-\mathbf{e}_{l}\right)\nabla^{2}u_{s}\left(\mathbf{h}\right)\left(\mathbf{e}^{\text{tr}}_{k}-\mathbf{e}^{\text{tr}}_{l}\right)\right)\right]+O\left(\frac{1}{N\sqrt{N}}\right) & \geq\\
\frac{1}{N}\mathbb{E}^{N}\left[\sigma^{N}\left(\mathbf{h}\right)\left(\frac{h_{l}}{f_{l}}\left(\mathbf{e}_{k}-\mathbf{e}_{l}\right)\nabla u_{s}\left(\mathbf{h}\right)^{\text{tr}}+\frac{1}{2}\left(\mathbf{e}_{k}-\mathbf{e}_{l}\right)\nabla^{2}u_{s}\left(\mathbf{h}\right)\left(\mathbf{e}^{\text{tr}}_{k}-\mathbf{e}^{\text{tr}}_{l}\right)\right)\right]+O\left(\frac{1}{N\sqrt{N}}\right)
\end{align*}
Hence, as $N$ converges to infinity, the above multiplied by $N$
converges to
\begin{align*}
\mathbb{E}\left[\sigma\left(\mathbf{h}\right)\left(\frac{h_{k}}{f_{k}}-\frac{h_{l}}{f_{l}}\right)\left(\frac{\partial u_{s}\left(\mathbf{h}\right)}{\partial h_{k}}-\frac{\partial u_{s}\left(\mathbf{h}\right)}{\partial h_{l}}\right)\right] & \geq\\
\mathbb{E}\left[\sigma\left(\mathbf{h}\right)\left(\frac{\partial^{2}u_{s}\left(\mathbf{h}\right)}{\partial h^{2}_{k}}+\frac{\partial^{2}u_{s}\left(\mathbf{h}\right)}{\partial h^{2}_{l}}-2\frac{\partial^{2}u_{s}\left(\mathbf{h}\right)}{\partial h_{l}\partial h_{k}}\right)\right]
\end{align*}
The only if proof follows closely that of Theorem \ref{thm:A-recommendation-mechanism}.
Specifically, we use the power bound from Assumption \ref{assu:The-payoff-functions}
and apply the uniform convergence used in proof
of Theorem \ref{thm:A-recommendation-mechanism}. This concludes the proof.
\end{proof}

\subsection{Proof of Proposition \ref{prop:Suppose-that-}}
\begin{proof}
If we apply the characterization result of Proposition \ref{prop:A-recommendation-mechanism-gen}
to this setup, it implies that a mechanism $\sigma\left(\mathbf{h}\right)$
is ICL if and only if it satisfies
\begin{align}
\mathbb{E}\left[\sigma\left(\mathbf{h}\right)\left(\mathbf{t}_{s}\cdot\mathbf{h}\frac{h_{k}}{f_{k}}-t_{s,k}\right)\right] & =U,\forall k\label{eq: ICLhetlin}\\
\mathbb{E}\left[\sigma\left(\mathbf{h}\right)\frac{h_{k}}{f_{k}}\right] & \geq\mathbb{E}\left[\sigma\left(\mathbf{h}\right)\frac{h_{l}}{f_{l}}\right],\forall k>l\label{eq:monhetlin}
\end{align}
The fact that $\sum_{k}f_{k}t_{s,k}=0$ implies that $U=0$.

The rest of the proof closely follows that of Proposition \ref{prop:The-receiver-optimal}.
Namely, if $\sigma$ is ICL, i.e., satisfies (\ref{eq: ICLhetlin})
and (\ref{eq:monhetlin}), then by multiplying (\ref{eq: ICLhetlin})
by $f_{k}t_{s,k}$ and summing over $k$, we arrive at
\begin{equation}
\mathbb{E}\left[\sigma\left(\mathbf{h}\right)\left(\omega^{2}_{s}-1\right)\right]=0\label{eq:ICLagggen}
\end{equation}
where $\omega_{s}=\mathbf{t}_{s}\cdot\mathbf{h}$. Similarly, by repeating
this for $f_{k}t_{r,k}$ , we arrive at 
\begin{equation}
\mathbb{E}\left[\sigma\left(\mathbf{h}\right)\left(\omega_{s}\omega_{r}-\gamma\right)\right]=0\label{eq:ICLagggen2}
\end{equation}
Finally, note that properties of the normal distribution implies that
\begin{align*}
\mathbb{E}\left[h_{k}|\omega_{s},\omega_{r}\right] & =\left[\begin{array}{cc}
\omega_{s} & \omega_{r}\end{array}\right]\left[\begin{array}{cc}
\mathbb{E}\left[\omega^{2}_{s}\right] & \mathbb{E}\left[\omega_{s}\omega_{r}\right]\\
\mathbb{E}\left[\omega_{s}\omega_{r}\right] & \mathbb{E}\left[\omega^{2}_{r}\right]
\end{array}\right]^{-1}\left[\begin{array}{c}
\mathbb{E}\left[\omega_{s}h_{k}\right]\\
\mathbb{E}\left[\omega_{r}h_{k}\right]
\end{array}\right]
\end{align*}
where
\begin{align*}
\mathbb{E}\left[\omega^{2}_{s}\right] & =\sum_{k,l}\mathbb{E}\left[t_{s,k}t_{s,l}h_{k}h_{l}\right]=\sum_{k,l}t_{s,k}t_{s,l}f_{k}\left(\mathbf{1}\left[k=l\right]-f_{l}\right)\\
 & =\sum_{k}f_{k}t^{2}_{s,k}=1\\
\mathbb{E}\left[\omega^{2}_{r}\right] & =1,\mathbb{E}\left[\omega_{r}\omega_{s}\right]=\sum_{k}f_{k}t_{r,k}t_{s,k}=\gamma\\
\mathbb{E}\left[\omega_{s}h_{k}\right] & =f_{k}t_{s,k},\mathbb{E}\left[\omega_{r}h_{k}\right]=f_{k}t_{r,k}
\end{align*}
and hence
\begin{align*}
\mathbb{E}\left[h_{k}|\omega_{s},\omega_{r}\right] & =f_{k}\left[\begin{array}{cc}
\omega_{s} & \omega_{r}\end{array}\right]\left[\begin{array}{cc}
1 & \gamma\\
\gamma & 1
\end{array}\right]^{-1}\left[\begin{array}{c}
t_{s,k}\\
t_{r,k}
\end{array}\right]\\
 & =\frac{f_{k}}{1-\gamma^{2}}\left[\begin{array}{cc}
\omega_{s} & \omega_{r}\end{array}\right]\left[\begin{array}{cc}
1 & -\gamma\\
-\gamma & 1
\end{array}\right]\left[\begin{array}{c}
t_{s,k}\\
t_{r,k}
\end{array}\right]\\
 & =\frac{f_{k}}{1-\gamma^{2}}\left[\left(t_{s,k}-\gamma t_{r,k}\right)\omega_{s}+\left(t_{r,k}-\gamma t_{s,k}\right)\omega_{r}\right]
\end{align*}
Replacing the above in the (\ref{eq:monhetlin}) leads to
\begin{equation}
\mathbb{E}\left[\sigma\left(\mathbf{h}\right)\left(\left(t_{s,k}-\gamma t_{r,k}\right)\omega_{s}+\left(t_{r,k}-\gamma t_{s,k}\right)\omega_{r}\right)\right]\text{: increasing in \ensuremath{k}}\label{eq: monagggen}
\end{equation}
Hence, similar to the proof of Proposition \ref{prop:The-receiver-optimal},
we can focus on a relaxed problem where instead of ICL we impose (\ref{eq:ICLagggen}),
(\ref{eq:ICLagggen2}), and (\ref{eq: monagggen}). The solution of
the relaxed problem should then satisfy the conditions provided in
proposition \ref{prop:Suppose-that-}. Finally, we must show that
the solution of the relaxed problem is indeed ICL. This follows steps
similar to those in proof of Proposition \ref{prop:The-receiver-optimal}.
\end{proof}

\section{Numerical Solution of Small Economy with Large Bias\label{sec:Numerical-Solution-of}}

\begin{figure}
\begin{centering}
\includegraphics[scale=0.15]{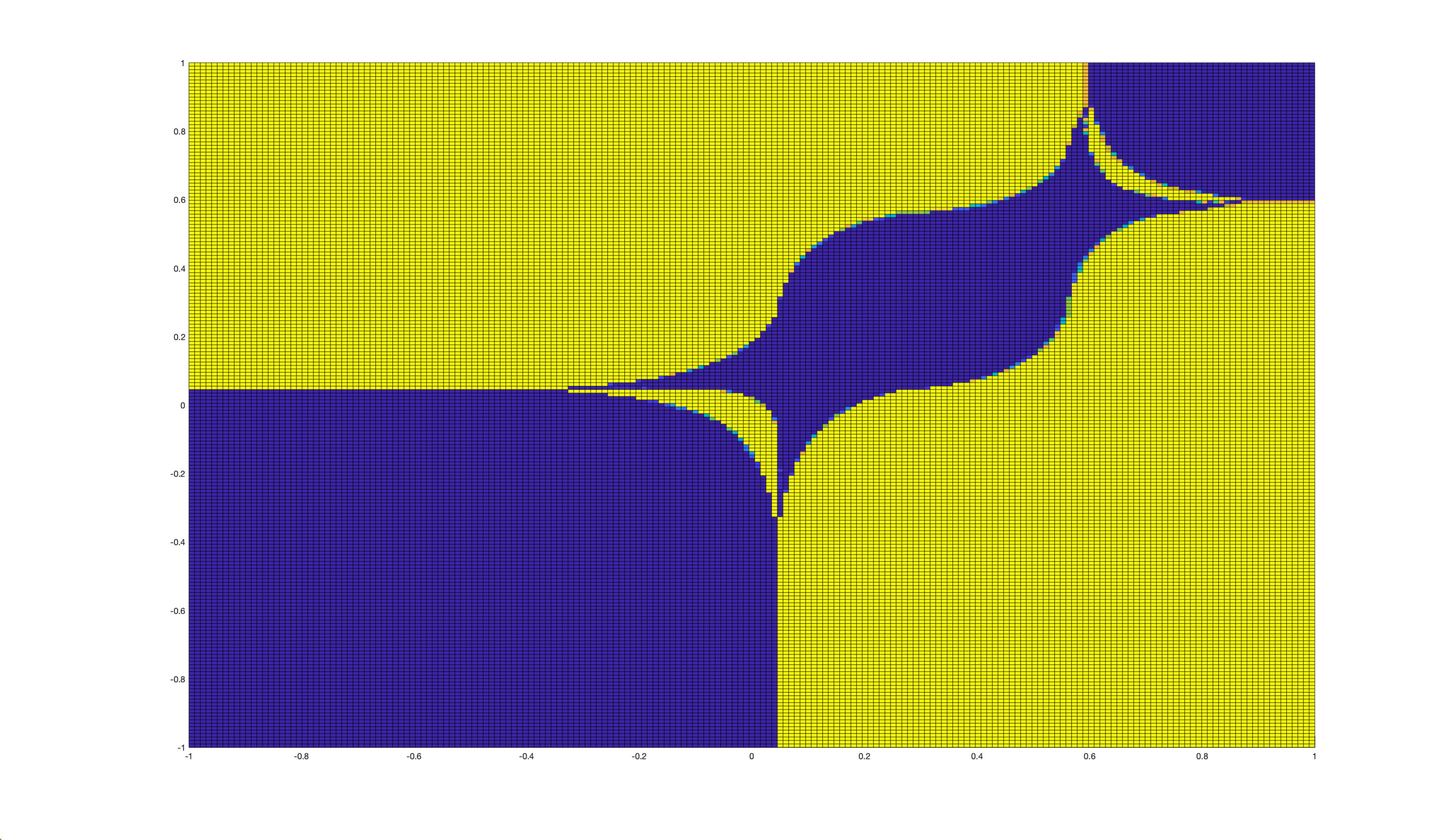}
\par\end{centering}
\caption{Numerical simulation of the optimal mechanism \label{fig:Numerical-simulation-of}}
\end{figure}

The problem of finding the optimal mechanism, (\ref{eq:P}), is a
linear program. Here we plot the numerical solution of this linear
program when $N=2$, $b=0.6\sqrt{2}$, $s_{i}\sim U\left[-1,1\right]$,
and $r=0$. For the numerical solution, we discretize $\left[-1,1\right]$
into 200 subintervals and solve for optimal $\sigma$ for each square
generated by all such subintervals, i.e., we solve for a vector of
40000 values. We use the \texttt{linprog} function in MATLAB to solve
this linear program. Figure \ref{fig:Numerical-simulation-of} depicts
the solution. The yellow area represents the values for which $\sigma=1$
and the blue area is associated with $\sigma=0$.

\end{document}